\documentclass[usenatbib]{mn2e}
\usepackage{psfig,graphicx,color}

\catcode`\@=11
\def\gsim{\ifmmode{\,\mathrel{\mathpalette\@versim>\,}}
    \else{$\,\mathrel{\mathpalette\@versim>}\,$}\fi}
\def\lsim{\ifmmode{\,\mathrel{\mathpalette\@versim<\,}}
    \else{$\,\mathrel{\mathpalette\@versim<}\,$}\fi}
\def\@versim#1#2{\lower 2.9truept \vbox{\baselineskip 0pt \lineskip
    0.5truept \ialign{$\m@th#1\hfil##\hfil$\crcr#2\crcr\sim\crcr}}}
\catcode`\@=12  
\def\km{{\rm \,km}}
\def\Sersic{{S\'ersic }}
\def\sminus{{\rm \,s^{-1}}}
\def\Mpcminus{{\rm \,Mpc^{-1}}}
\def\kpcminust{{\rm \,kpc^{-3}}}
\def\kpc{{\rm \,kpc}}

\def\gr{g_r}
\def\Conc{C_{200}}
\def\gammastar{\gamma_*}

\def\Mstar{M_*}

\def\LV{L_{\rm V}}
\def\LVsun{L_{{\rm V}\odot}}

\def\Msun{M_{\odot}}
\def\Metdim{M^{\rm dim}_{\rm e2}}
\def\Mstarp{M_{\rm *}^{\rm p}}
\def\Mp{M^{\rm p}}
\def\Mtotp{M_{\rm tot}^{\rm p}}
\def\Metp{M_{\rm e2}^{\rm p}}
\def\Mtot{M_{\rm tot}}

\def\Mdm{M_{\rm dm}}
\def\fdm{f_{\rm dm}}
\def\fdmp{f_{\rm dm}^{\rm p}}

\def\rhostar{\rho_*}
\def\rhocrittilde{\tilde{\rho}_{\rm crit}}
\def\rhocrit{\rho_{\rm crit}}
\def\varrhostar{\varrho_*}
\def\rhotot{\rho_{\rm tot}}
\def\rhodm{\rho_{\rm dm}}

\def\Ie{\langle I\rangle _{\rm e}}
\def\Re{R_{\rm e}}
\def\Resq{R_{\rm e}^2}
\def\Retilde{\tilde{R}_{\rm e}}

\def\rs{r_{\rm s}}
\def\rvir{r_{\rm vir}}

\def\rstar{r_{*}}
\def\fstar{f_*}

\def\ra{r_{\rm a}}
\def\am{a_{\rm m}}
\def\bm{b_{\rm m}}

\def\sgee{\sigma_{\rm e8}}
\def\sgeesq{\sigma_{\rm e8}^2}
\def\sgSIS{\sigma_{SIS}}

\def\sget{\sigma_{\rm e2}}
\def\sgetsq{\sigma_{\rm e2}^2}

\def\cet{c_{\rm e2}}
\def\Sget{\Sigma_{\rm e2}}
\def\sgr{\sigma_r}
\def\sgrsq{\sigma_r^2}

\def\sgpsq{\sigma_{\rm p}^2}
\def\sgasq{\sigma_{\rm a}^2}
\def\sgvsq{\sigma_{\rm v}^2}
\def\sgth{\sigma_{\vartheta}}
\def\sgthsq{\sigma_{\vartheta}^2}
\def\sgph{\sigma_{\varphi}}
\def\sgphsq{\sigma_{\varphi}^2}
\def\Sgstar{\Sigma_*}
\def\Sgstarz{\Sigma_{*,0}}
\def\Sgtot{\Sigma_{\rm tot}}
\def\gm{\gamma}

\def\const{\mathit{const}}

\title[The Mass Plane of early-type galaxies] 
{Mass distribution and orbital anisotropy of early-type galaxies:
  constraints from the Mass Plane}

\author[C. Nipoti, T. Treu,
A. S. Bolton]{C.~Nipoti$^1$\thanks{email:carlo.nipoti@unibo.it},
T.~Treu$^2$, A.~S.~Bolton$^3$\\ $^{1}$Dipartimento di Astronomia,
Universit\`a di Bologna, via Ranzani 1, 40127 Bologna, Italy\\
$^2$Department of Physics, University of California, Santa Barbara, CA
93106-9530, USA; Sloan Fellow, Packard Fellow\\ $^3$Institute for
Astronomy, University of Hawaii, 2680 Woodlawn Dr., Honolulu, HI 96822
USA; B. W. Parrent Fellow}

\date{Accepted 2008 July 22.  Received 2008 July 21; in original form 2008 April 16}

\pagerange{\pageref{firstpage}--\pageref{lastpage}}
\pubyear{2008}

\begin{document}

\maketitle

\label{firstpage}

\begin{abstract}
Massive early-type galaxies are observed to lie on the Mass Plane
(MP), a two-dimensional manifold in the space of effective radius
$\Re$, projected mass $\Metp$ (measured via strong gravitational
lensing) and projected stellar velocity dispersion $\sget$ within
$\Re/2$.  The MP is less `tilted' than the traditional Fundamental
Plane, and the two have comparable associated scatter. This means that
the dimensionless structure parameter $\cet=2G\Metp/(\Re\sgetsq)$ is a
nearly universal constant in the range $\sget=175-400$ km
s$^{-1}$. This finding can be used to constrain the mass distribution
and internal dynamics of early-type galaxies: in particular, we
explore the dependence of $\cet$ on light profile, dark-matter
distribution, and orbital anisotropy for several families of spherical
galaxy models.  We find that a relatively wide class of models has
values of $\cet$ in the observed range, because $\cet$ is not very
strongly sensitive to the mass distribution and orbital anisotropy.
The degree of fine tuning required to match the small intrinsic
scatter of $\cet$ depends on the considered family of models: if the
total mass distribution is isothermal ($\propto r^{-2}$), a broad
range of stellar luminosity profile and anisotropy is consistent with
the observations, while Navarro, Frenk \& White dark-matter halos
require more fine tuning of the stellar mass fraction, luminosity
profile and anisotropy. 
If future data can cover a broader range of masses, the MP could be
seen to be tilted by the known non-homology of the luminosity profiles
of early-type galaxies, and the value of any such tilt would provide a
discriminant between models for the total mass-density profile of the
galaxies.
\end{abstract}


\section{Introduction}

The origin of empirical scaling laws is a key open issue in
observational cosmology. Galaxies do not come in all sizes, shapes,
colours, but rather tend to to live in lower-dimensional manifolds, which
represent a stringent testing ground for theories of galaxy formation
and evolution.

Early-type galaxies obey a particularly tight scaling law: the
so-called Fundamental Plane \cite[FP;][]{Djo87,Dre87}.  In the space
of effective radius $\Re$, central velocity dispersion $\sgee$ and
effective surface brightness $\Ie{\equiv}L/(2\pi\Resq)$ (where $L$ is
the total luminosity of the galaxy), they lie on the following
relation with remarkably small scatter \citep[$\lsim 20 \%$ in
$\Re$;][]{Ber03b}:
\begin{equation}
\log\Re=a\log\sgee+b\log\Ie+\const,
\label{eqFP}
\end{equation}
where the numerical value of $a$ and $b$ depends somewhat upon the
wavelength of observations and upon the sample and the fitting method
\citep{Pah98,Ber03b}. The FP is said to be `tilted', in the sense that
the coefficients $a$ and $b$ differ significantly from the values
$a=2$ and $b=-1$ expected {\it for structurally and dynamically
homologous systems with luminosity-independent stellar mass-to-light
ratio and dark-matter distribution}. Several explanations have been
proposed for the tilt, including a systematic dependence of stellar
mass-to-light ratio or dark-matter content and distribution upon
luminosity (and hence presumably upon mass), structural non-homology
and orbital anisotropy
\citep[e.g.][]{Fab87,Ben92,RenC93,CioLR96,Bor03}.

Recently, \cite{Bol07,Bol08b}, using a sample of strong gravitational
lenses, have shown that early-type galaxies lie on a Mass Plane (MP)
\begin{equation}
\log\Re=\am\log\sget+\bm\log\Sget+\const,
\label{eqMP}
\end{equation}
where $\sget$ is the projected velocity dispersion within an aperture
radius $\Re/2$ and $\Sget$ is the surface mass density within $\Re/2$,
with $\am=1.82 \pm 0.19$, $\bm=-1.20 \pm 0.12$ and RMS orthogonal
scatter of 1.24 when normalized by the observational errors.  The fact
that $(\am,\bm)$ are close to $(2,-1)$ and that the scatter is small
can be expressed in terms of structural and dynamical homology of the
lenses, by defining the dimensionless structure parameter
\begin{equation}
\cet\equiv{2G\Metp\over \Re\sget^2},
\label{eqcet}
\end{equation}
where $\Metp$ is the total projected mass within $\Re/2$.  For their
sample of lens early-type galaxies from the Sloan Lens ACS (SLACS)
Survey \cite{Bol08b} find on average
\begin{equation}
\langle\log \cet\rangle=0.53\pm0.057,
\label{eqobsrange}
\end{equation}
which throughout the paper we will refer to as the ``observed range''
of $\cet$. We note that the observed scatter on $\langle\log
\cet\rangle$ is 0.08, but here we consider the estimated intrinsic
scatter 0.057 \citep[see][]{Bol08b}.

As discussed in several papers \citep{Bol06,Tre06,Bol08a,Tre08} the
SLACS lenses are found to be indistinguishable from control samples of
Sloan Digital Sky Survey (SDSS) galaxies with the same stellar
velocity dispersion and size, in terms of luminosity/surface
brightness, location on the FP, and environment.  This inspires some
confidence that the results found for the lens sample, including the
MP, are generic properties of the overall class of early-type
galaxies.

Independent of its origin and theoretical interpretation, the
existence of the MP is a powerful empirical tool to estimate galaxy
mass by using information on size and velocity dispersion only
\citep{Bol07}. In addition, it is clear that the very existence of the
MP may be used to improve our understanding of galaxy formation and
evolution.  What is the origin of such a strong correlation among
measurable galaxy quantities? Or, in other words, what kinds of galaxy
models can be ruled out by the existence of a tight MP? Although this
question has been asked before in regards to the traditional FP, the
MP provides an additional powerful tool. In fact, there are a few
differences between the FP (equation~\ref{eqFP}) and the MP
(equation~\ref{eqMP}):

\begin{enumerate}

\item The FP is sensitive to the galaxy {\it stellar} mass-to-light
ratio, while the MP is not.  This implies that, e.g., the role of
stellar populations in establishing the tilt and scatter of the FP can
be disentangled by looking at the MP.\footnote{Strictly speaking the
MP depends on the properties of the stellar populations through $\Re$
and $\sget$. However, $\Re$ and $\sget$ do not depend on the value of
the stellar mass-to-light ratio, but only on its radial variation.
This variation is expected to be small based on observed colour
gradients and is generally neglected in dynamical studies
\citep[e.g.][]{Kro00,Cap06}. For simplicity, in this study we assume
uniform stellar mass-to-light ratios within each galaxy.}

\item The FP is traditionally based on the central velocity dispersion
$\sgee$ (measured within $\Re/8$), while the MP has been constructed
using $\sget$, which is measured within $\Re/2$.  This is a
consequence of the fixed spatial observing aperture of the SDSS
spectrograph; an MP based upon $\sgee$ could be constructed using
spatially resolved spectroscopy of the SLACS lens
sample.\footnote{In general, the
larger the aperture radius $R$ used to measure the aperture velocity
dispersion $\sgasq$, the less $\sgasq$ is sensitive to the orbital
anisotropy. We recall that for any stationary, non-rotating,
spherically symmetric system with constant mass-to-light ratio
$\sgasq(R)\to\sgvsq/3$ for $R\to\infty$, where $\sgvsq$ is the virial
velocity dispersion \citep[e.g.][]{Cio94}.}

\item The FP combines quantities evaluated on different scales
($\Re$, $\Re/8$), while MP combines quantities evaluated within the
same radius $\Re/2$.  Again, this is partially due
to the fixed SDSS spectroscopic aperture, though the
apertures of the lensing mass measurements are fixed by the
cosmic configuration of the individual strong-lens systems.

\end{enumerate}

Each of the points above can contribute to make the MP less tilted
(and presumably with less scatter) than the FP. For example, given the
relatively large spectroscopic aperture used to define $\cet$, we
expect it to be robust with respect to changes in the detailed
properties of galaxy structure, internal dynamics, and dark-matter
content. Similarly, by replacing surface brightness with surface mass
density we expect that tilt and scatter due to diversity of chemical
composition or star formation history be reduced in MP. Furthermore,
having all but removed the effects of stellar population the MP is
potentially a cleaner diagnostic than the FP of the structural and
dynamical properties of early-type galaxies.

In this paper, we exploit the existence of the MP to constrain
important properties of early-type galaxies, such as orbital
anisotropy and dark-matter distribution. We achieve this goal by
constructing observationally and cosmologically motivated families of
galaxy models and finding the range of parameter spaces consistent
with the observed range of $\cet$.  For the sake of simplicity, in the
present investigation we limit ourselves to spherically symmetric
models. As with the FP \citep{Fab87,Sag93,Pru94,Lan03,Ric05}, deviation from
spherical symmetry is expected to increase the scatter of the MP
because of projection effects. Thus, a natural follow-up of the
present work would be the extension to non-spherical models.

The paper is organized as follows. In Section~\ref{secmod} we describe
the models, in Sections~\ref{secres} and~\ref{secser} we present our
results, and in Section~\ref{seccon} we conclude.

\section{Models}
\label{secmod}

\subsection{Methodology and general properties}

We consider spherical galaxy models with stellar density distribution
$\rhostar(r)$ and total density distribution $\rhotot(r)$.  The radial
component $\sgr(r)$ of the velocity dispersion tensor is given by solving
the Jeans equation \citep[e.g.][]{Bin08}
\begin{equation}
{d \rhostar \sgrsq \over d r}+{2\beta\rhostar\sgrsq\over r}=-\rhostar\gr,
\end{equation}
where $\gr(r)=d\Phi(r)/d r$, $\Phi(r)$ is the total gravitational
potential generated by $\rhotot (r)$, and
\begin{equation}
\beta(r) \equiv 1 -{\sgthsq+\sgphsq\over 2\sgrsq}
\end{equation}
is the anisotropy parameter ($\sgth$ and $\sgph$ are, respectively,
the $\vartheta$ and $\varphi$ components of the velocity-dispersion
tensor).

The line-of-sight velocity dispersion is \citep{Bin82}
\begin{equation}
\sgpsq(R)={2\over\Sgstar(R)}\int_R^\infty \left[{1-\beta(r){R^2\over
r^2}}\right]{\rhostar(r)\sgrsq r dr\over\sqrt{r^2-R^2}},
\end{equation}
where
\begin{equation}
\Sgstar(R)=2\int_R^\infty {\rhostar(r)rdr\over\sqrt{r^2-R^2}}
\end{equation}
is the stellar surface density (we assume that the stellar
mass-to-light ratio is independent of radius). The aperture velocity
dispersion within a projected radius $R$, the closest analog to the
measured stellar velocity dispersion, is determined via
\begin{equation}
\sgasq(R)={2\pi\over\Mstarp(R)}\int_0^R \Sgstar(R')\sgpsq(R') R'\,dR',
\end{equation}
where
\begin{equation}
\Mstarp(R)=2\pi\int_0^R \Sgstar(R')R'\,dR'
\end{equation}
is the projected stellar mass within $R$. So,
$\sgeesq\equiv\sgasq(\Re/8)$ and $\sgetsq\equiv\sgasq(\Re/2)$.
Note that the mass weighting expressed here is equivalent to luminosity
weighting for the case of a spatially uniform stellar
mass-to-light ratio.

Gravitational lensing analysis allows one to measure the total
projected mass density within the Einstein radius.  The
total projected mass within a radius $R$ of a spherical system is
\begin{equation}
\Mtotp(R)=2\pi\int_0^R \Sgtot(R')R'\,dR',
\end{equation}
where 
\begin{equation}
\Sgtot(R)=2\int_R^\infty {\rhotot(r)rdr\over\sqrt{r^2-R^2}}
\end{equation}
is the total surface density. The mass within the Einstein radius,
measured by gravitational lensing is obtained by setting $R$ equal to
the Einstein radius. The size of the Einstein radius depends on the
geometry of the lensing system, through the angular diameter distances
between the observer, lens and background source, as well as on the
mass distribution of the lens. Typically, for galaxy-size lenses,
Einstein radii are of order of one arcsecond, or $\sim$5 kpc
for lenses at moderate redshift.  For the
SLACS lens sample, the Einstein radii are typically about half the
effective radius of the lens galaxy.

For a given projected radius $R$ we define the structure parameter
\begin{equation}
c(R)\equiv{G\Mtotp(R)\over R\sgasq(R)}.
\end{equation}
So, by definition (equation~\ref{eqcet}) $\cet=c(\Re/2)$, because
$\Metp\equiv\Mtotp(\Re/2)$.

\subsection{Stellar density distribution}

We consider two families of stellar density distributions---$\gamma$
models and \Sersic models---that are known to match well the observed
surface brightness profiles of early-type galaxies over the range of
interest $\sim$1-10 kpc for constant stellar mass-to-light ratios.
The density profile of the $\gm$-models \citep{Deh93,Tre94} is given
by
\begin{equation}
\rhostar (r)= {3-\gm\over 4\pi}{\Mstar\rstar \over r^{\gm} (\rstar+r)^{4-\gm}}  
\qquad  (0\leq \gm <3),
\end{equation}
%
%
where $\Mstar$ is the total stellar mass. The cases $\gamma=1$ and
$\gamma=2$ are the \citet{Her90} and \citet{Jaf83} models,
respectively.

In the case of the \Sersic models, the projected density distribution
follows the \cite{Ser68} $R^{1/m}$ law:
\begin{equation}
\Sgstar(R)=\Sgstarz\,\exp\left[-b(m)\left(\frac{R}{\Re}\right)^{1/m}\right]
\end{equation}
where $b(m)\sim 2m-1/3+4/(405m)$ \citep{CioB99}. For $m=4$ the de
Vaucouleurs' (1948) $R^{1/4}$ law is obtained.  By deprojecting
$\Sgstar$ one obtains the corresponding intrinsic density distribution
\citep{Bin08}
\begin{equation}
\rhostar(r)=-{1\over\pi}\int_r^\infty{d\Sgstar\over dR}{dR\over
    \sqrt{R^2-r^2}}.
\end{equation}

\subsection{Total  density distribution}
\label{sectot}

\begin{figure}
\centerline{\psfig{file=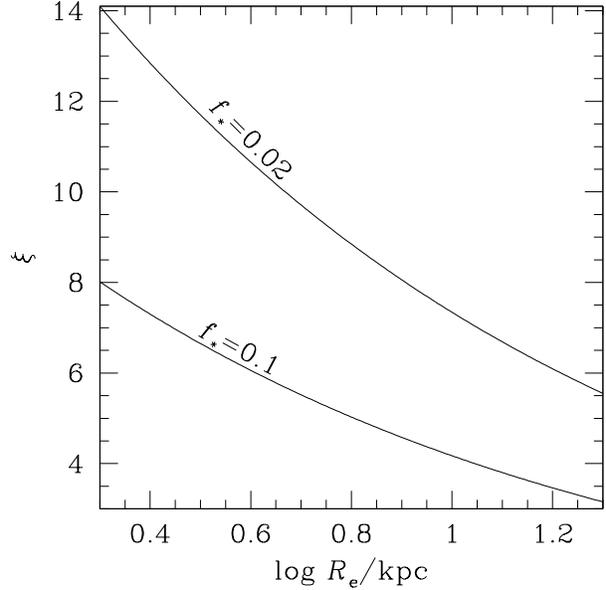,width=\hsize}}
\caption{The ratio $\xi=\rs/\Re$ as a function of the effective radius
  $\Re$ according to equation~(\ref{eqxi}) for NFW plus stars and adiabatically contracted NFW plus stars models (see text for details).}
\label{figrexi}
\end{figure}

\begin{figure}
\centering
\includegraphics[width=0.49\textwidth]{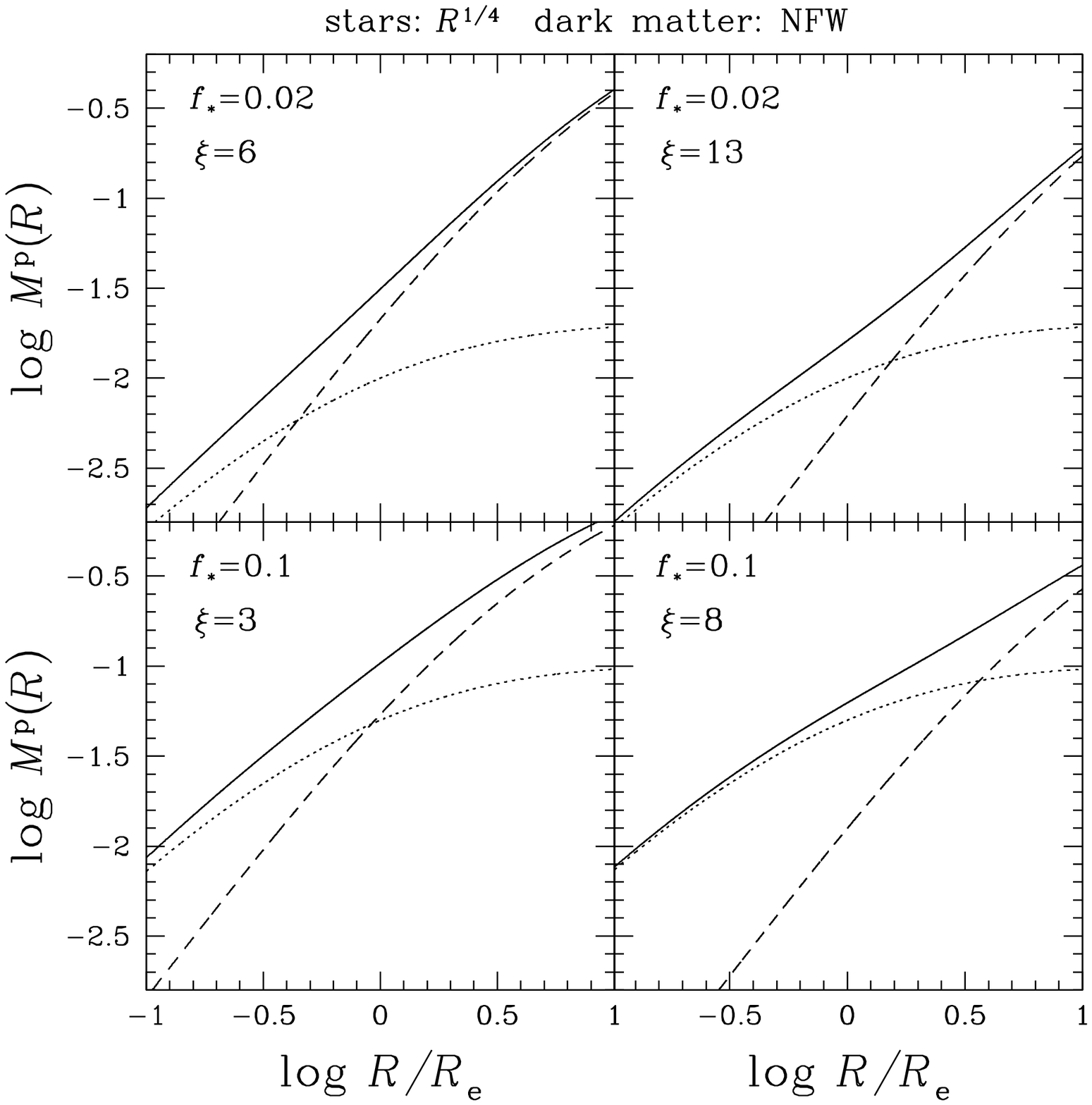}\\
\includegraphics[width=0.49\textwidth]{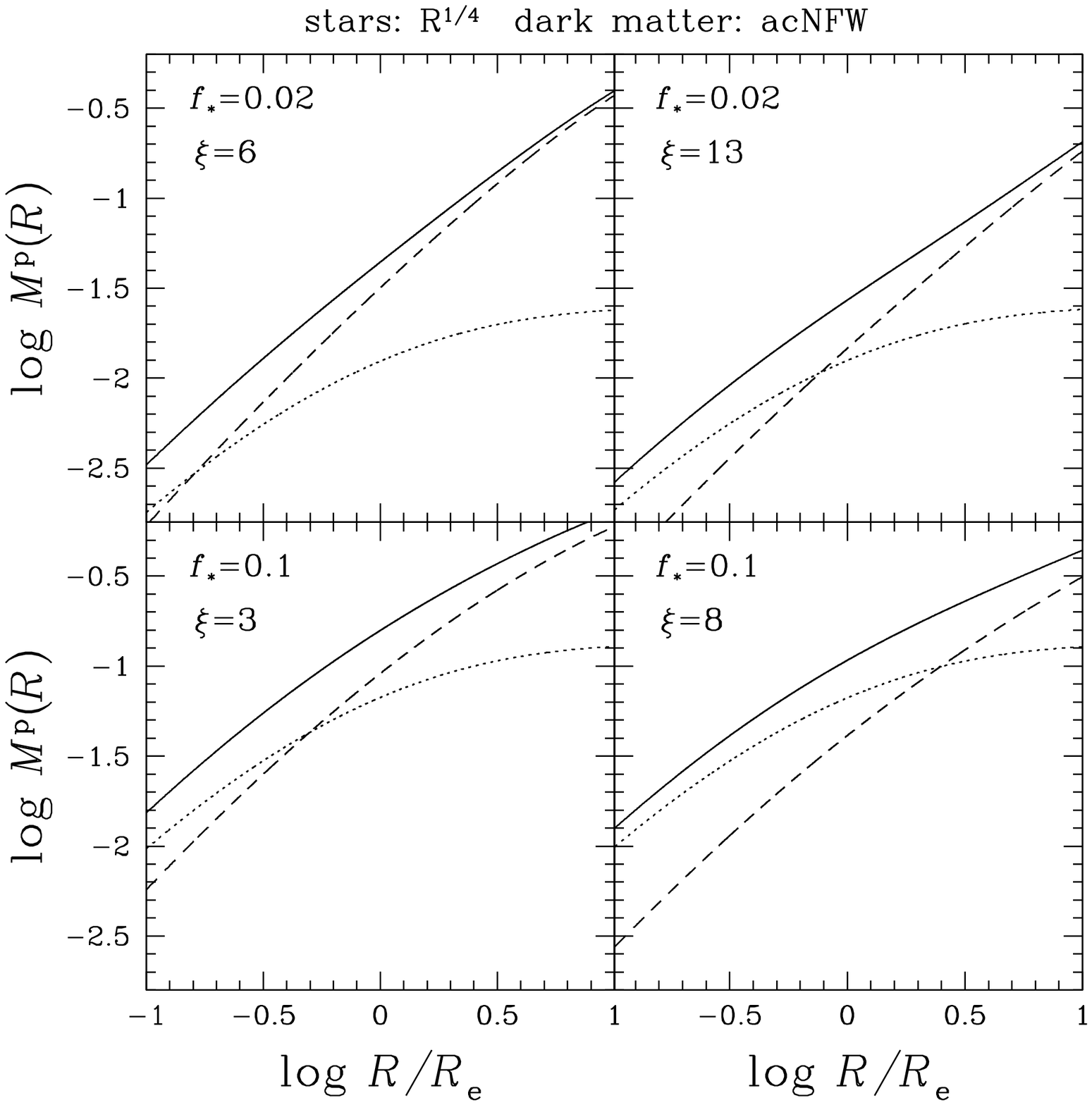}\\
\caption{Total (solid lines), stellar (dotted lines), and dark-matter
  (dashed lines) projected mass distributions $\Mp(R)$ for $m=4$
  \Sersic models with non-contracted (top) and adiabatically
  contracted (bottom) NFW dark-matter halos of given $\xi=\rs/\Re$ and
  stellar-to-total mass fraction $\fstar$. The projected mass $\Mp(R)$
  is in units of $\Mstar+\Mdm$. All models have concentration
  $\Conc=8$.}
\label{figmass_N}
\end{figure}

We consider four different models for the total density distribution.
First we consider a singular isothermal sphere (SIS) model, which
provides a generally good description of the lensing properties of
early-type galaxies \citep[e.g.][]{Koc94,Tre04,Koo06}, as well as of
their stellar kinematics \citep[e.g.][]{Ber94,Ger01}. Second, we
consider light-traces-mass (LTM) models. Although they are known to
fail observational constraints based on statistical analyses of strong
lenses \citep[e.g.,][]{Rus03, Rus05, Bol08b}, stellar dynamics
\citep[e.g.,][]{Ger01}, individual strong-lens modeling
\citep[e.g.,][]{Way05, Dye05, Dye07, Gav08}, weak-lensing analysis
\citep[e.g.,][]{Gav07}, and combined strong-lens/dynamical analyses
\citep[e.g.,][]{Koo02, Koo03, Tre02, Tre03, Tre04, Koo06}, they serve
as a useful ``straw man'' hypothesis to test against the MP\@. Lastly,
we consider two families of cosmologically motivated models based on
\citet[][NFW]{Nav96} halos with the addition of stars. In one case
(NFW plus stars models, hereafter NFW+S) the stars are added leaving
the halo unperturbed; in the other one (adiabatically contracted NFW
plus stars models, hereafter acNFW+S), the halo is assumed to
``respond'' to the sinking of baryons towards the centre of the galaxy
as prescribed by the adiabatic contraction recipe of \citet{Blu86}.
Several arguments suggest that the \citet{Blu86} model might
overestimate the compression of the halo
\citep[see][]{Gne04,Elz04,Nip04}.  In particular, \citet{Gne04}
argued that the standard adiabatic contraction model of \citet{Blu86},
based on some simplifying assumptions such as spherical symmetry and
circularity of the particle orbits, tends to overpredict the increase
of dark matter density in the central regions.  However, considering
both NFW+S and acNFW+S we should bracket the realistic range of NFW
halo models \citep[see also][]{Jia07}.

We now define for each model the total (stellar plus dark
matter) density profile $\rhotot(r)$. As we limit to spherically
symmetric distributions, in all cases the modulus of the gravitational
field is
\begin{equation}
{\gr(r)}={G\Mtot(r)  \over r^2},
\end{equation}
where $\Mtot(r)=4\pi \int_0^{r}\rhotot(r')r'^2dr'$ the total
mass within $r$.

In the case of SIS models, the total (stars plus dark matter) density
profile is
\begin{equation}
\rhotot (r)= {\sgSIS^2\over 2\pi G r^2},
\end{equation}
%
%
where $\sgSIS$ is the one-dimensional velocity dispersion of isotropic
SIS.  For LTM models the total density profile is
$\rhotot(r)=K\rhostar(r)$, where $K$ is a dimensionless constant (note
that $\cet$ is independent of $K$).  In NFW+S models the dark-matter
halo is described by a NFW model, so the dark-matter density
distribution is
\begin{equation}
\rhodm (r)={\Mdm\over 4 \pi f(\Conc)} {1\over r(r+\rs)^2},
\end{equation}
where $\rs$ is the scale radius, and the distribution is truncated at
the virial radius $\rvir$. The average dark-matter density within
$\rvir$ equals 200 times the critical density of the Universe. In the
equation above, $\Conc\equiv\rvir/\rs$ is the concentration parameter,
\begin{equation}
f(\Conc)=\ln (1+\Conc)-{\Conc\over 1+\Conc},
\end{equation}
and $\Mdm$ is the total dark-matter mass.  
In this case the total (stars plus dark matter) density 
profile is 
\begin{equation}
\rhotot (r)=\rhodm(r)+ \rhostar(r).
\end{equation}
Formally, these NFW+S models have three free parameters: concentration
$\Conc$, the stellar mass fraction $\fstar\equiv\Mstar/(\Mstar+\Mdm)$
and the ratio $\xi=\rs/\Re$.  Cosmological simulations suggest values
of $\Conc\sim 7-9$ for low-redshift galaxy-size halos \citep{Net07}.
Thus, we fix $\Conc=8$, and we explore different combinations of
values of $\xi$ and $\fstar$. There are indications that $\fstar$ is
typically of few percent in early-type galaxies \citep[e.g.][]{Jia07},
so here we consider the cases $\fstar=0.02$ and $\fstar=0.1$.  
The value of $\xi$ for given $\fstar$ is not strongly constrained by
models and observations, but for the present investigation it is
sufficient to individuate a realistic range of values of $\xi$. For
this purpose we can use the observed correlation between effective
radius and total stellar mass for early-type galaxies:
\begin{equation}
{\Re \over \kpc}=B\left({\Mstar \over \Msun}\right)^A,
\label{eqshen}
\end{equation}
where $A=0.56$ and $B=2.88\times10^{-6}$ \citep{She03}.  
By definition
of the virial radius, 
\begin{equation}
{3\Mdm\over 4\pi\rvir^3}=200\rhocrit,
\label{eqrhocrit}
\end{equation}
where $\rhocrit=3H_0^2/(8\pi G)$ is the critical density of the Universe
and $H_0$ is the Hubble constant (here we assume $H_0=70 \km \sminus
\Mpcminus$). Combining equations~(\ref{eqshen}) and (\ref{eqrhocrit}),
we get the following relation between $\xi$ and $\Re$
\begin{equation}
\xi^3=
{3 \over800\pi B^{1/A}}
{1-\fstar \over \fstar \Conc^3}
{\Retilde^{(1/A)-3}
\over
\rhocrittilde },
\label{eqxi}
\end{equation}
where $\Retilde\equiv \Re/\kpc$ and $\rhocrittilde \equiv
\rhocrit/(\Msun \kpcminust)$: $\xi$ as a function of $\Re$ is plotted
in Fig.~\ref{figrexi}.  Early-type galaxies of the SLACS sample have
effective radii in the range $0.3\lsim \log \Re/\kpc \lsim 1.3$
\citep{Bol08a}. For each value of $\fstar$ we show results for two
values of $\xi$ roughly corresponding to the upper and lower limits of
this range. In particular, we consider $\xi=6$ and $\xi=13$ when
$\fstar=0.02$, and $\xi=3$ and $\xi=8$ when $\fstar =0.1$.  The
resulting stellar, dark-matter and total projected mass profiles are
plotted in Fig.~\ref{figmass_N} (top) when the stellar profile is a de
Vaucouleurs (or $m=4$ S\'ersic) model.

We also consider acNFW+S models, in which the dark-matter halo is
adiabatically contracted following the standard recipe
\citep{Blu86,Kee01}.  In the considered models, the dark-matter
profile is obtained by adiabatically compressing an initial NFW
profile with the same values of the parameters $\Conc$, $\xi$ and
$\fstar$ as for the non-compressed models. The final dark-matter
distribution is computed numerically for each given stellar
distribution. Projected mass profiles of acNFW+S models with de
Vaucouleurs' stellar density distribution are shown in
Fig.~\ref{figmass_N} (bottom) to allow a direct comparison with
  the corresponding non contracted models shown in the top panels.

\subsection{Orbital anisotropy}
\label{secani}

We consider two parameterizations of radial anisotropy in the stellar
distribution: constant anisotropy and Osipkov-Merritt
\citep[OM;][]{Osi79,Mer85}.  In the first case the value of the anisotropy
parameter is the same at all radii:
\begin{equation}
\beta(r)=\const, 
\end{equation}
and the radial component of the velocity dispersion tensor is \citep{Bin08}
\begin{equation}
\sgrsq(r)= {1\over r^{2\beta}\rhostar(r)}\int_r^\infty r'^{2\beta}\rhostar(r'){d \Phi(r') \over d r'}dr'.
\end{equation}
In the case of OM anisotropic models, the anisotropy in the stellar
orbital distribution is introduced by using the following
parameterization: the radial dependence of the anisotropy parameter is
\begin{equation}
\beta(r)= {r^2\over r^2+\ra^2}, 
\end{equation}
where the quantity $\ra$ is the so--called ``anisotropy radius''. For
$r \gg\ra$ the velocity dispersion tensor is radially anisotropic,
while for $r \ll \ra $ the tensor is nearly isotropic.  Isotropy is
realized at the model centre, independently of the value of $\ra$. In
the case of OM models, the radial component of the velocity dispersion
tensor is given by
\begin{equation}
\rhostar(r) \sgrsq(r)= {\ra^2 \over \ra^2 + r^2}\int_r^\infty \varrhostar(r'){d \Phi(r') \over d r'}dr',
\end{equation}
where
\begin{equation}
\varrhostar (r)=\left(1+\frac{r^2}{\ra^2}\right)\rhostar (r)
\end{equation}
\citep{Mer85}.

We consider constant-anisotropy models because they are the simplest
possible anisotropic models, and span the full range of anisotropies,
from tangentially to radially biased.  However, OM models should be
more realistic, because observational indications suggest that typical
massive elliptical galaxies are, in the central regions, isotropic or
mildly radially anisotropic \citep[e.g.][]{Ger01,Cap07}, and different
theoretical models of galaxy formation predict that elliptical
galaxies should have anisotropy varying with radius, from almost
isotropic in the centre to radially biased in the outskirts
\citep[e.g.][]{Van82,Bar92,Her93,Nip06}.

\subsection{Consistency and stability}
\label{seccons}

A galaxy model is consistent if it has a positive distribution
function. Not all combinations of the parameters introduced in the
sections above generate consistent models. For instance, a necessary
(but not sufficient) condition for consistency of OM models is
\citep{CioP92}
\begin{equation}
{d\varrhostar(r)\over dr}\leq0,
\label{eqnec}
\end{equation}
which holds also in the presence of a dark-matter halo.  
For $\beta=const$ models, the necessary condition is $\gammastar\geq
2\beta$, where $\gammastar=-\lim_{r\to 0}(d\ln\rhostar(r)/d\ln r)$
\citep[][see also Richstone \& Tremaine 1984 and Tremaine et
al. 1994]{AnE06}. This condition holds not only for one-component
systems, but also for two-component systems if $\rhotot \sim
r^{-(2-\epsilon)}$ for $r \to 0$, with $\epsilon>0$ \citep{AnE06}.  In
the models here considered $\epsilon>0$, except for the limiting cases
of $\gamma=2$ models and/or SIS total density, in which $\epsilon=0$.
However, \citet{AnE06} show that the necessary condition is
$\gammastar\geq 2\beta+({1\over 2}-\beta)\eta$ if $\rhotot\propto
r^{-(2+\eta)}$, with $\eta>0$, so one can argue that the
condition $\gammastar\geq 2\beta$ must hold also if $\rhotot\propto
r^{-2}$ for $r\to 0$ (i.e., logarithmically divergent central
potential).

Thus, the requirement of consistency reduces the parameter space.  To
ensure physically meaningful models we computationally check
consistency and rule out regions of parameter space that would give
rise to non-consistent models. In particular, we exclude OM models
that do not satisfy the condition~(\ref{eqnec}), and $\beta=\const$
models with $\gammastar<2\beta$.  We recall here that for \Sersic
models $\gammastar=(m-1)/m$ \citep{Cio91}, while for $\gamma$ models
simply $\gammastar=\gamma$.

Additional constraints would come from the requirement of model
stability. In particular, strongly radially anisotropic systems are
expected to be radial-orbit unstable \citep{Fri84}.  However, while
there are robust estimates of the maximum amount of radial orbital
anisotropy allowed for stable one-component systems \citep[see,
e.g.,][]{MerA85,Ber89,Sah91,Mez97}, much less is known about the
stability of two-component systems, though there are indications that
the presence of a massive halo contributes to the systems' stability
\citep[e.g.][]{Sti91,Nip02}. As a consequence, we have not enough
information to exclude models on the basis of stability
arguments.\footnote{All the models considered in this paper are
two-component models. In LTM models we assume that the dark matter has
the same {\it density} distribution as the baryons, but not
necessarily the same {\it velocity} distribution.} Nevertheless, one
needs to bear in mind that models with extreme radial anisotropy, even
satisfying the necessary consistency condition, might be
non-consistent or radially unstable.

\section{Results: constraining model parameters with observations}  
\label{secres}

\label{secpar}

\begin{figure*}
\centering
\includegraphics[width=0.4\textwidth]{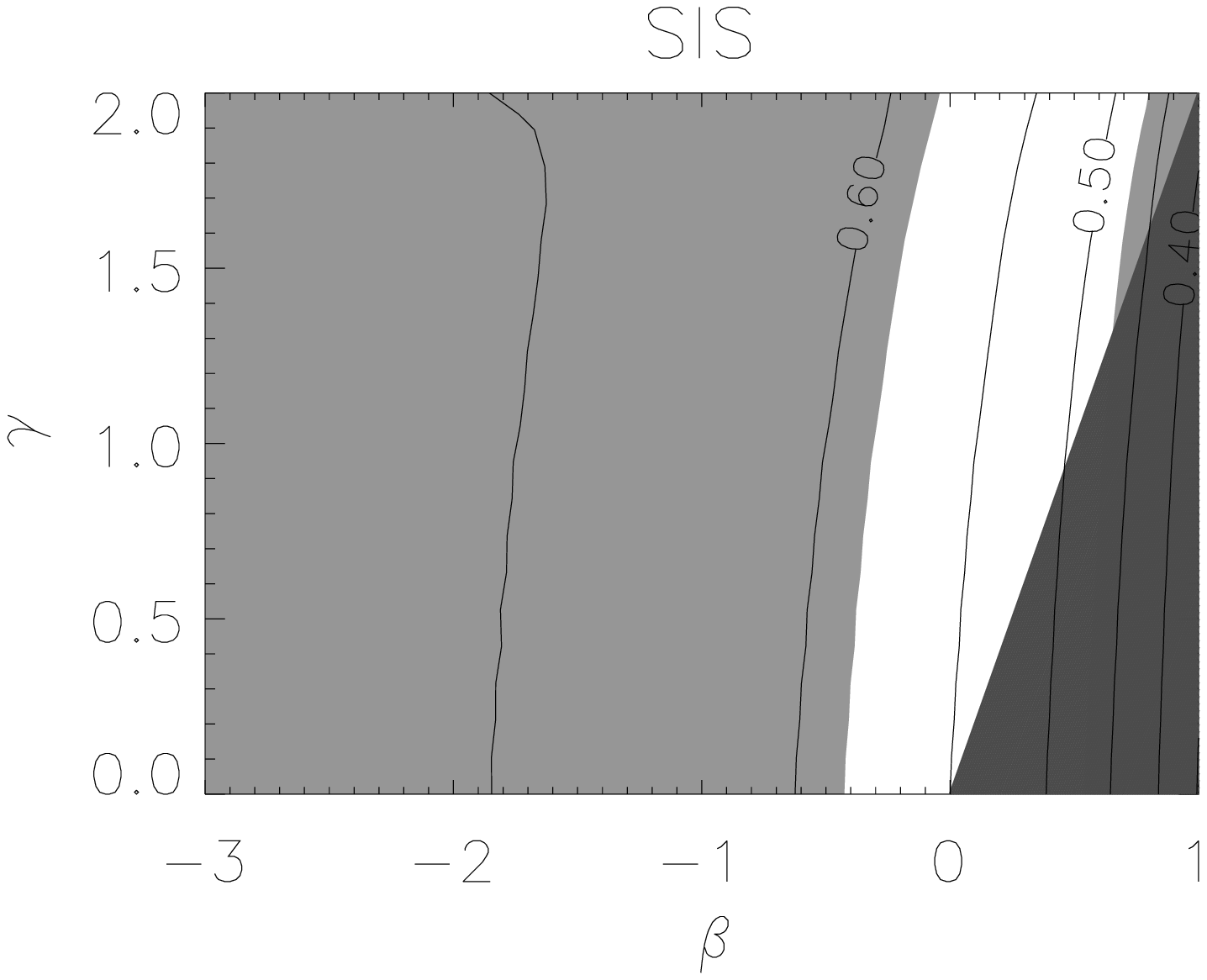}
\includegraphics[width=0.4\textwidth]{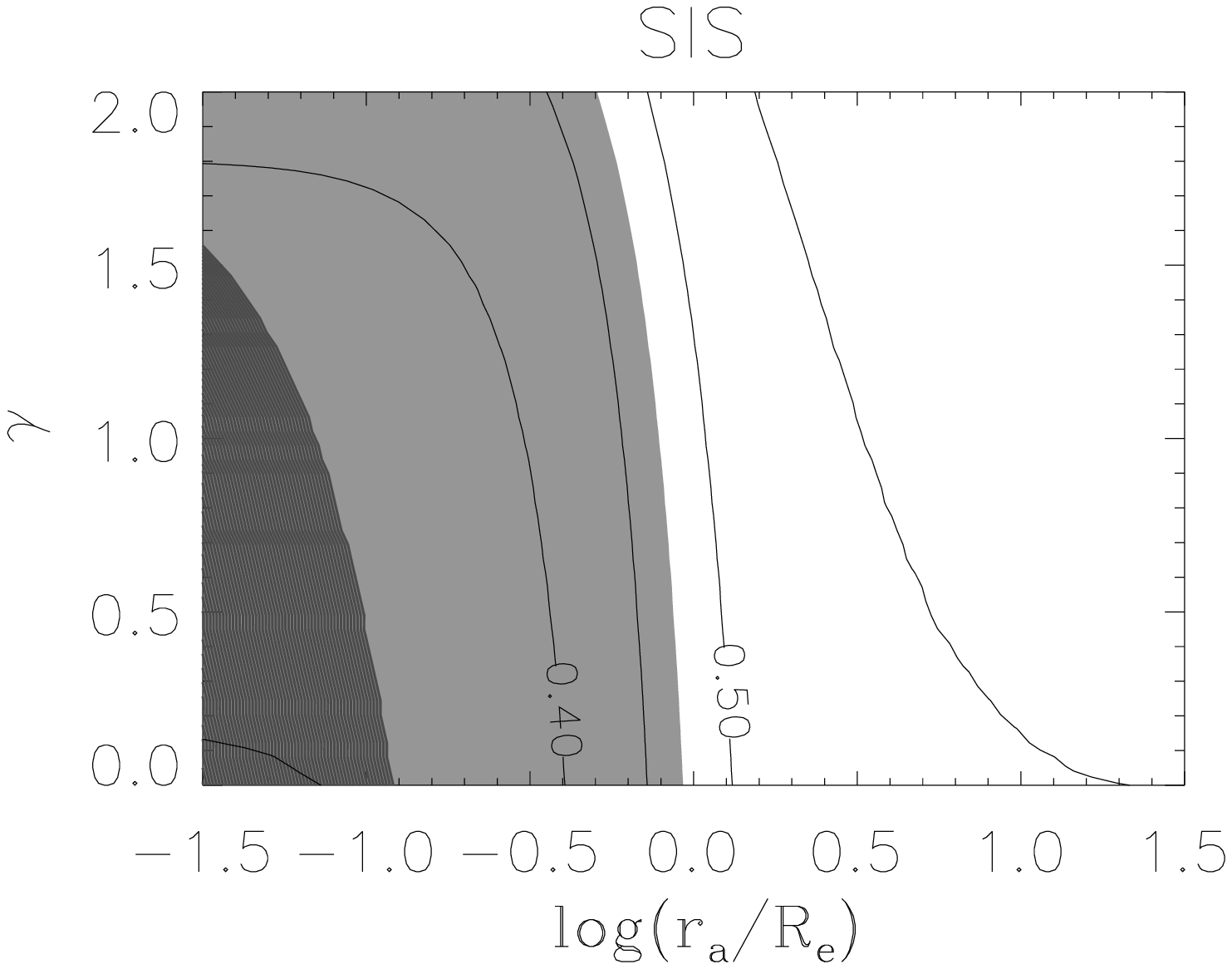}\\
\includegraphics[width=0.4\textwidth]{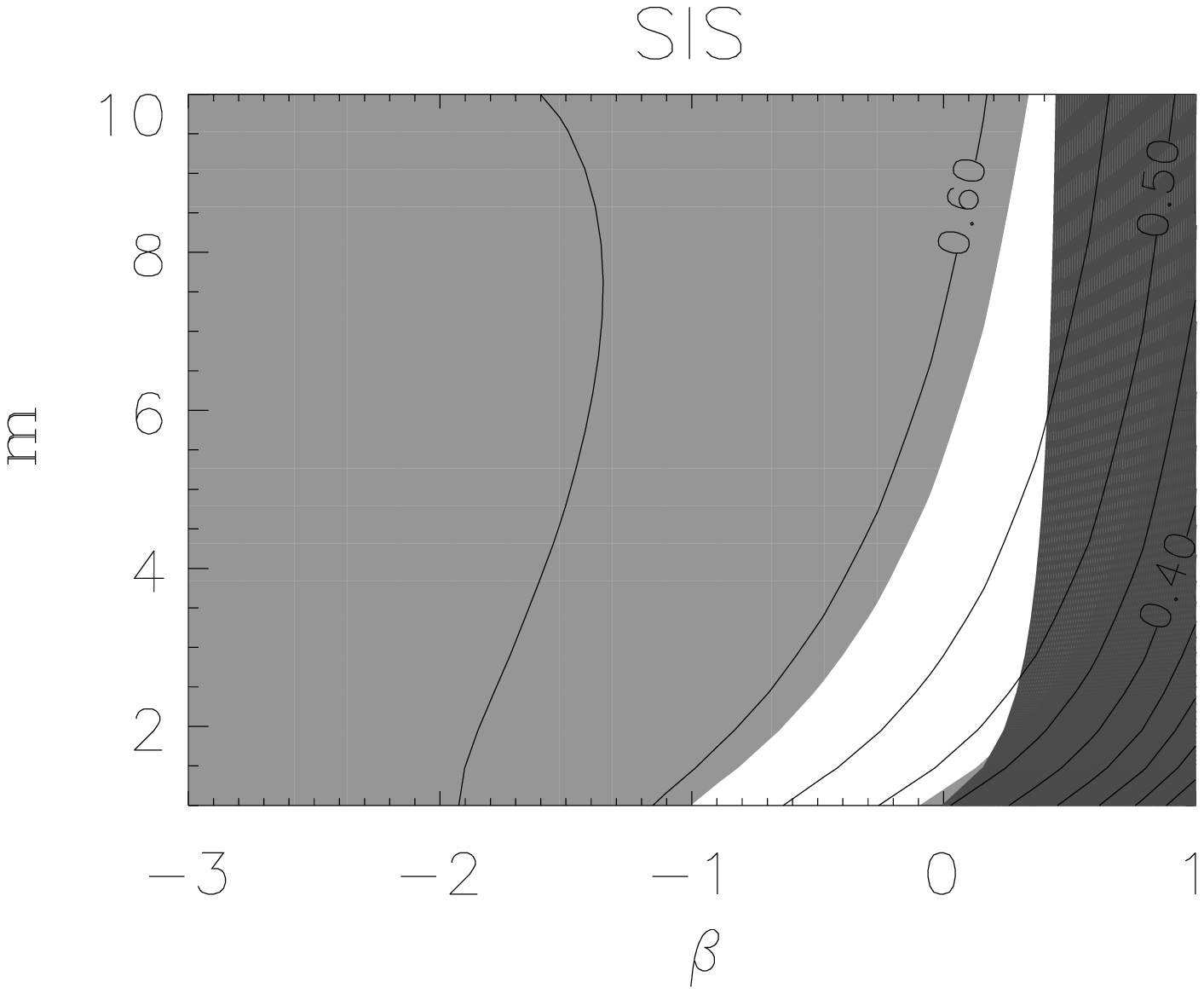}
\includegraphics[width=0.4\textwidth]{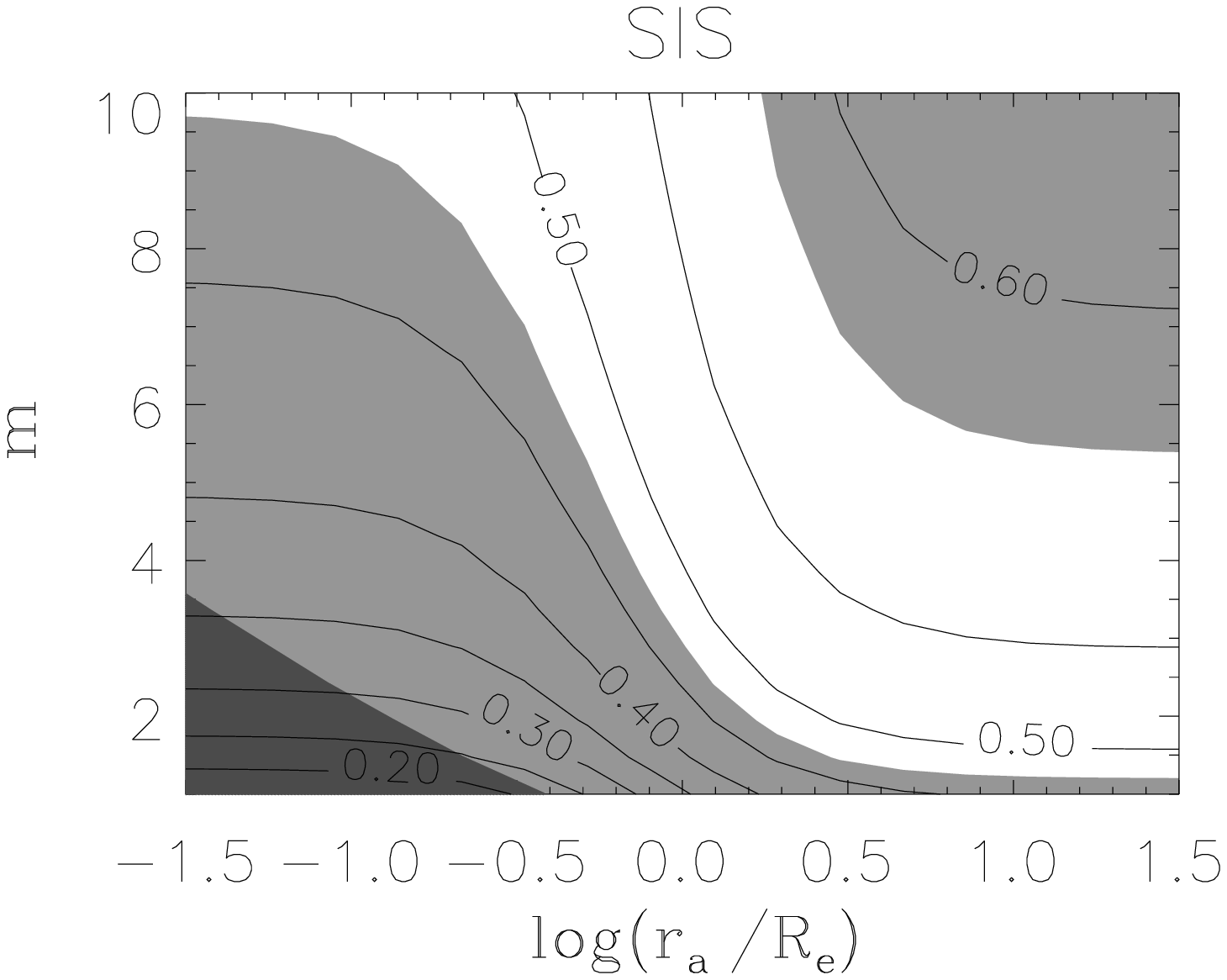}\\
\caption{Contours of constant $\log \cet$ in the planes
  $\beta$-$\gamma$, $\beta$-$m$ (left, for $\beta=const$ models), and
  $\log (\ra/\Re)$-$\gamma$, $\log (\ra/\Re)$-$m$ (right, for OM
  models), for systems with singular isothermal sphere total mass
  distribution. Dark-shaded regions correspond to non-consistent
  models.  Light-shaded regions are outside the observed range
  $\langle \log \cet\rangle=0.53\pm0.057$ \citep{Bol08b}.}
\label{figcsis}
\end{figure*}

\begin{figure*}
\centering
\includegraphics[width=0.4\textwidth]{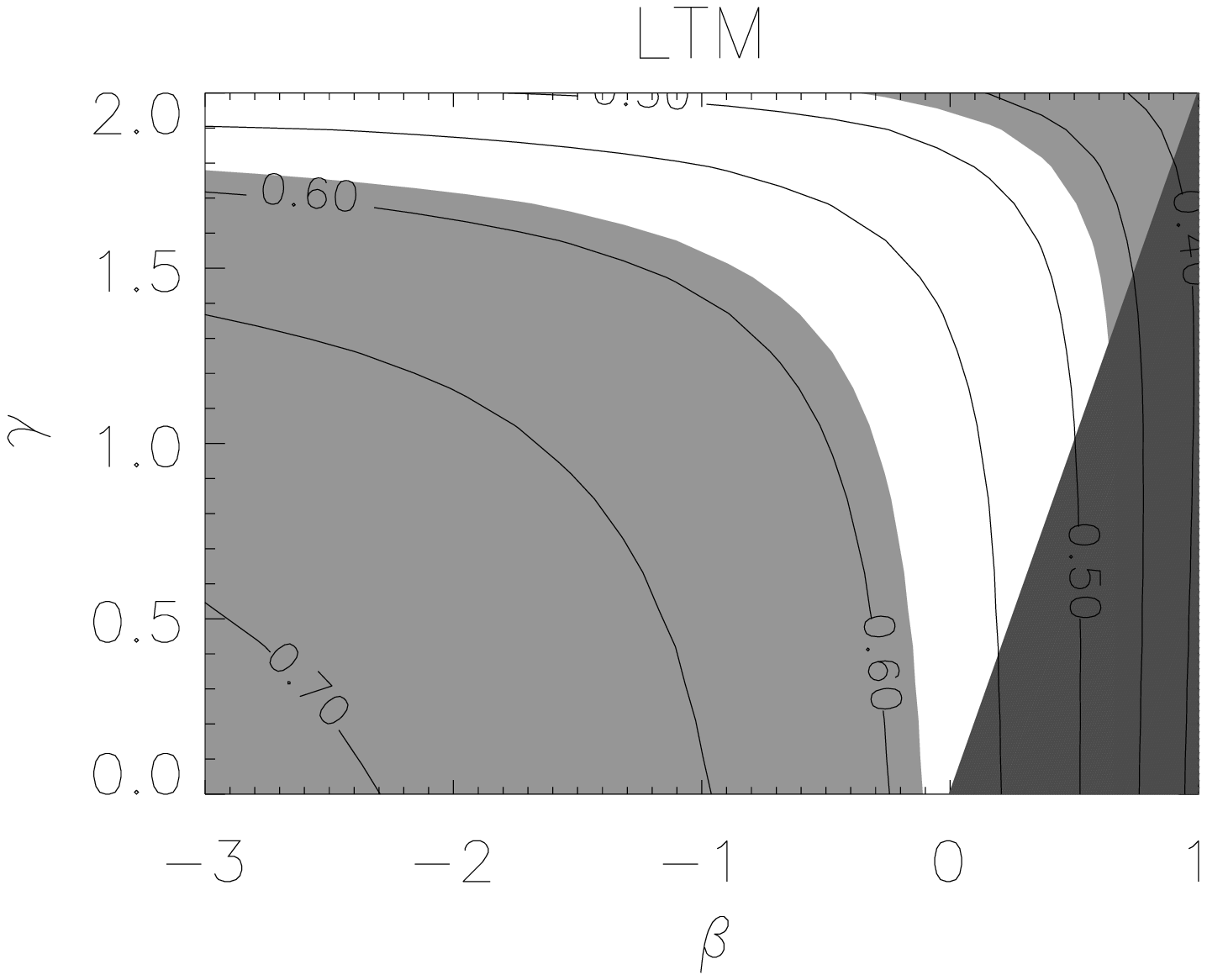}
\includegraphics[width=0.4\textwidth]{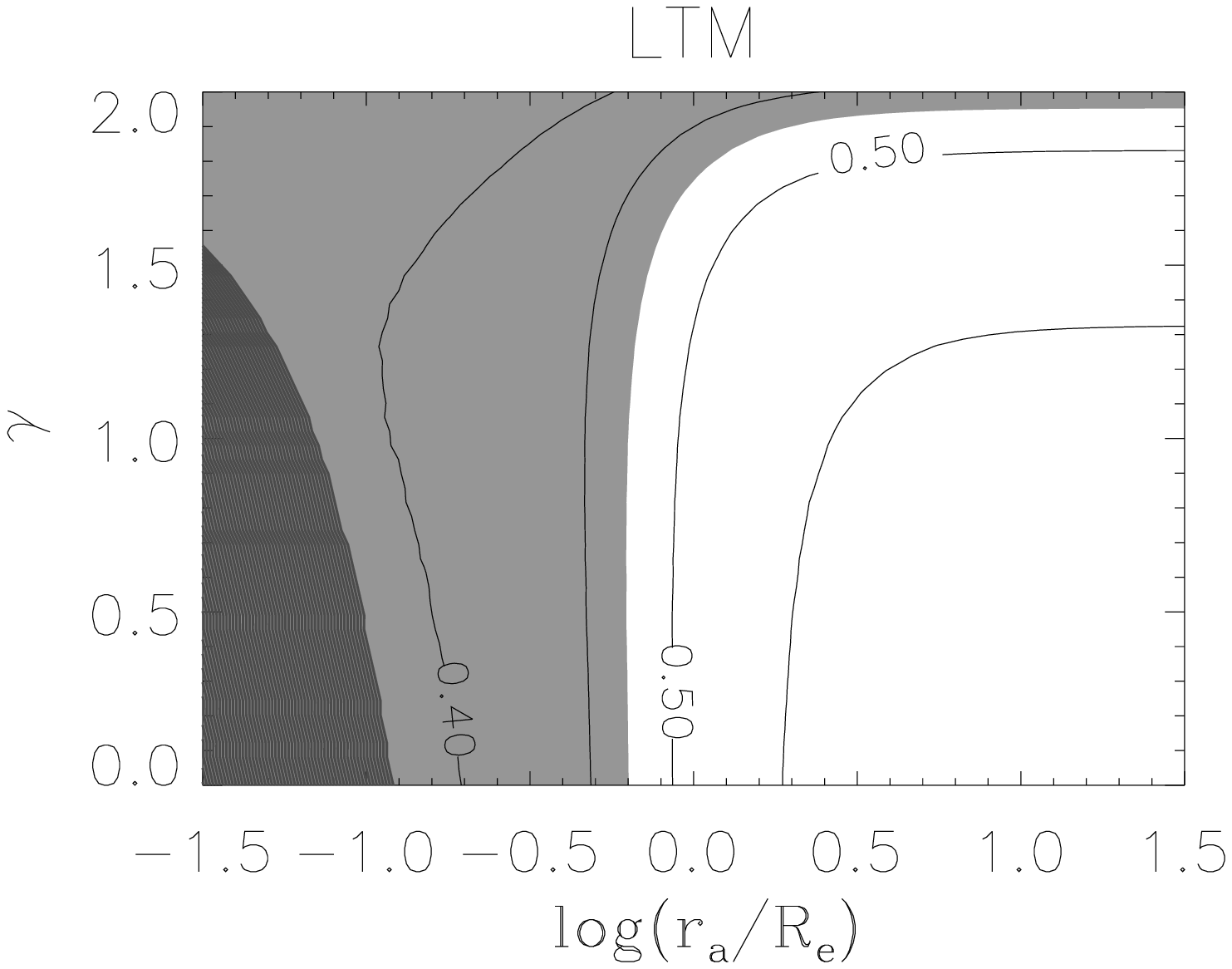}\\
\includegraphics[width=0.4\textwidth]{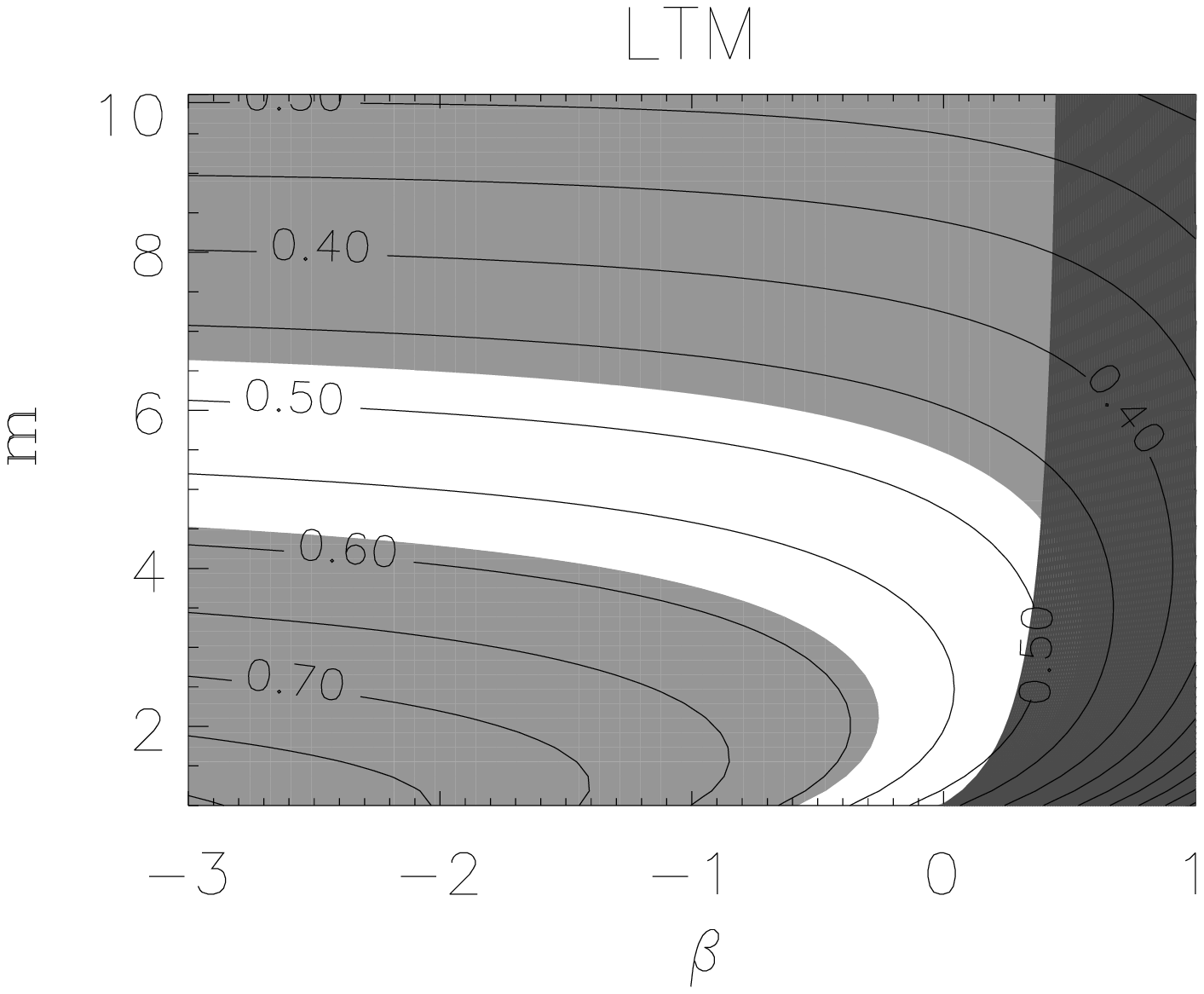}
\includegraphics[width=0.4\textwidth]{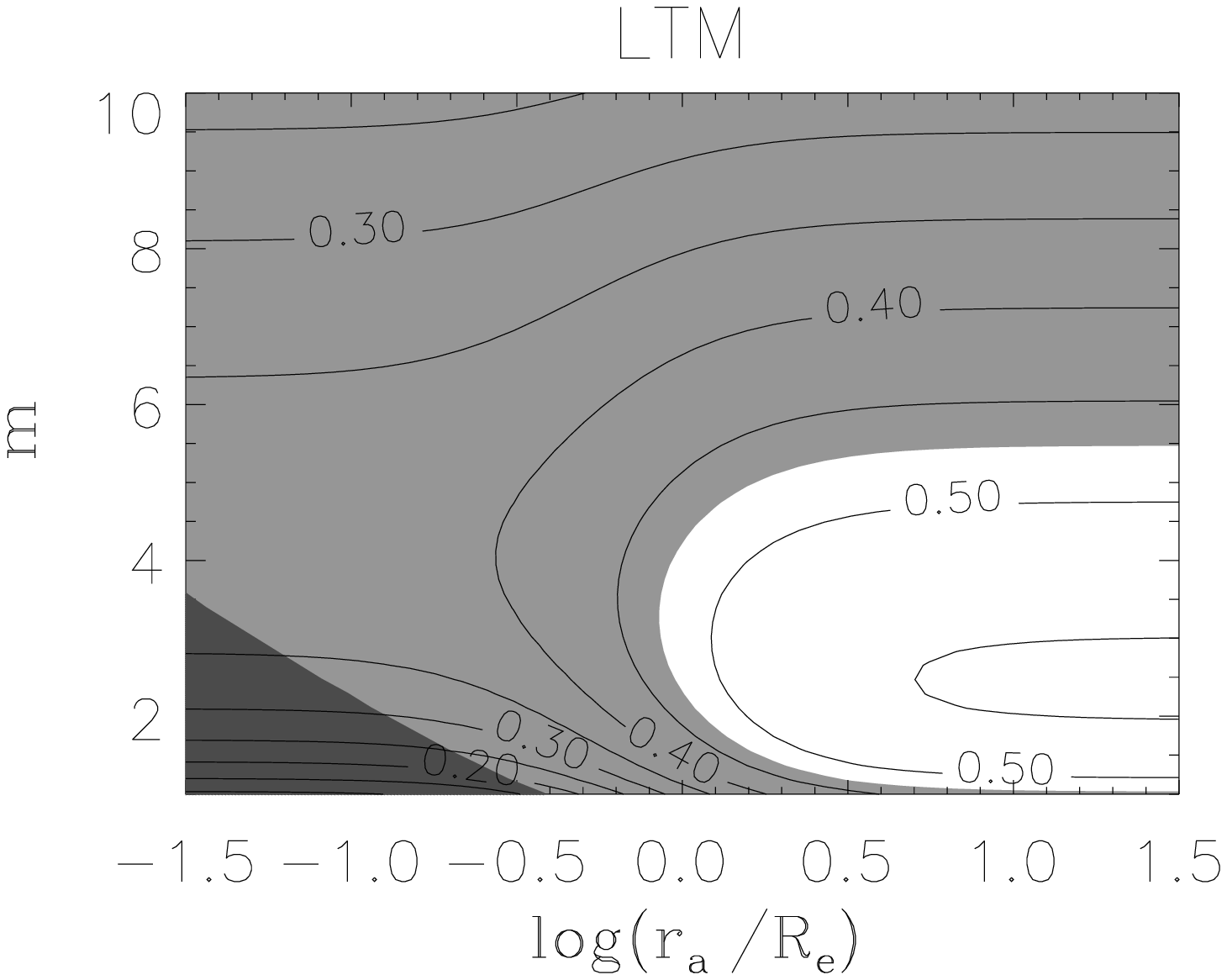}\\
\caption{Same as Fig.~\ref{figcsis}, but for light-traces-mass models.}
\label{figcltm}
\end{figure*}

\begin{figure*}
\centering
\includegraphics[width=0.4\textwidth]{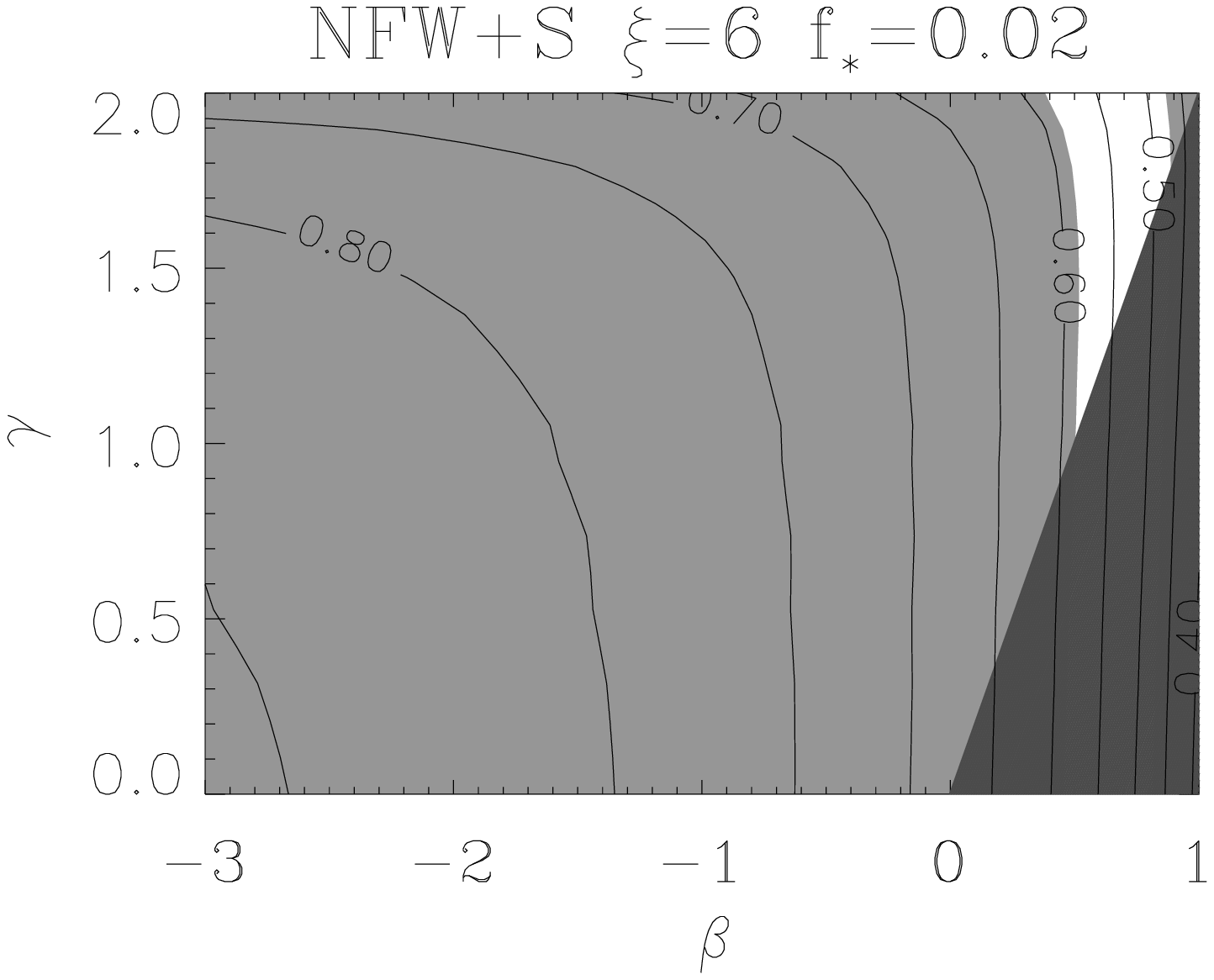}
\includegraphics[width=0.4\textwidth]{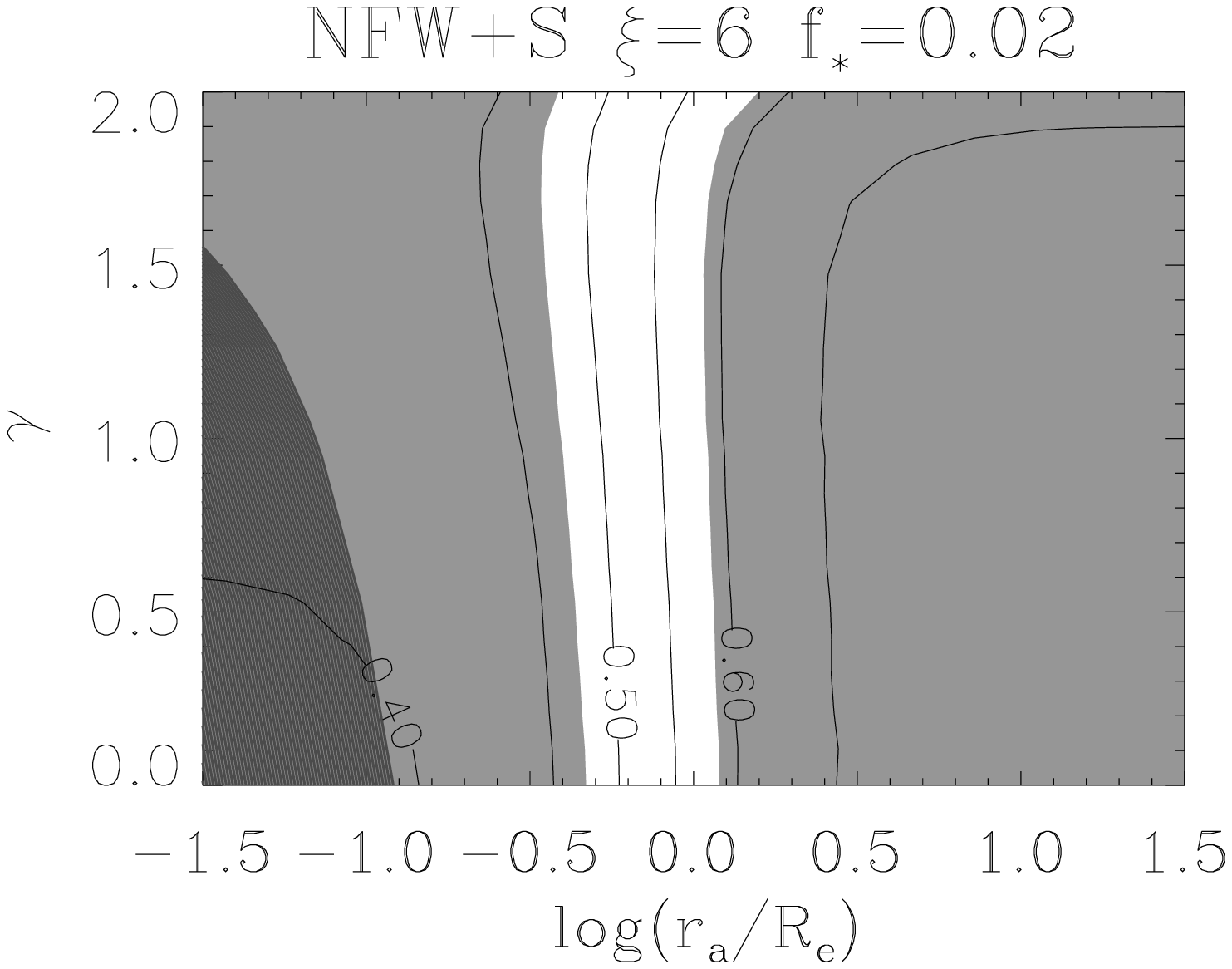}\\
\includegraphics[width=0.4\textwidth]{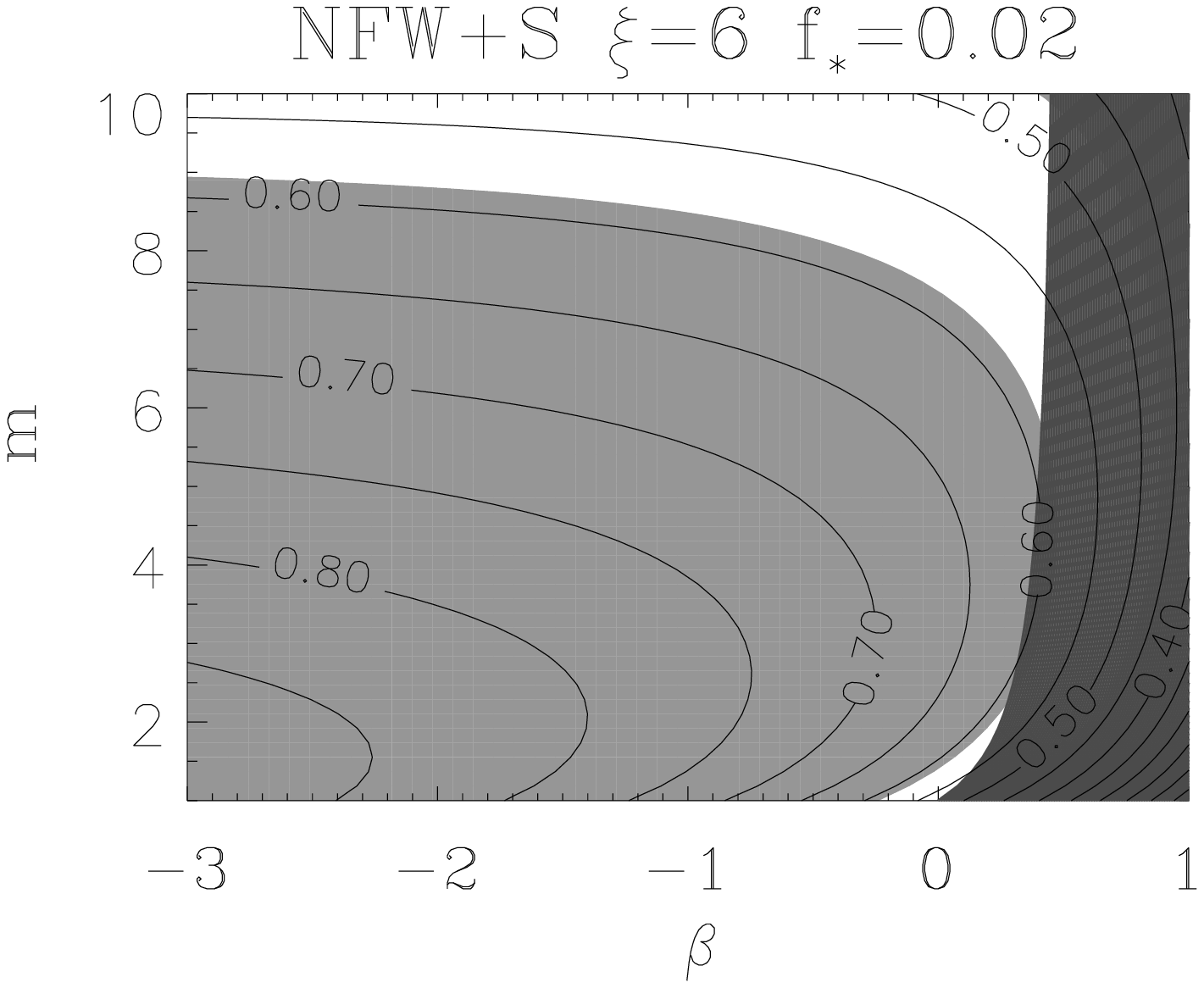}
\includegraphics[width=0.4\textwidth]{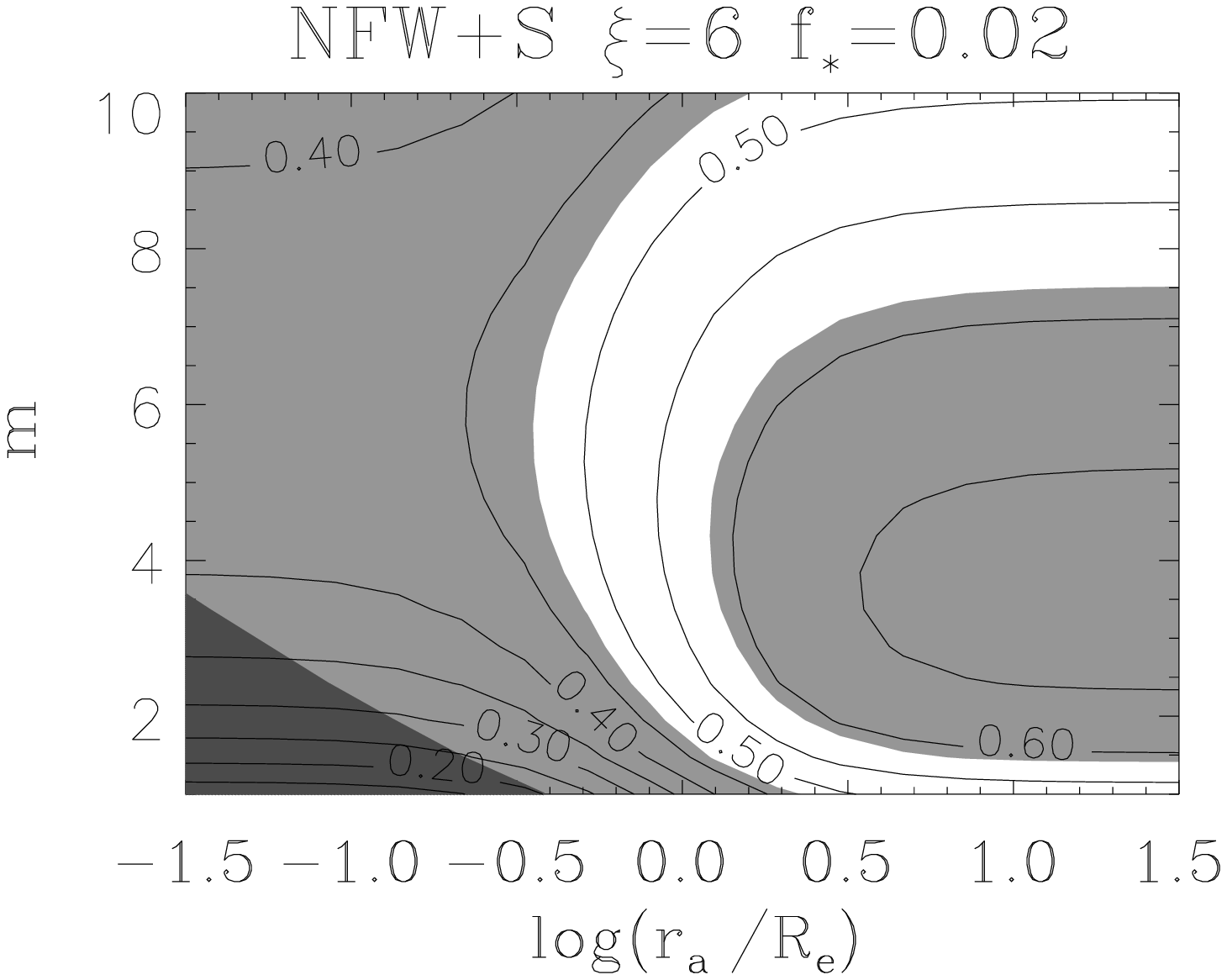}\\
\includegraphics[width=0.4\textwidth]{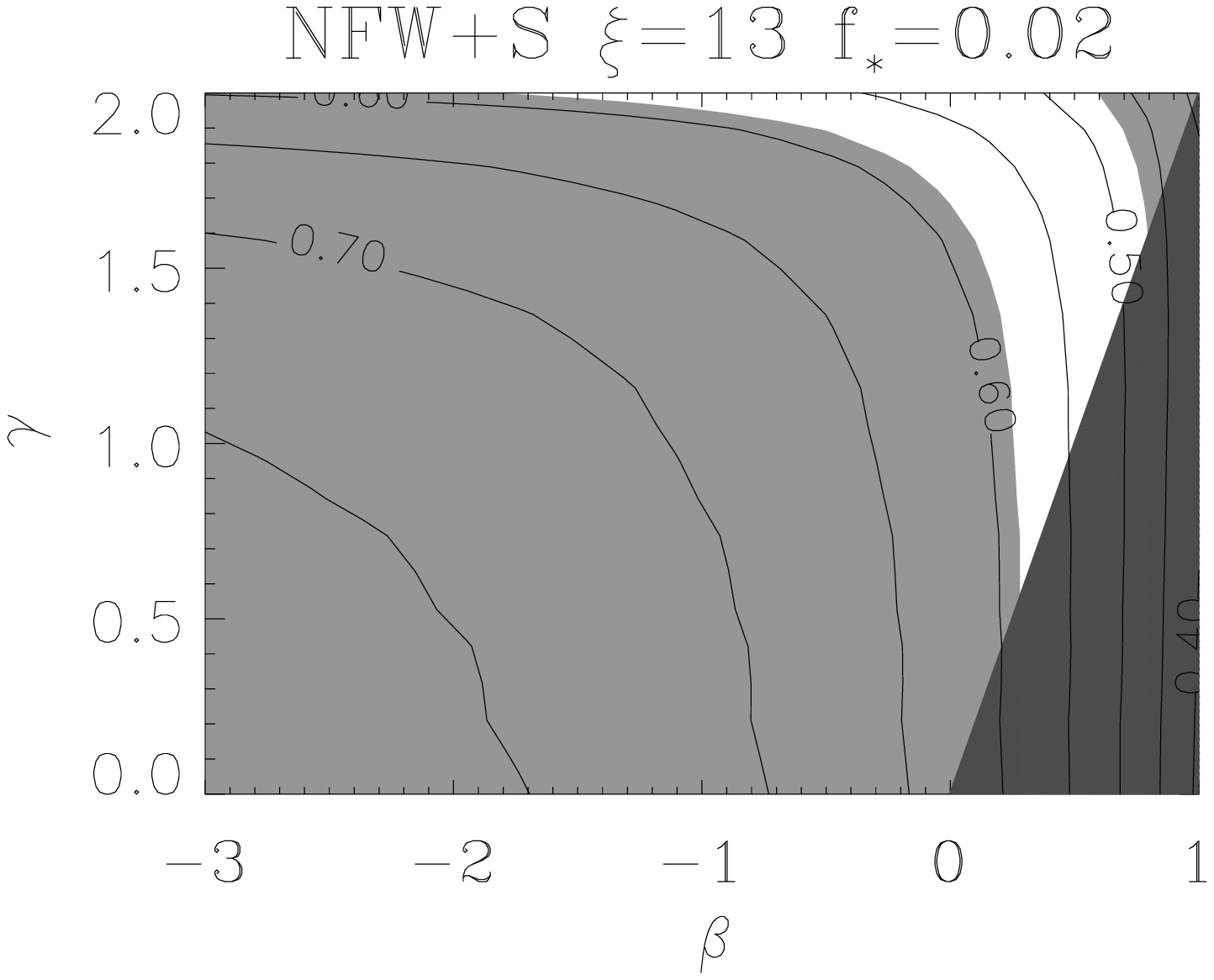}
\includegraphics[width=0.4\textwidth]{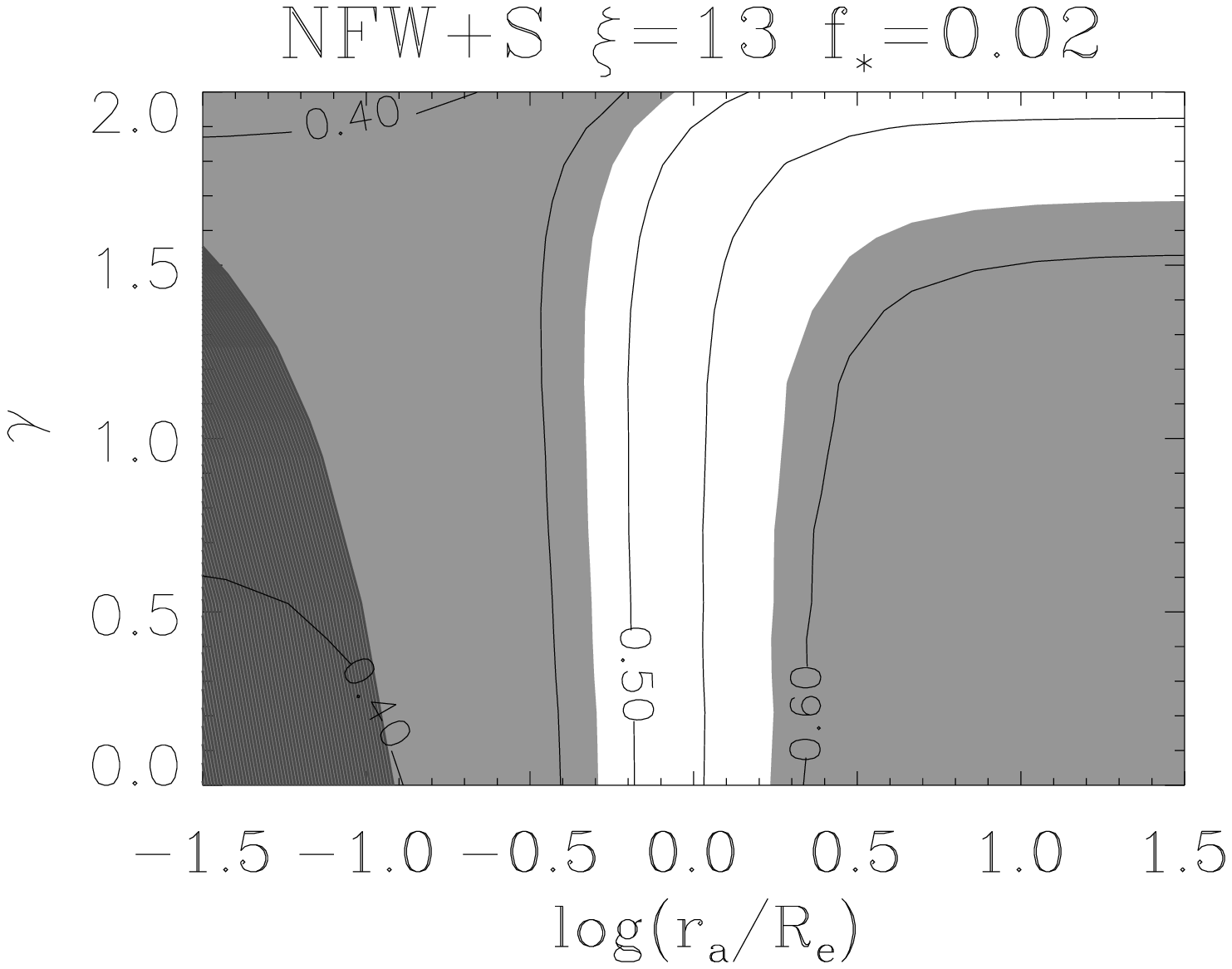}\\
\includegraphics[width=0.4\textwidth]{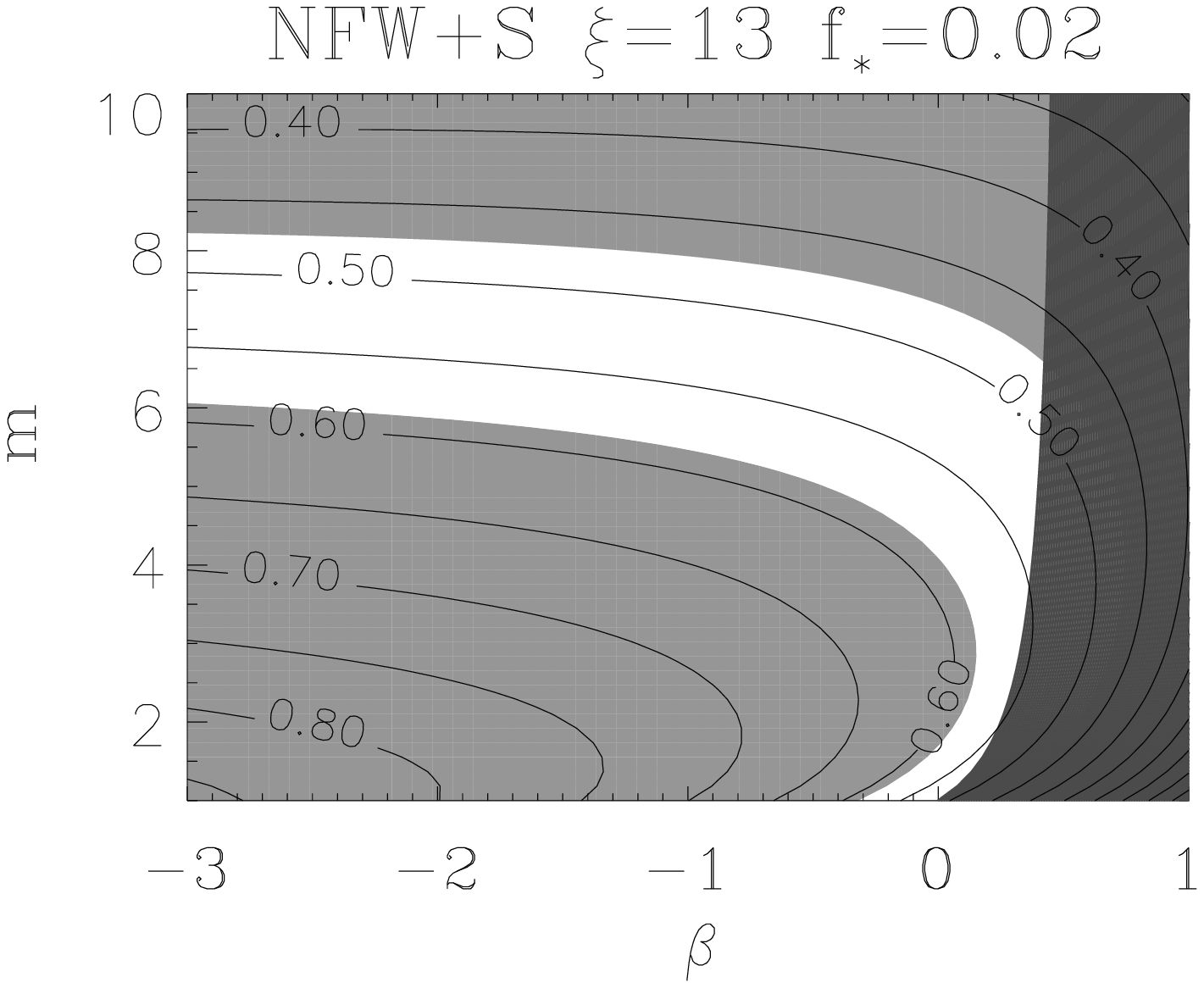}
\includegraphics[width=0.4\textwidth]{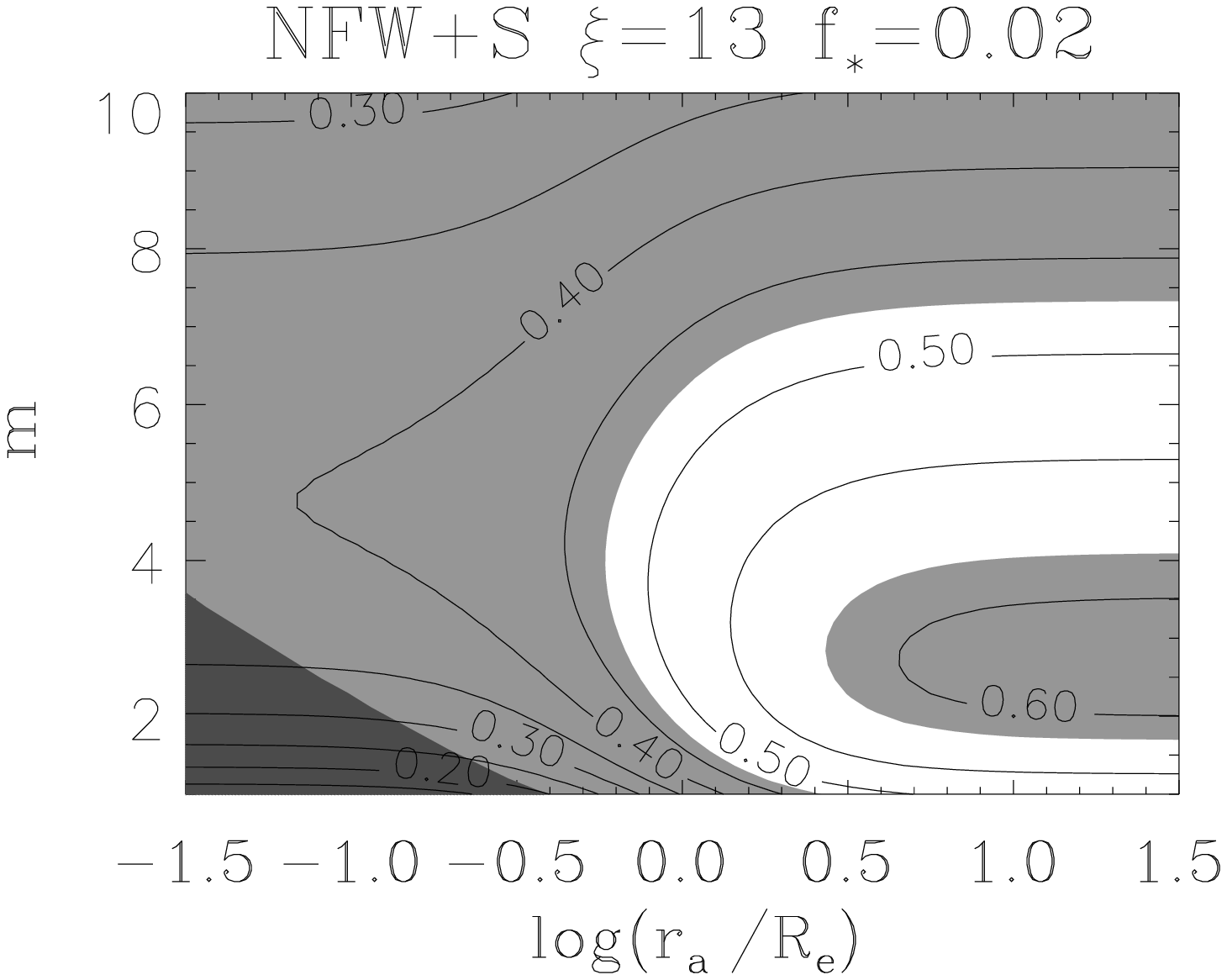}\\
\caption{Same as Fig.~\ref{figcsis}, but for NFW plus stars models
  with $\fstar=0.02$.}
\label{figcnfw02}
\end{figure*}

\begin{figure*}
\centering
\includegraphics[width=0.4\textwidth]{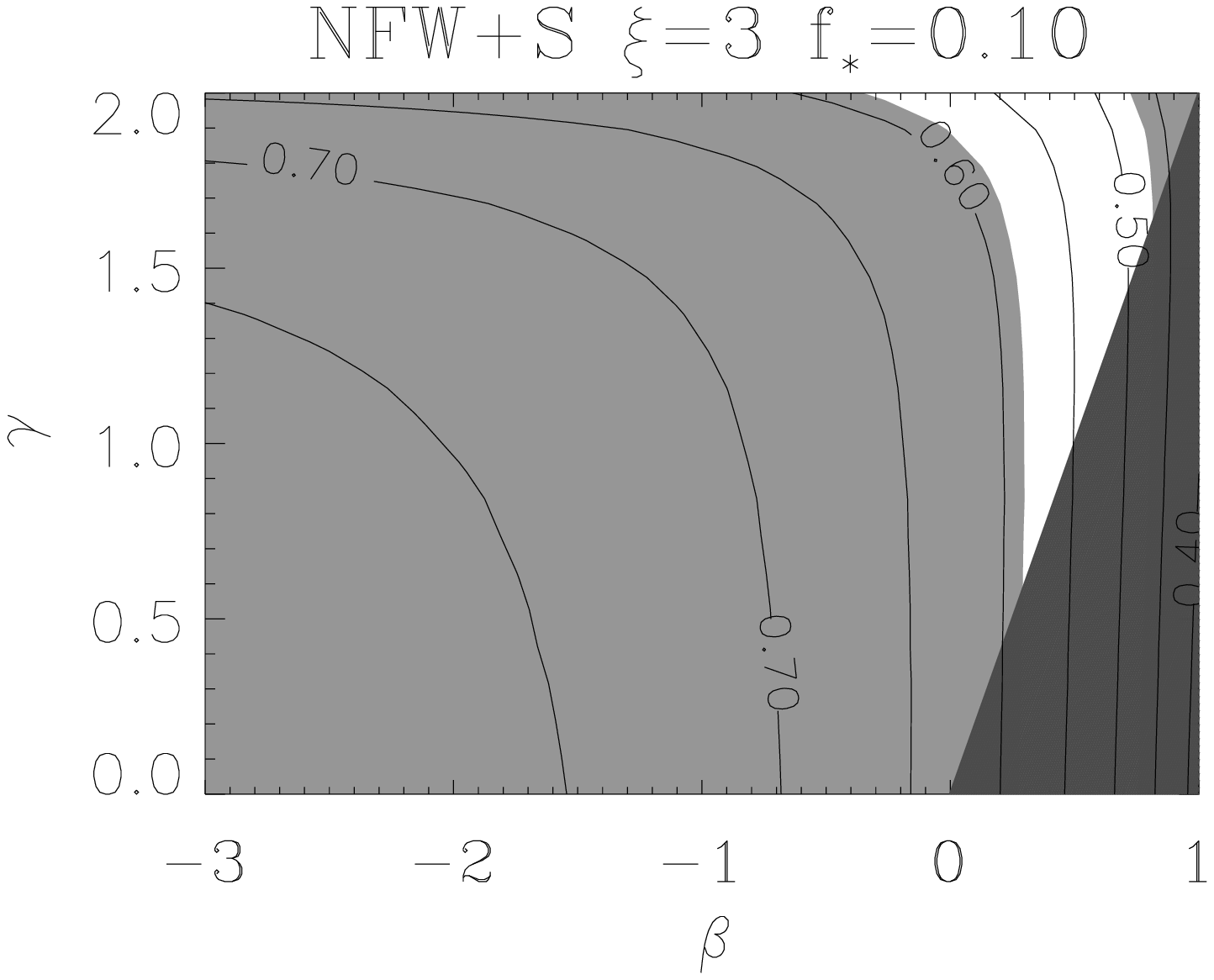}
\includegraphics[width=0.4\textwidth]{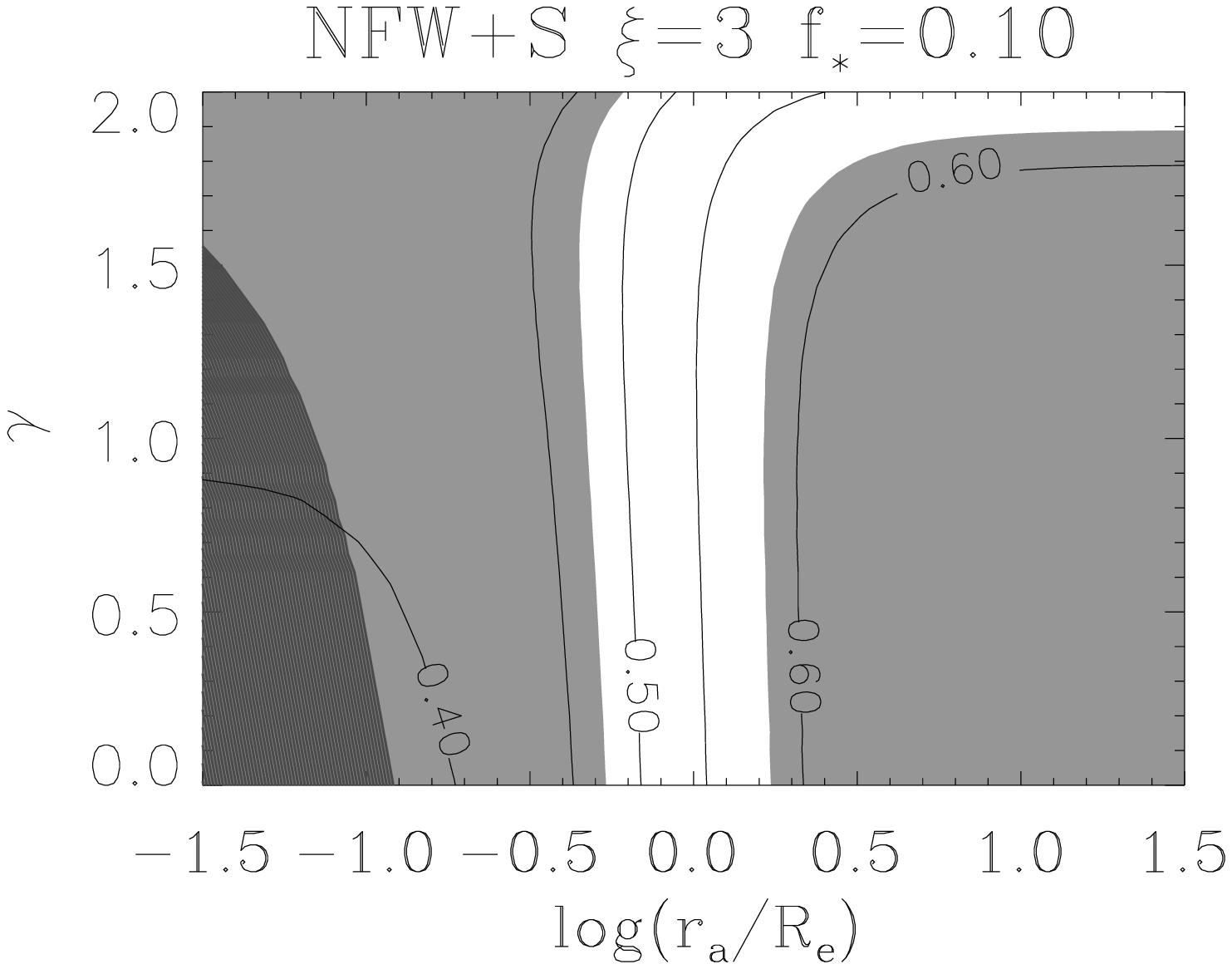}\\
\includegraphics[width=0.4\textwidth]{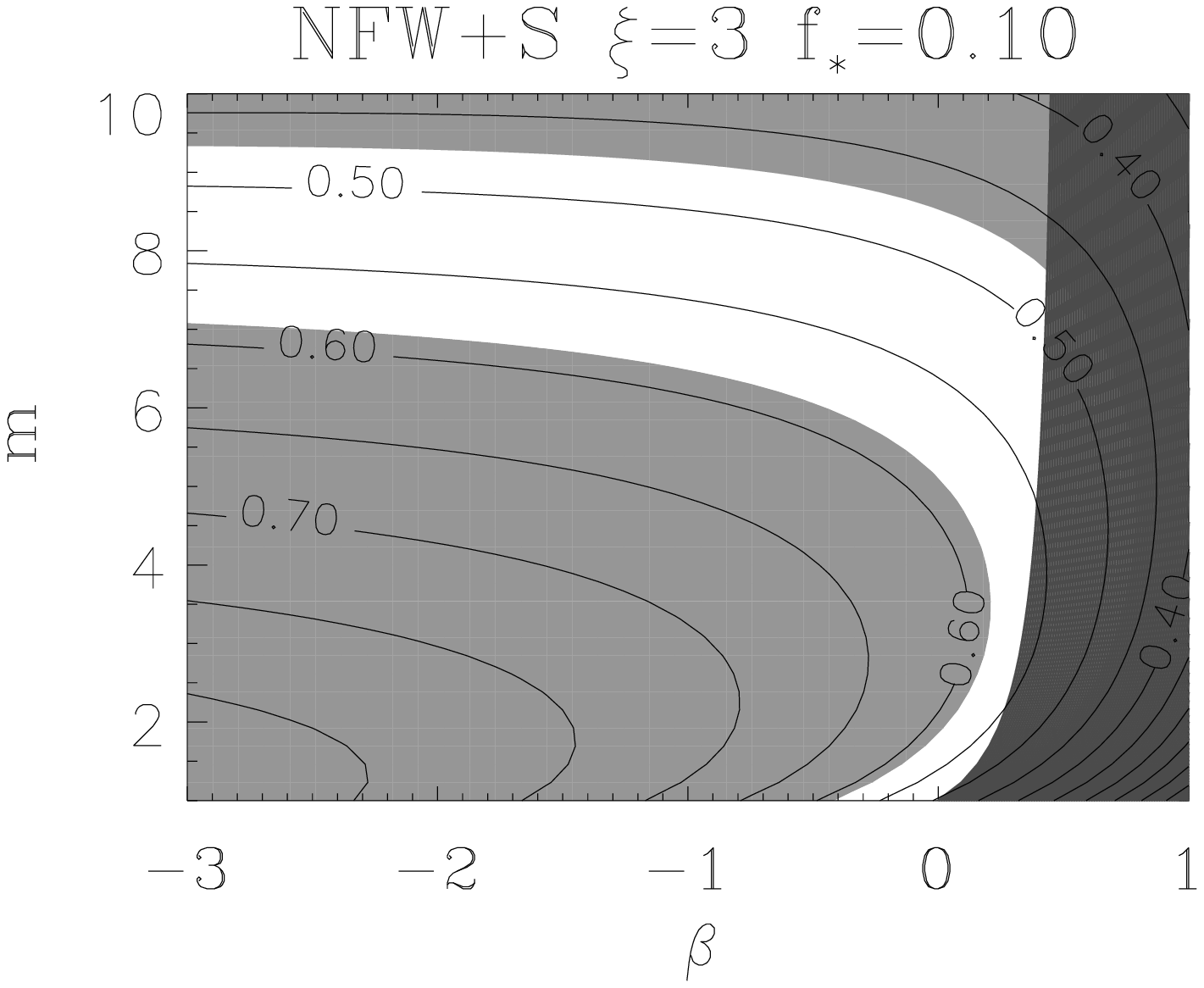}
\includegraphics[width=0.4\textwidth]{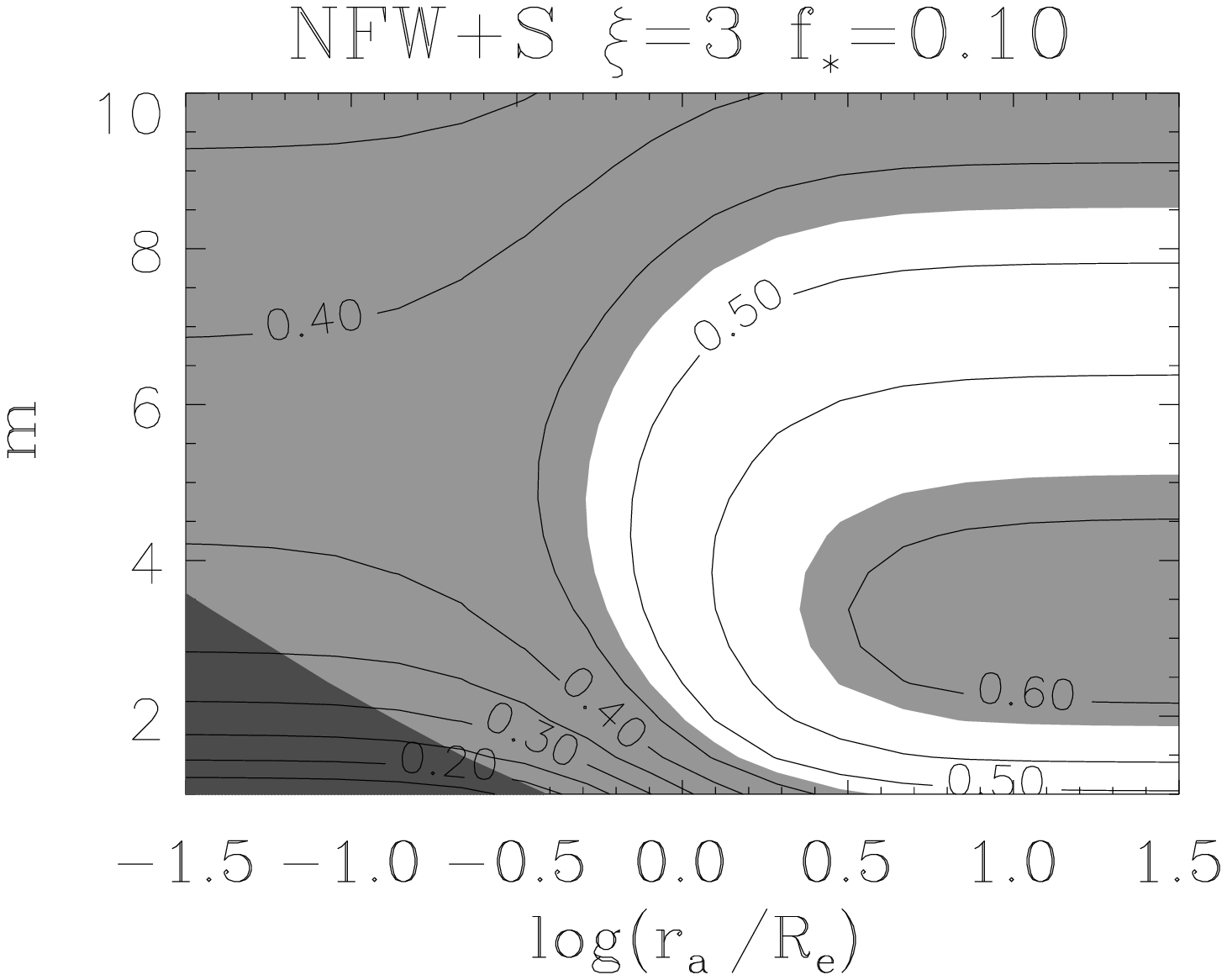}\\
\includegraphics[width=0.4\textwidth]{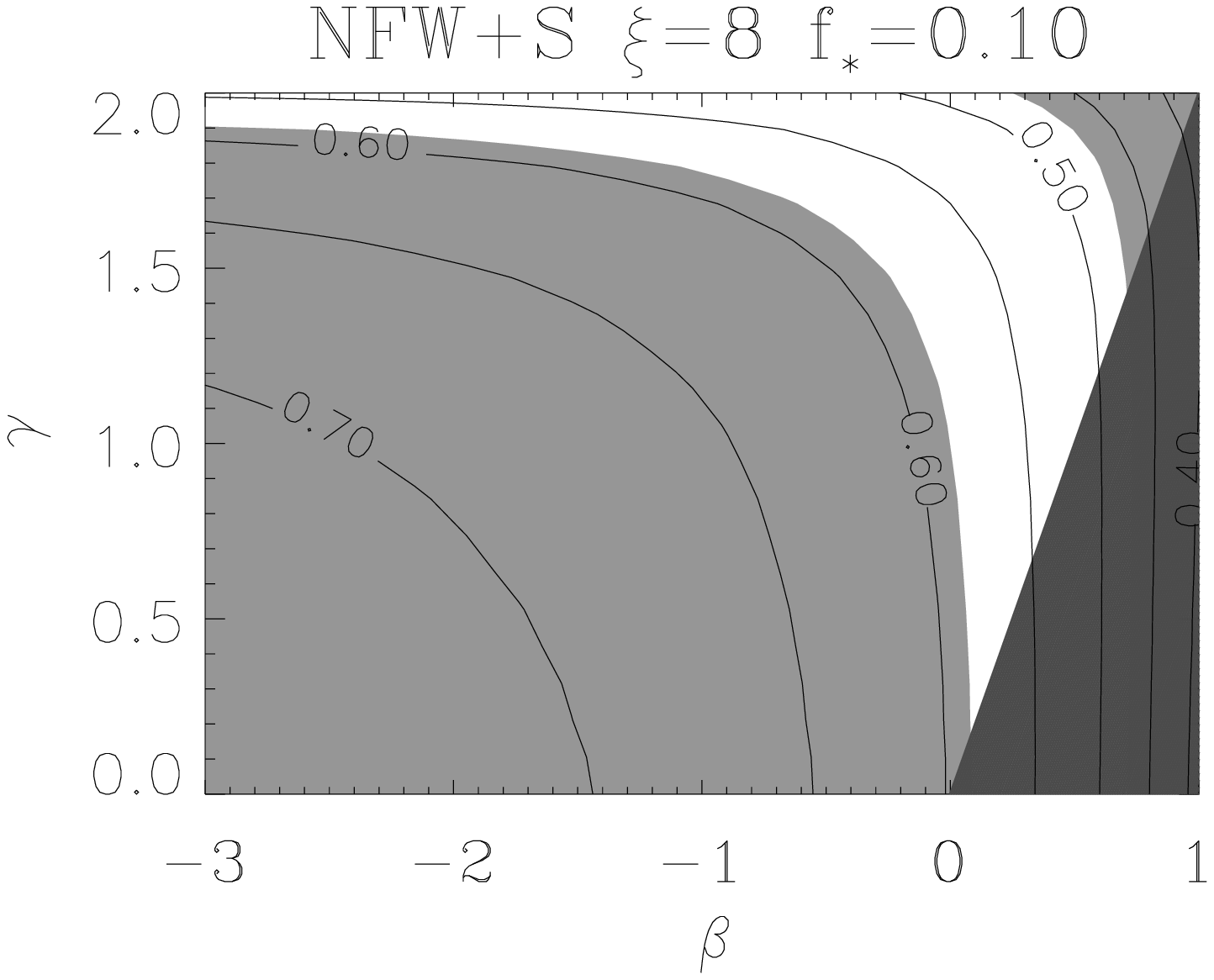}
\includegraphics[width=0.4\textwidth]{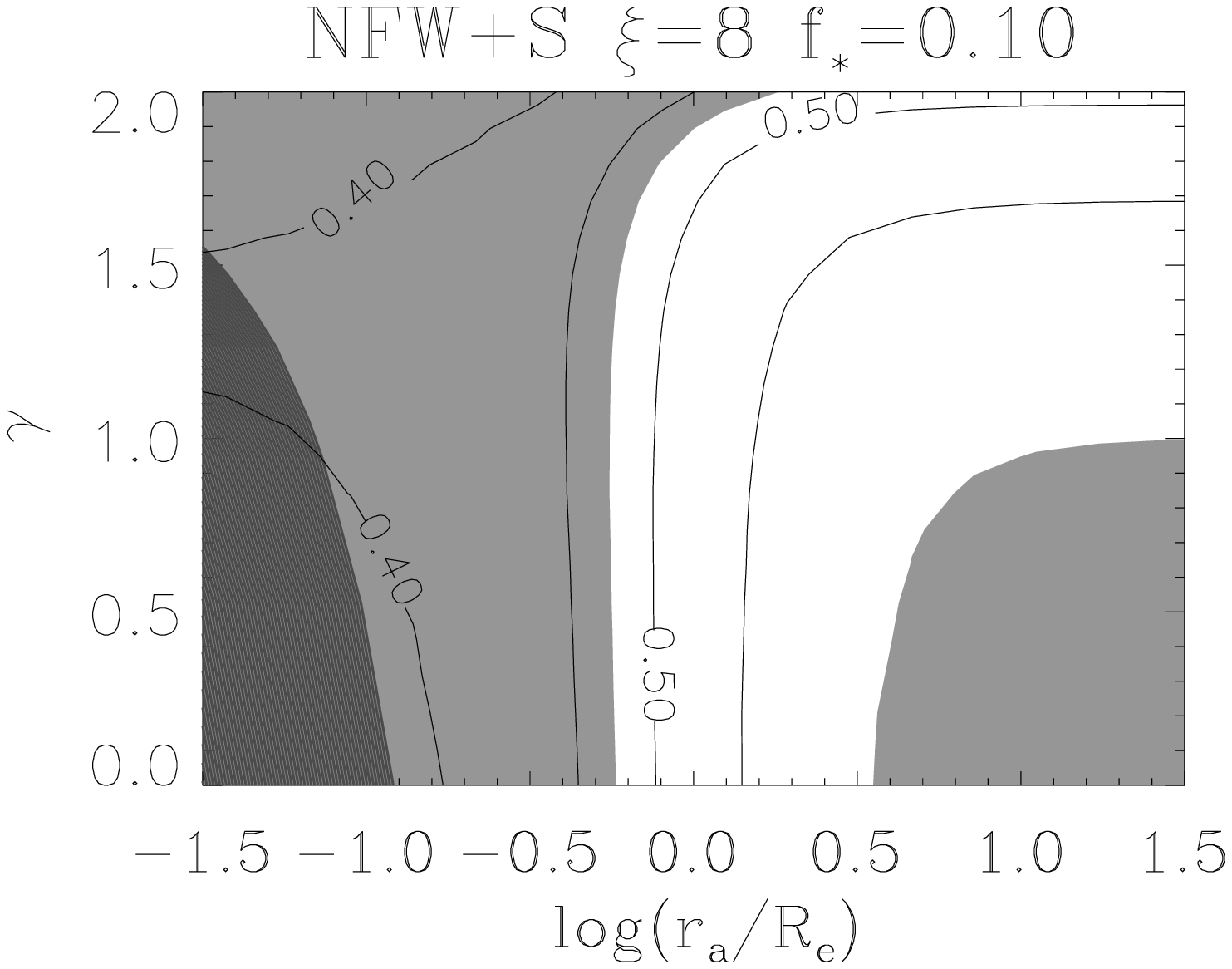}\\
\includegraphics[width=0.4\textwidth]{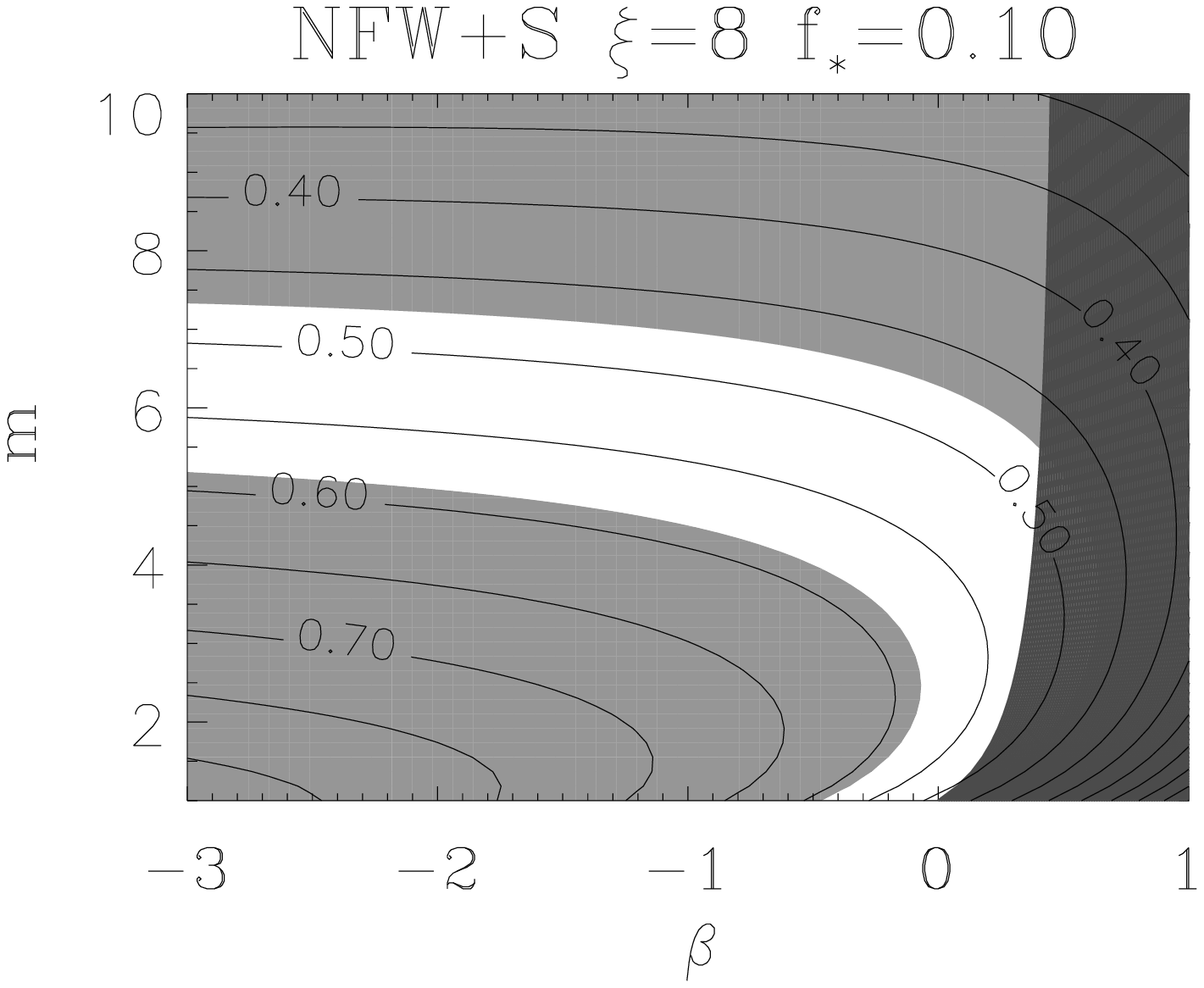}
\includegraphics[width=0.4\textwidth]{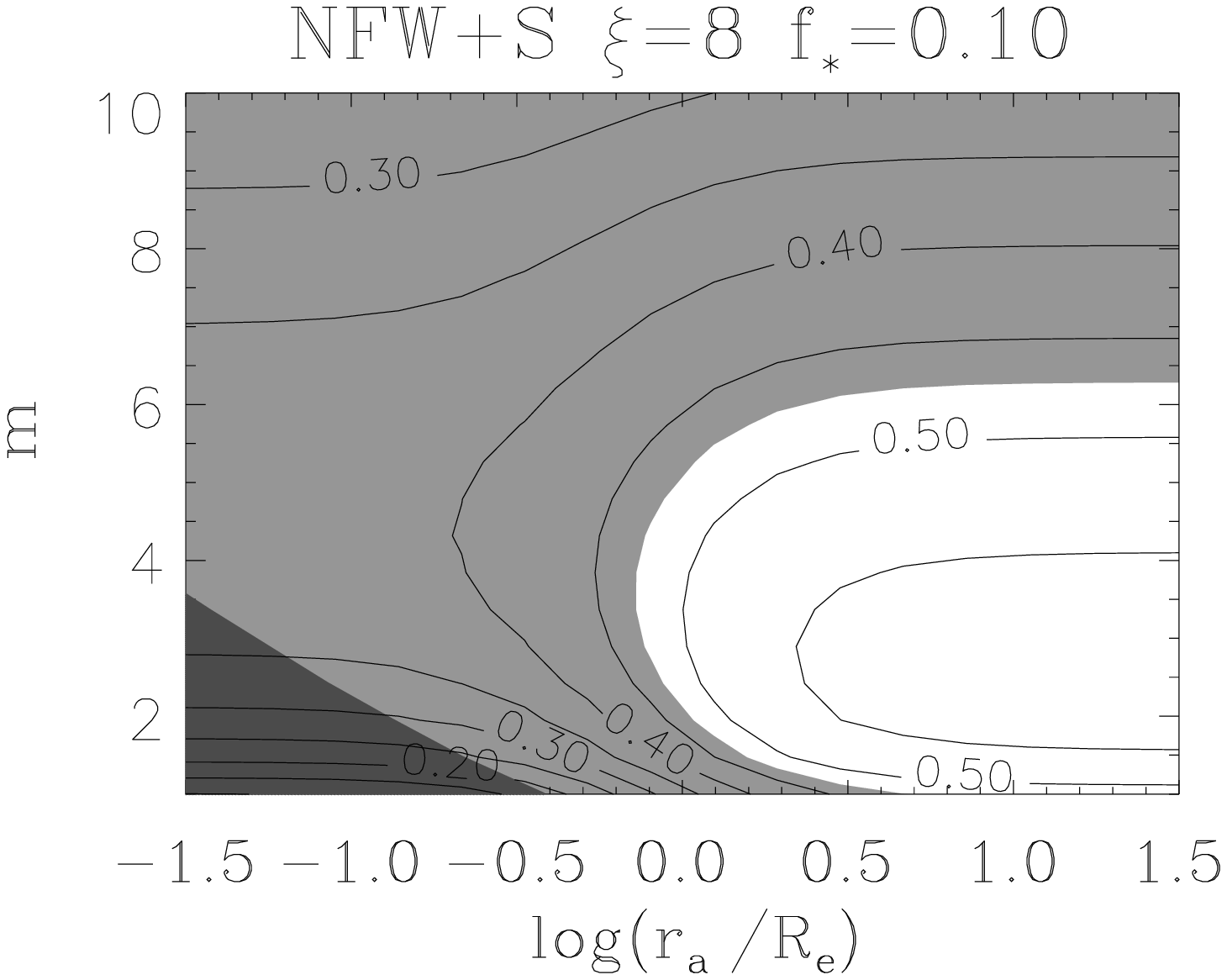}\\
\caption{Same as Fig.~\ref{figcsis}, but for NFW plus stars models with $\fstar=0.1$.}
\label{figcnfw10}
\end{figure*}

\begin{figure*}
\centering
\includegraphics[width=0.4\textwidth]{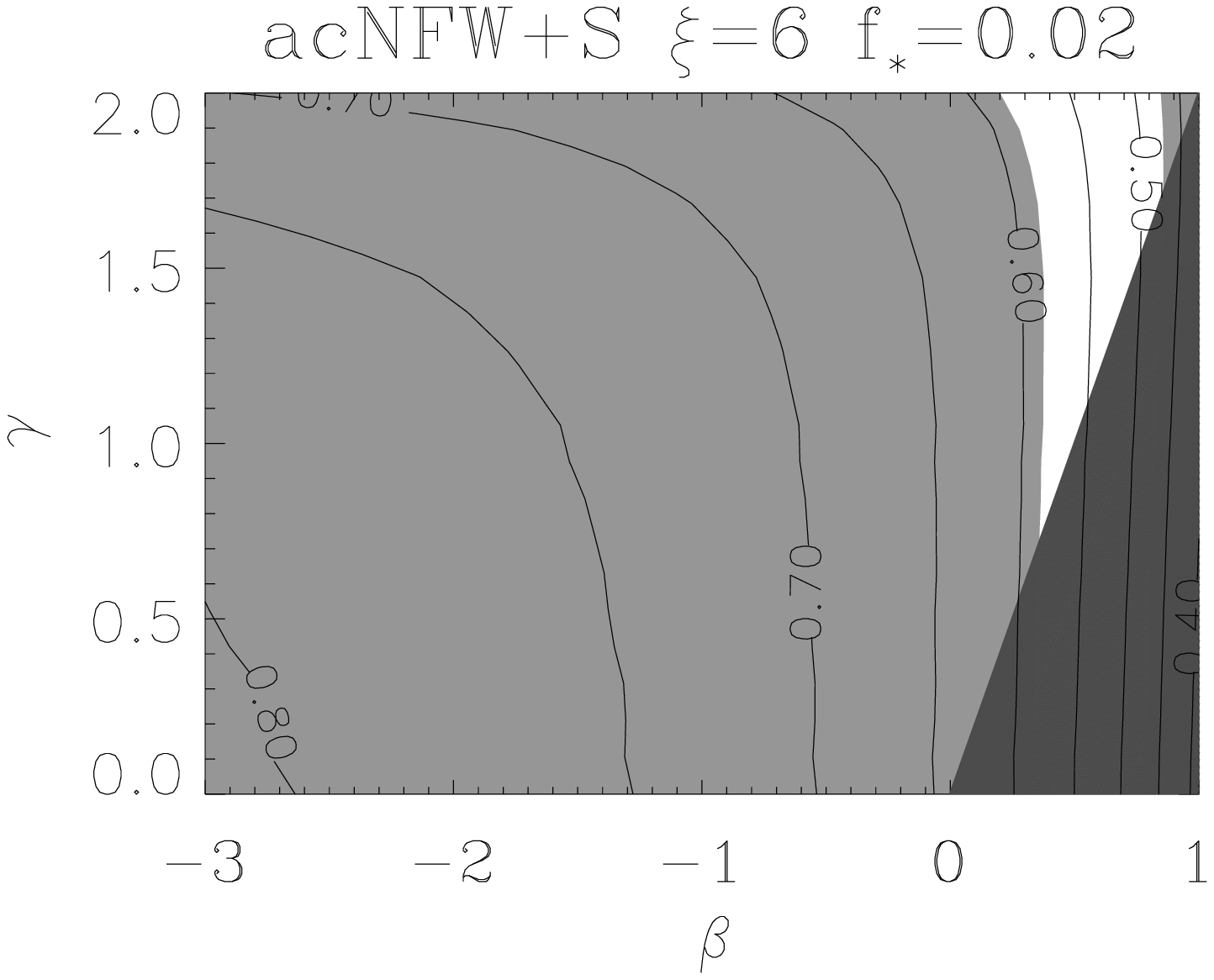}
\includegraphics[width=0.4\textwidth]{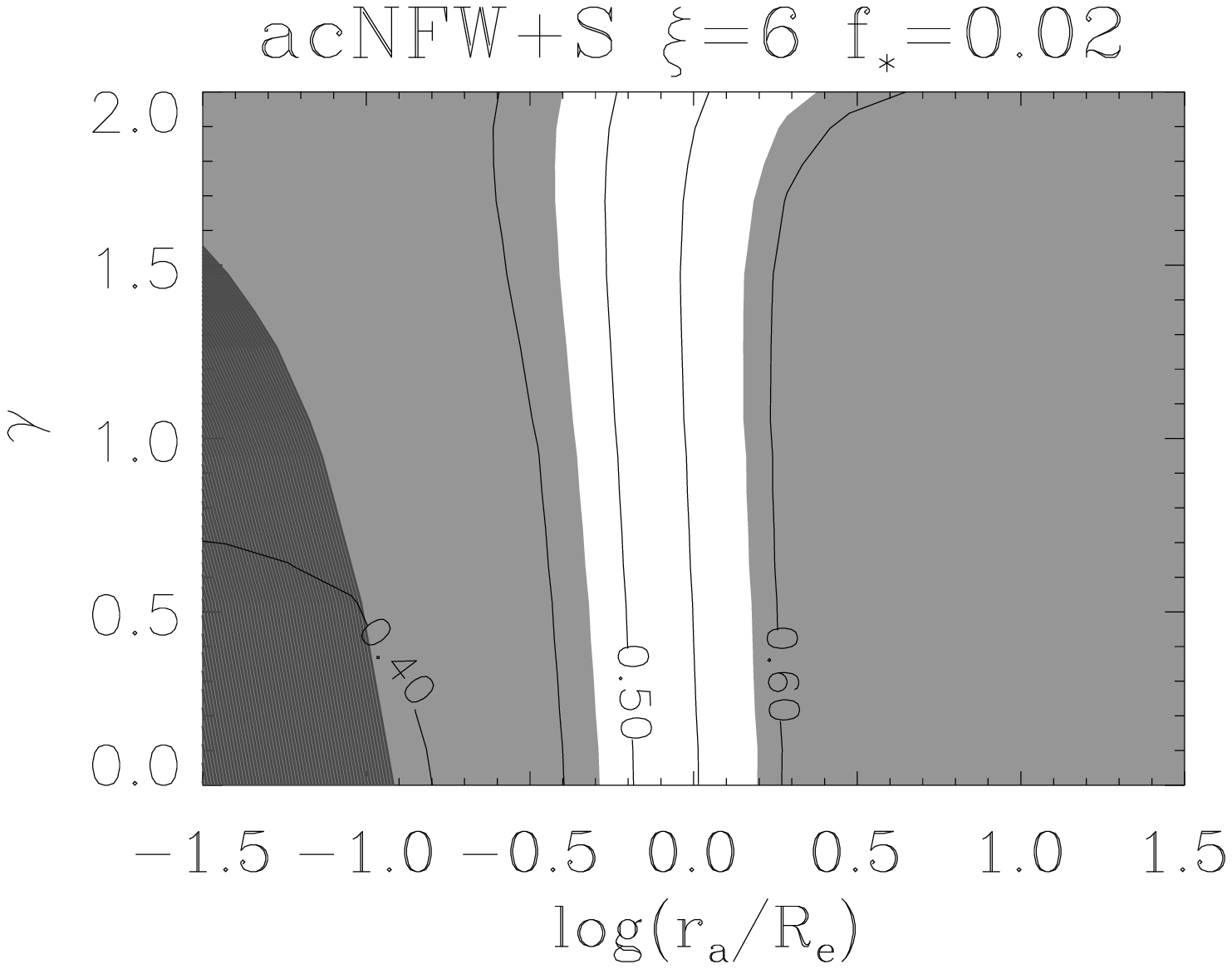}\\
\includegraphics[width=0.4\textwidth]{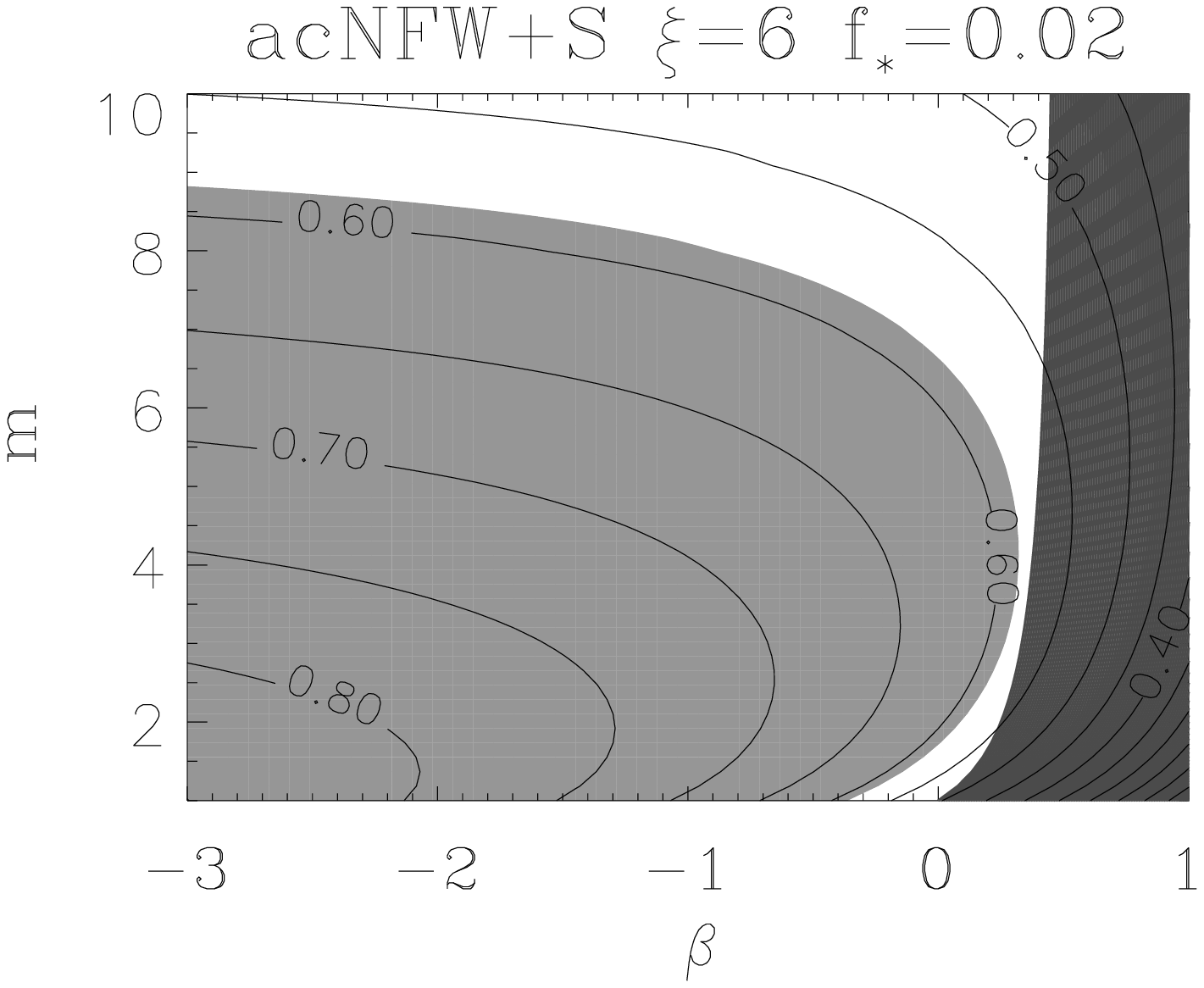}
\includegraphics[width=0.4\textwidth]{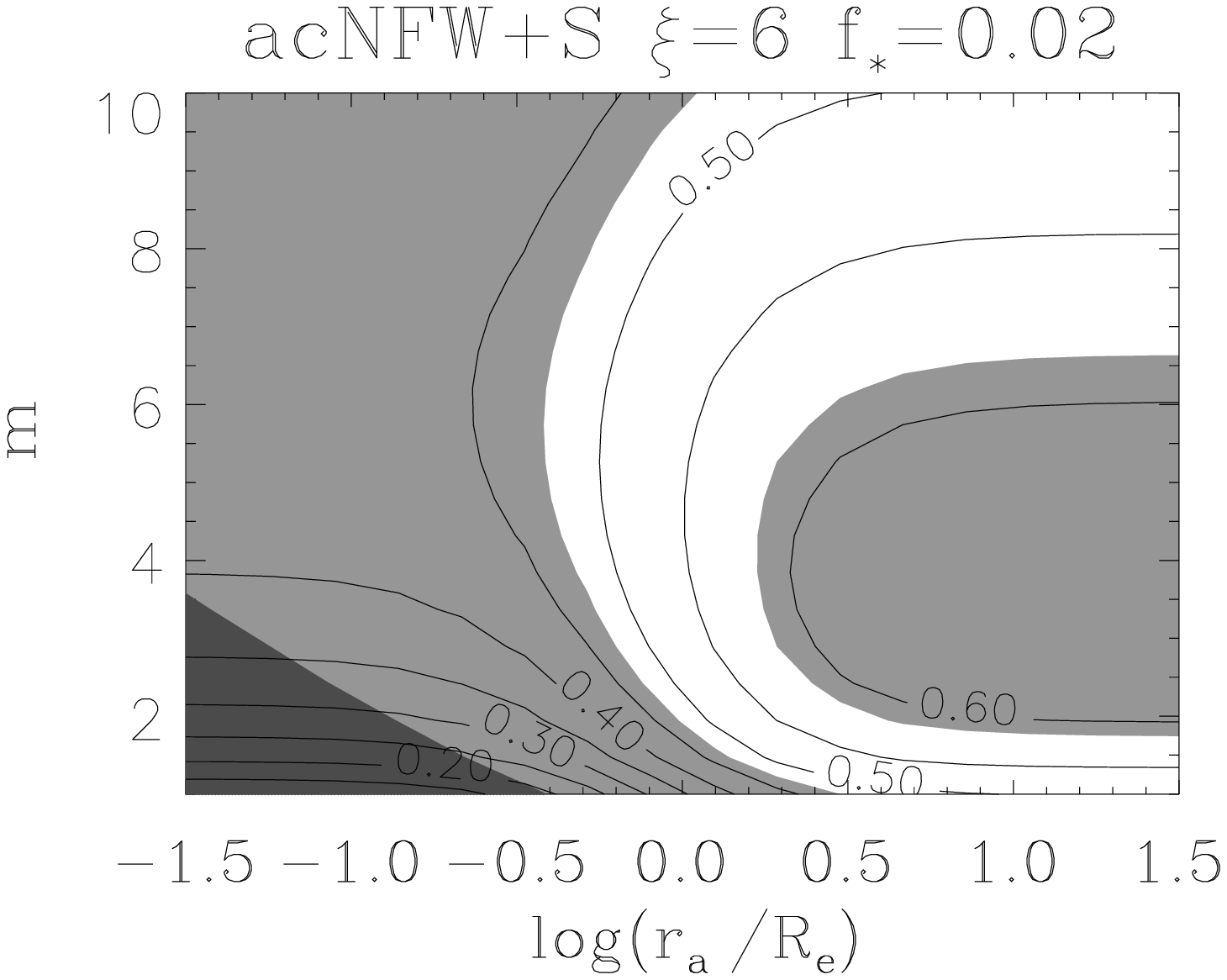}\\
\includegraphics[width=0.4\textwidth]{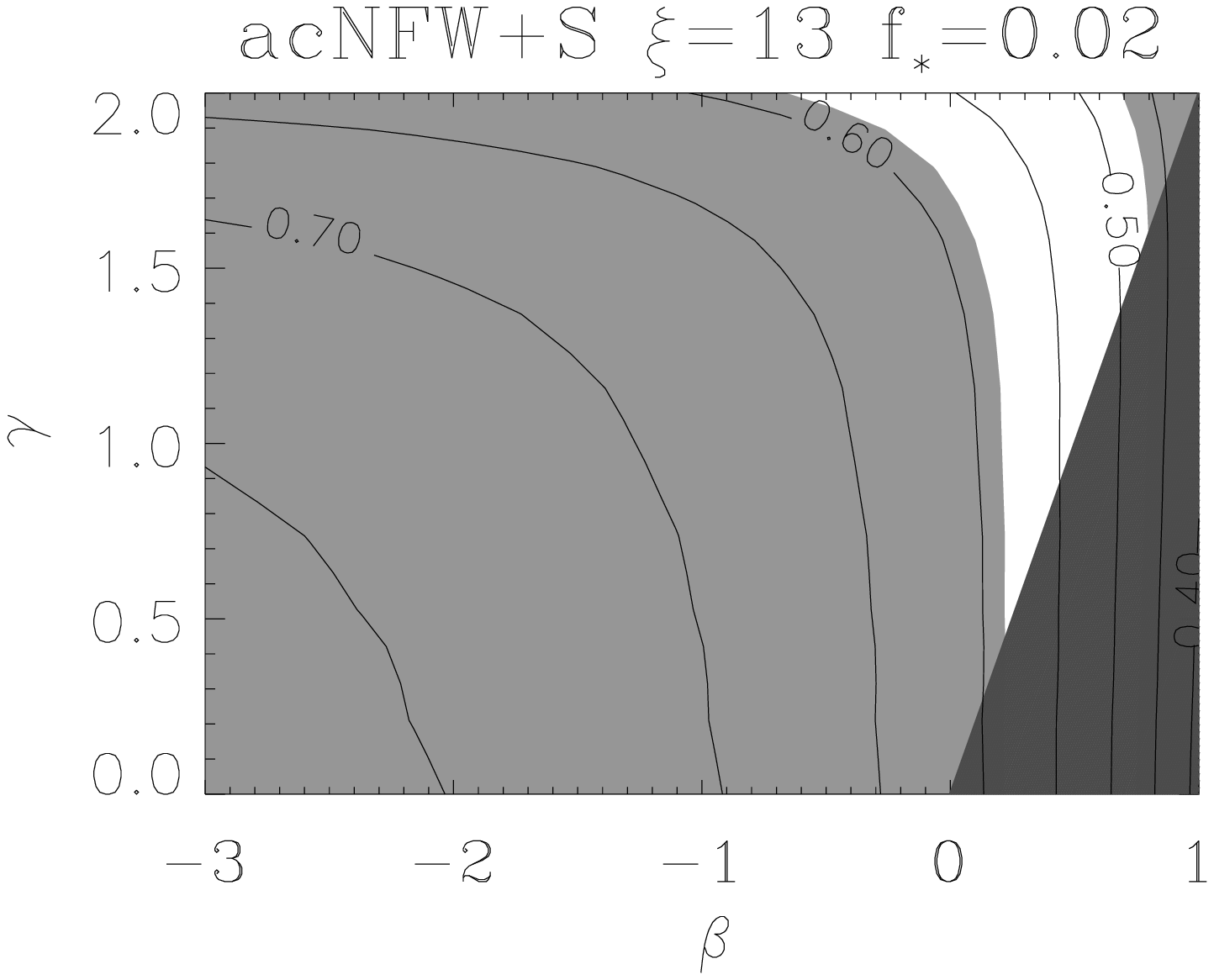}
\includegraphics[width=0.4\textwidth]{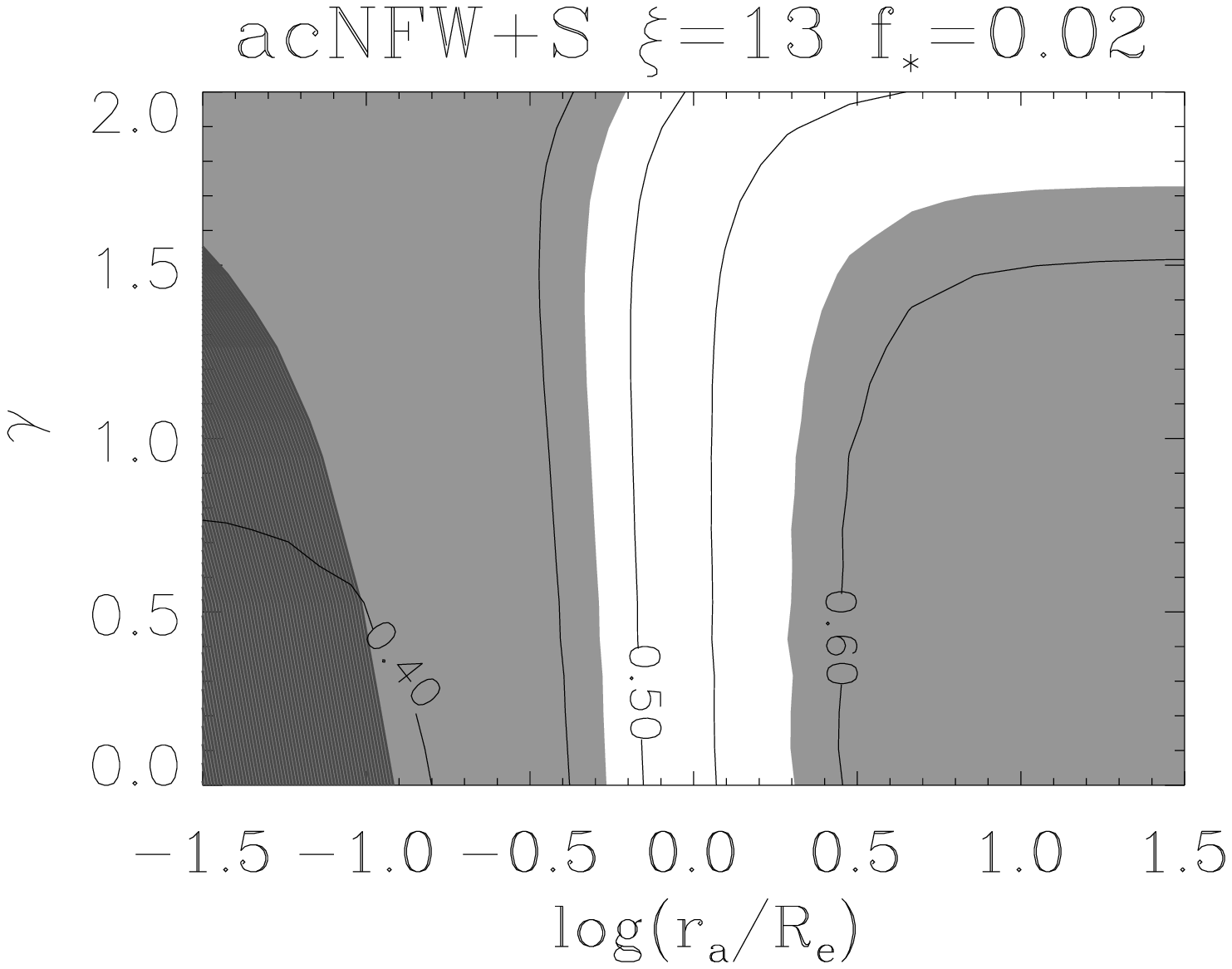}\\
\includegraphics[width=0.4\textwidth]{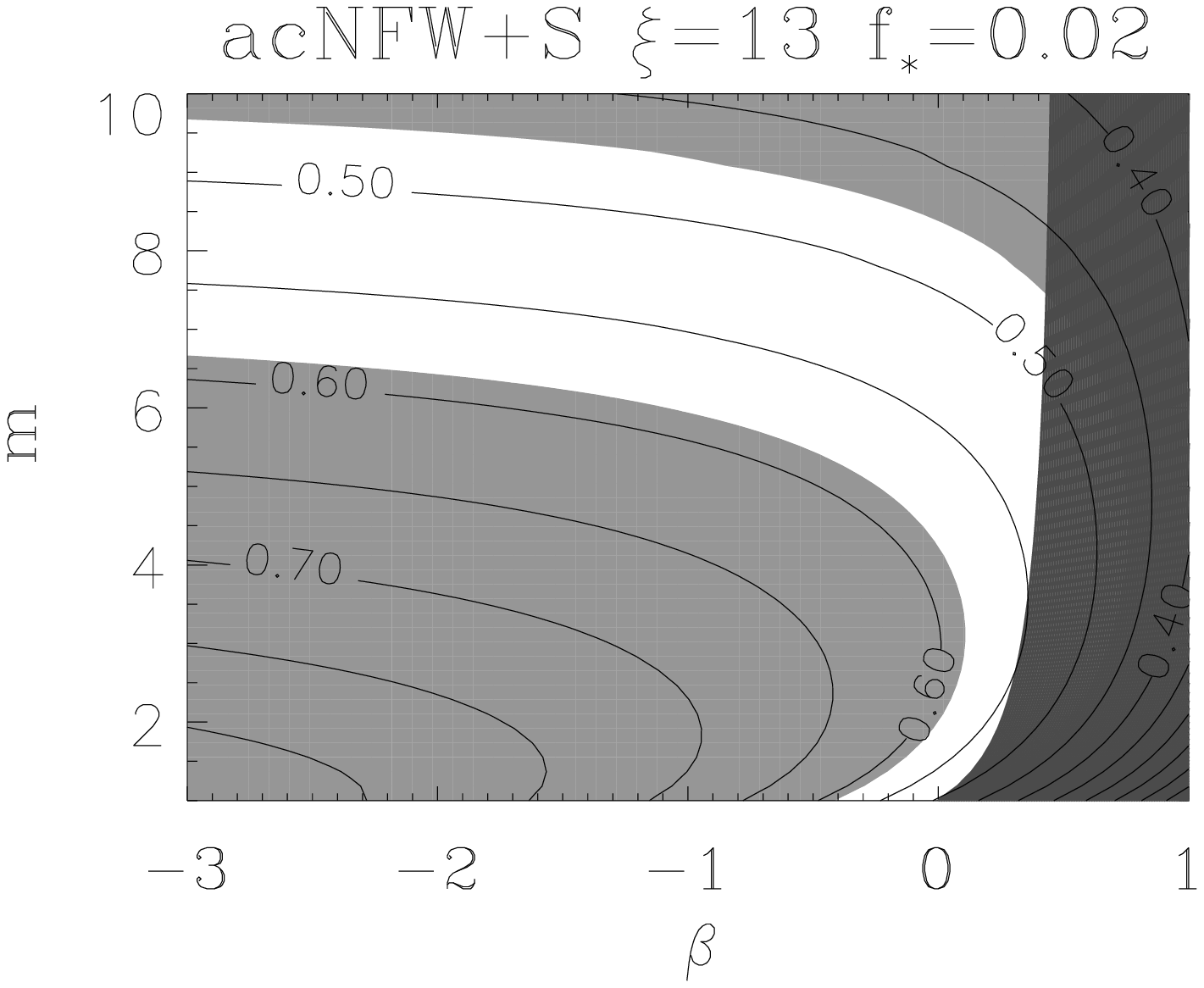}
\includegraphics[width=0.4\textwidth]{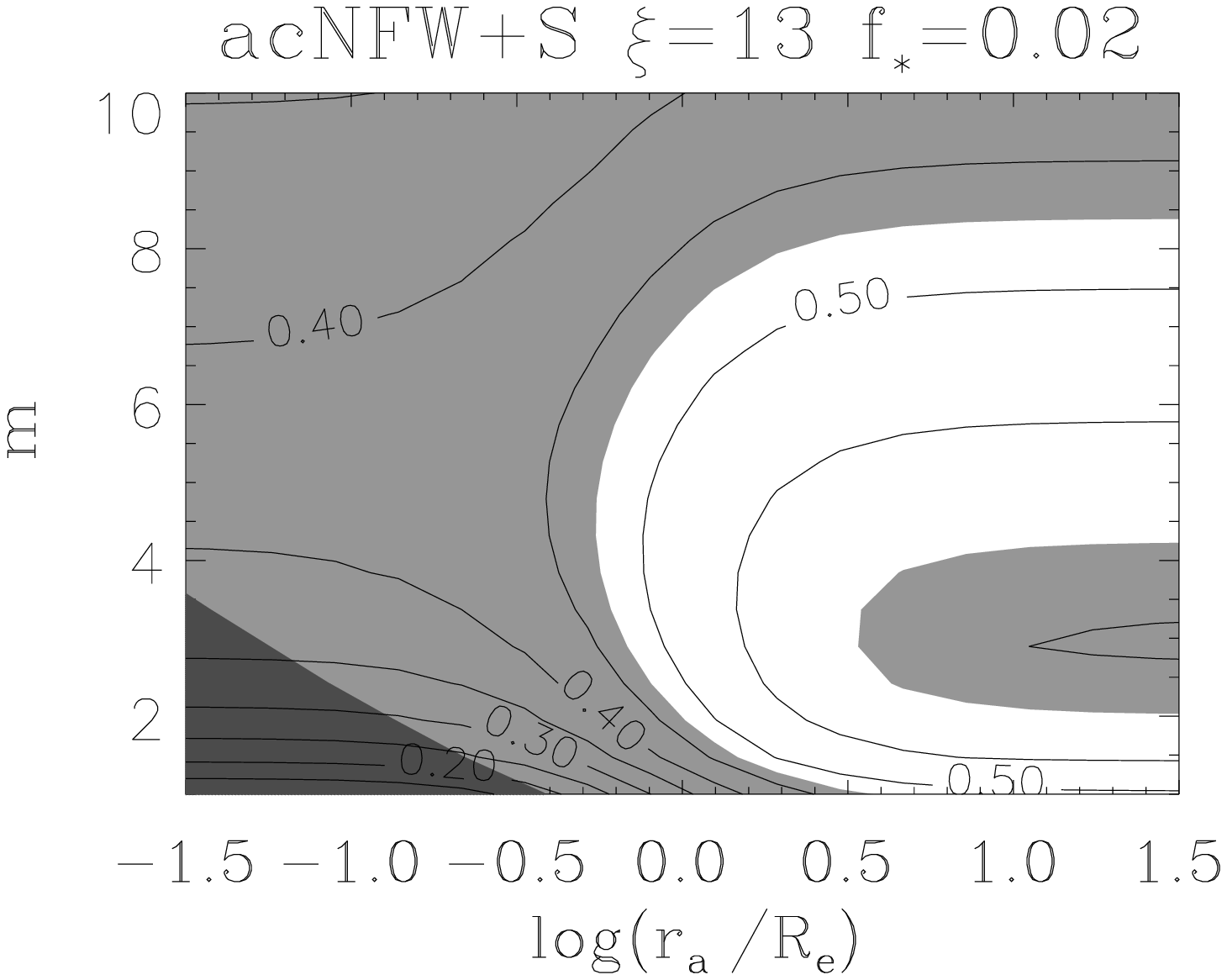}\\
\caption{Same as Fig.~\ref{figcsis}, but for adiabatically contracted NFW plus stars models
  with $\fstar=0.02$.}
\label{figcacnfw02}
\end{figure*}

\begin{figure*}
\centering
\includegraphics[width=0.4\textwidth]{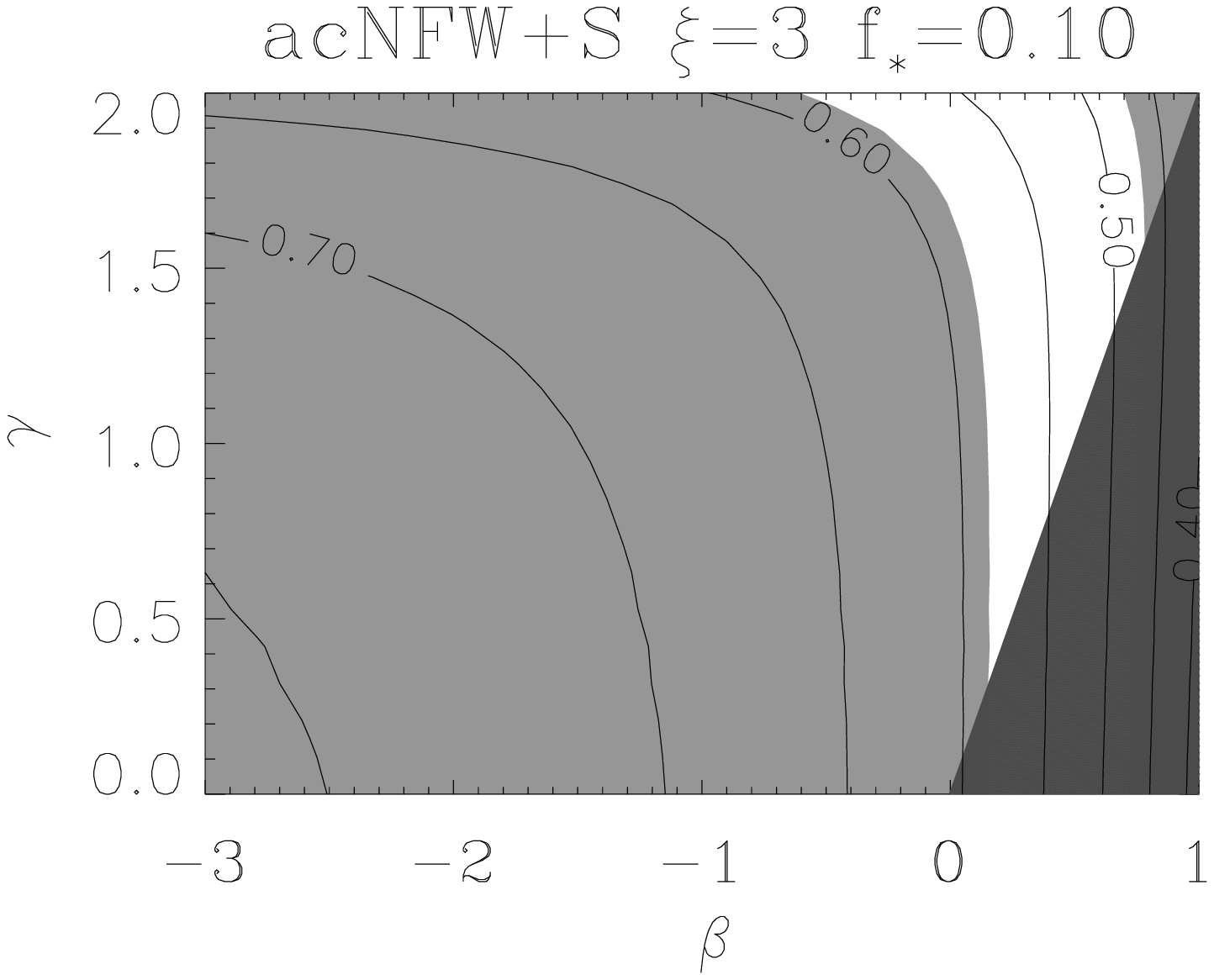}
\includegraphics[width=0.4\textwidth]{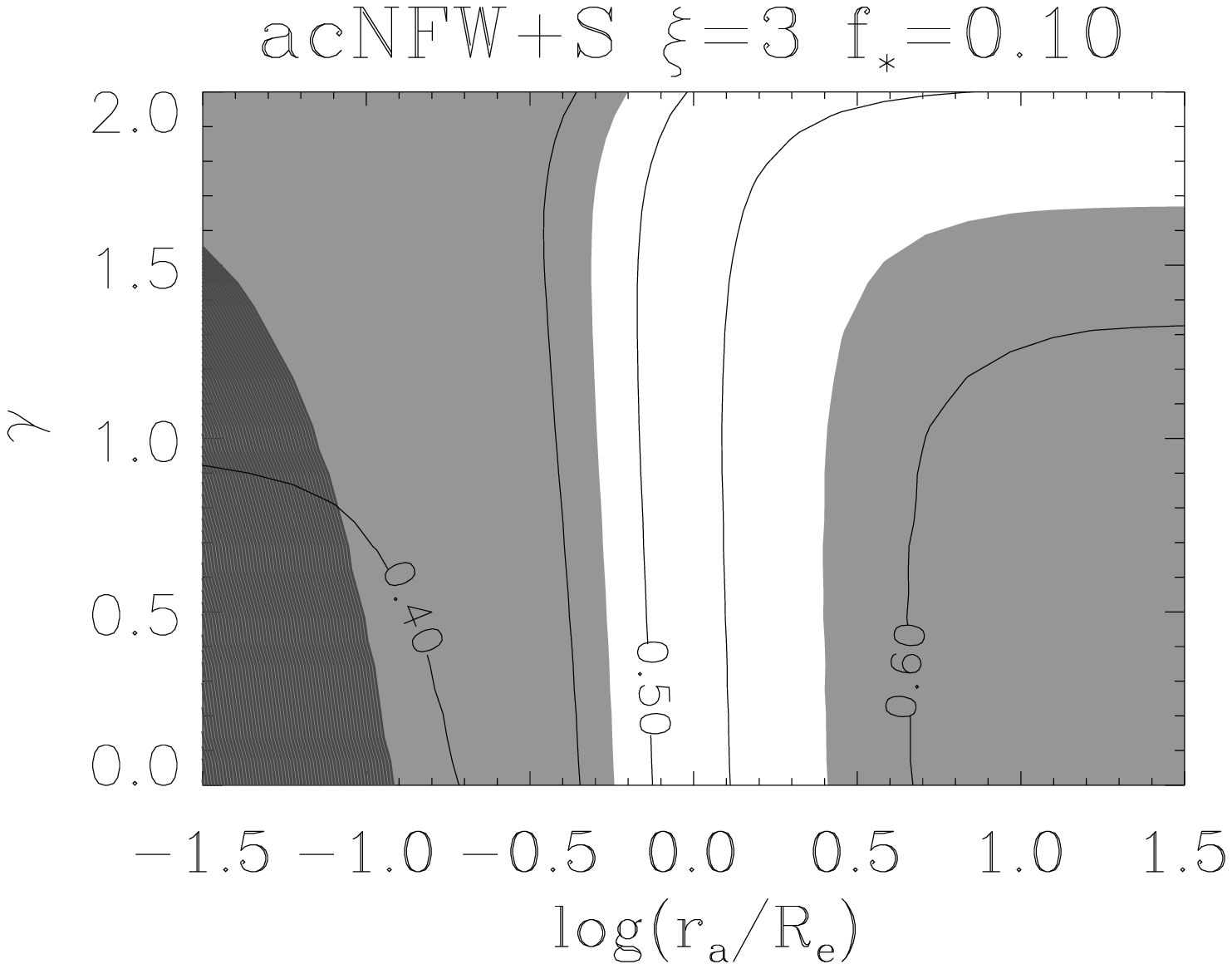}\\
\includegraphics[width=0.4\textwidth]{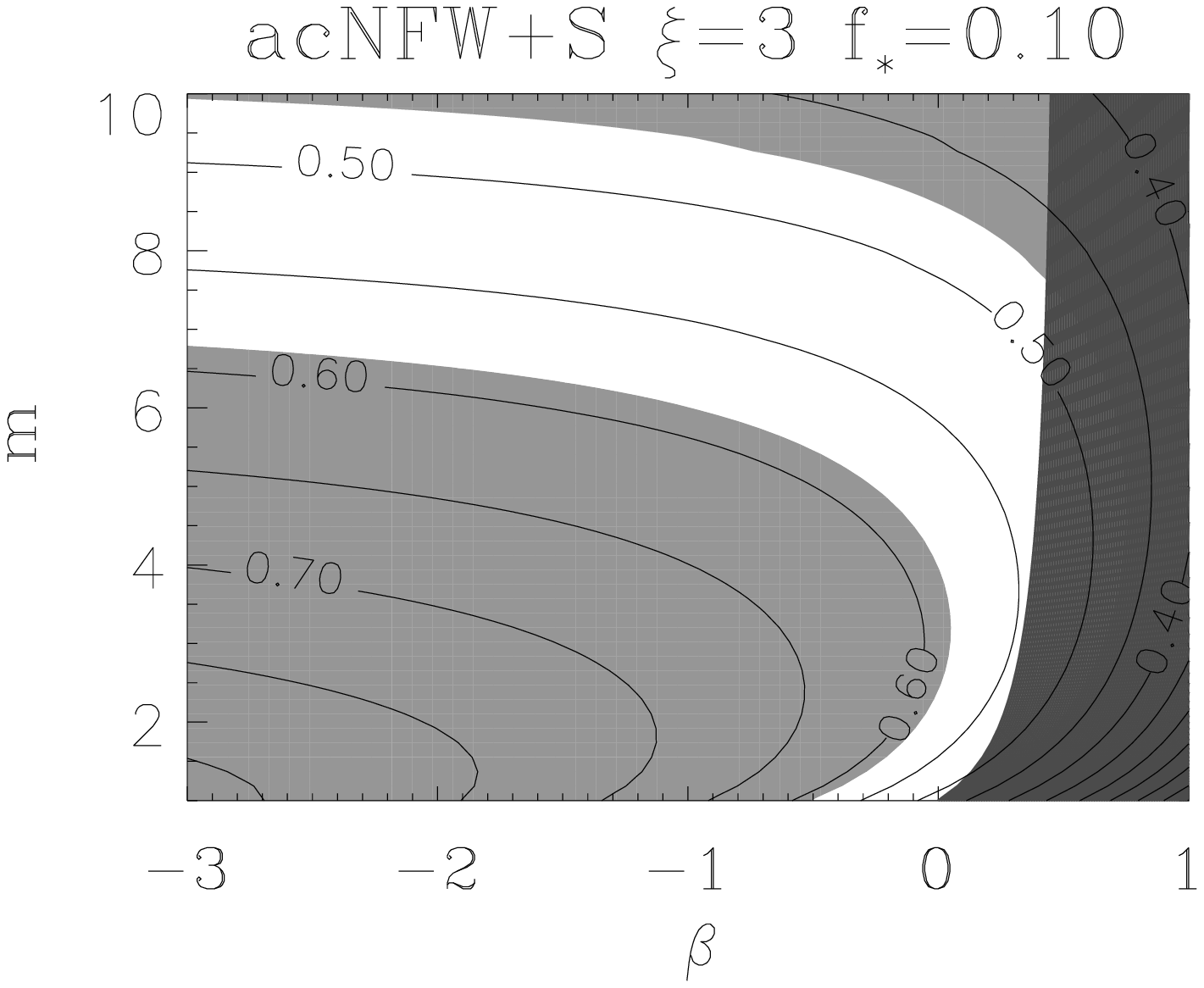}
\includegraphics[width=0.4\textwidth]{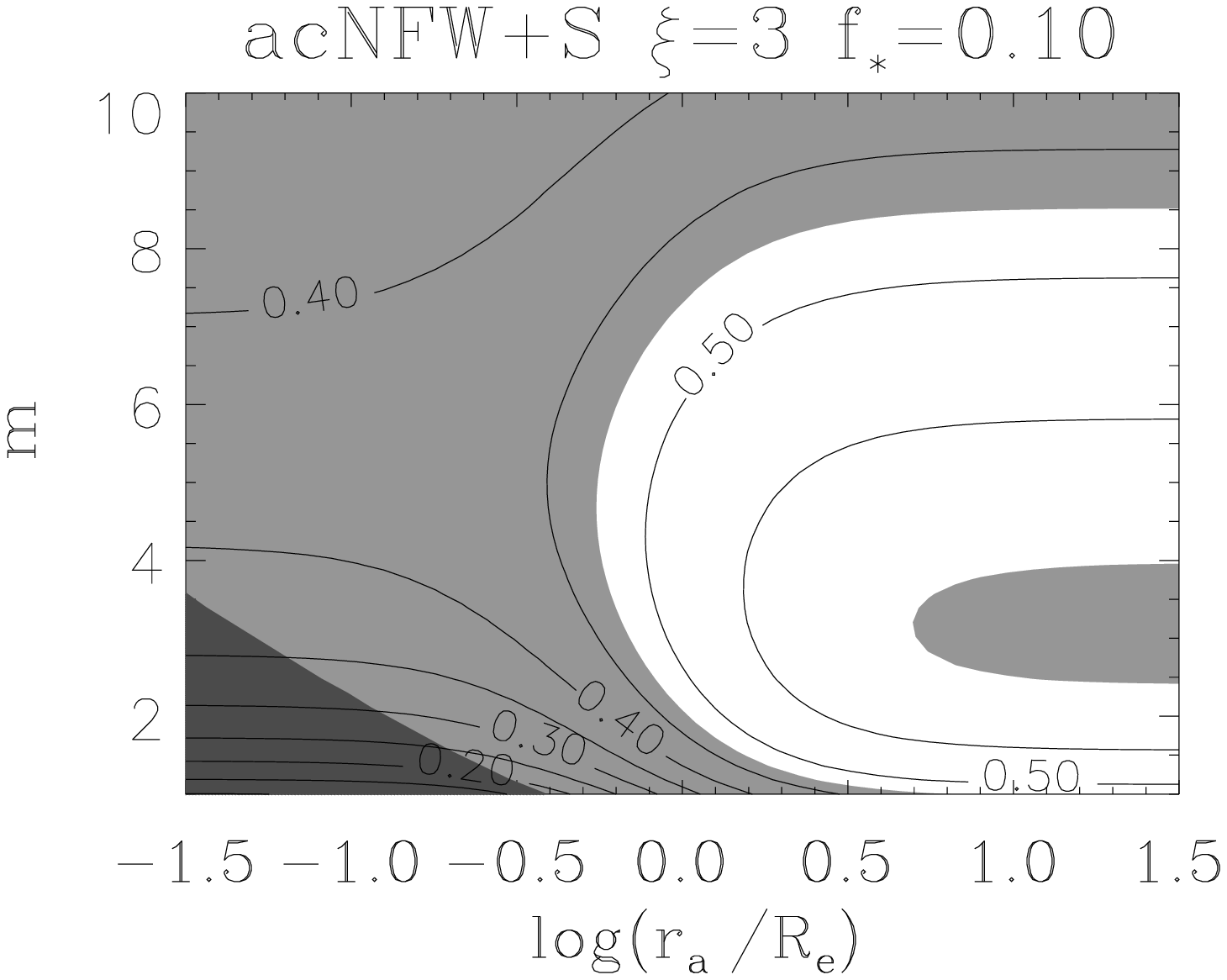}\\
\includegraphics[width=0.4\textwidth]{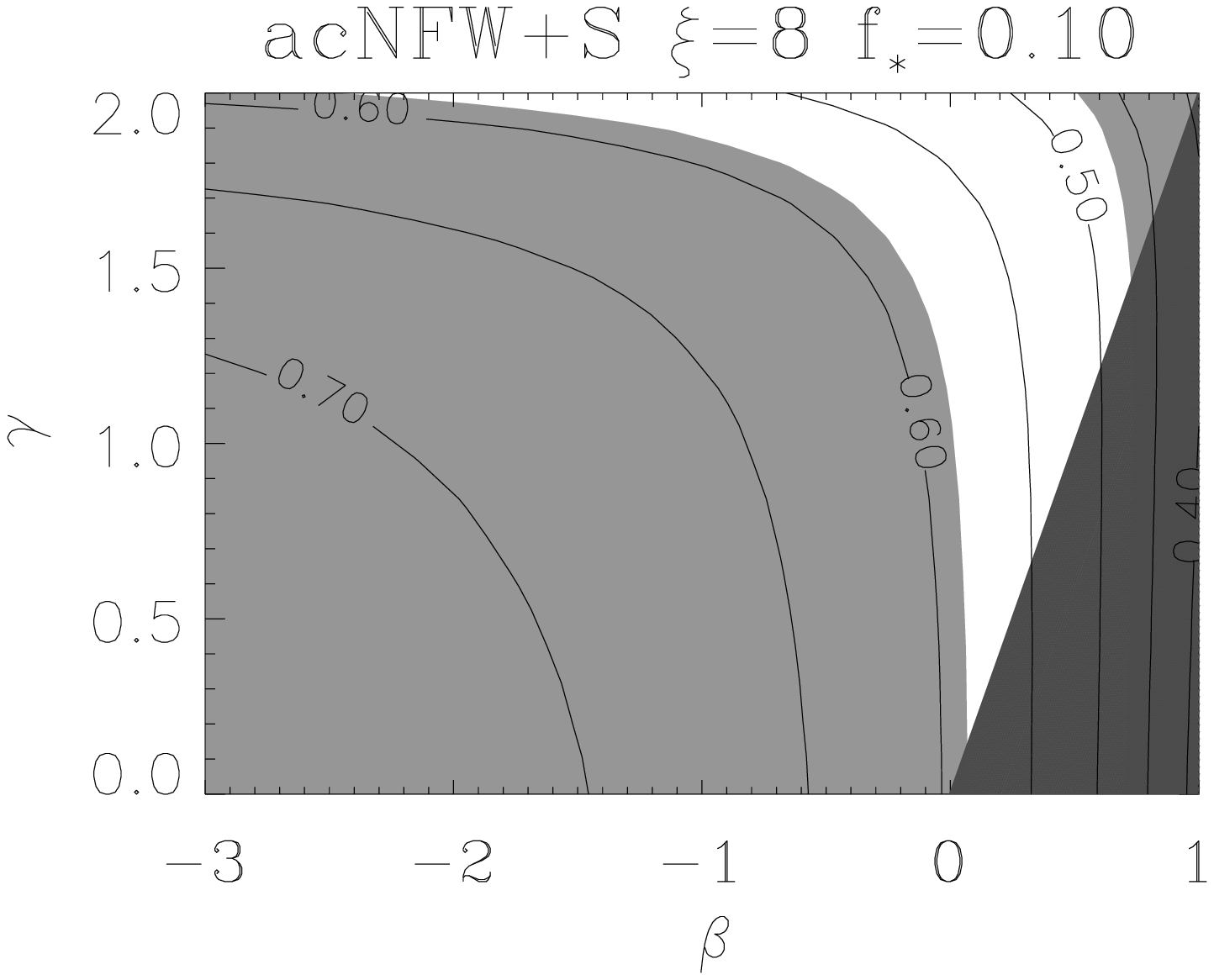}
\includegraphics[width=0.4\textwidth]{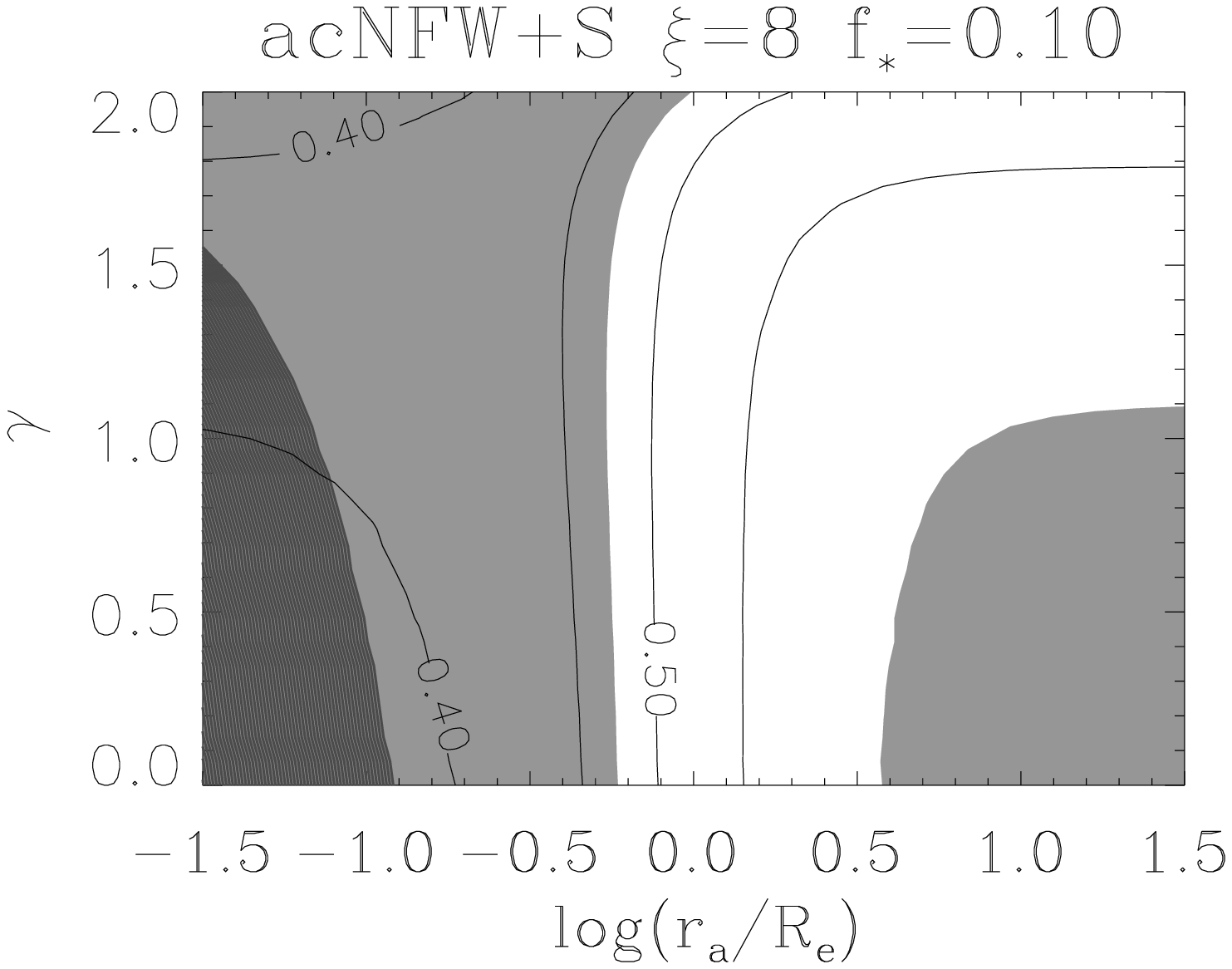}\\
\includegraphics[width=0.4\textwidth]{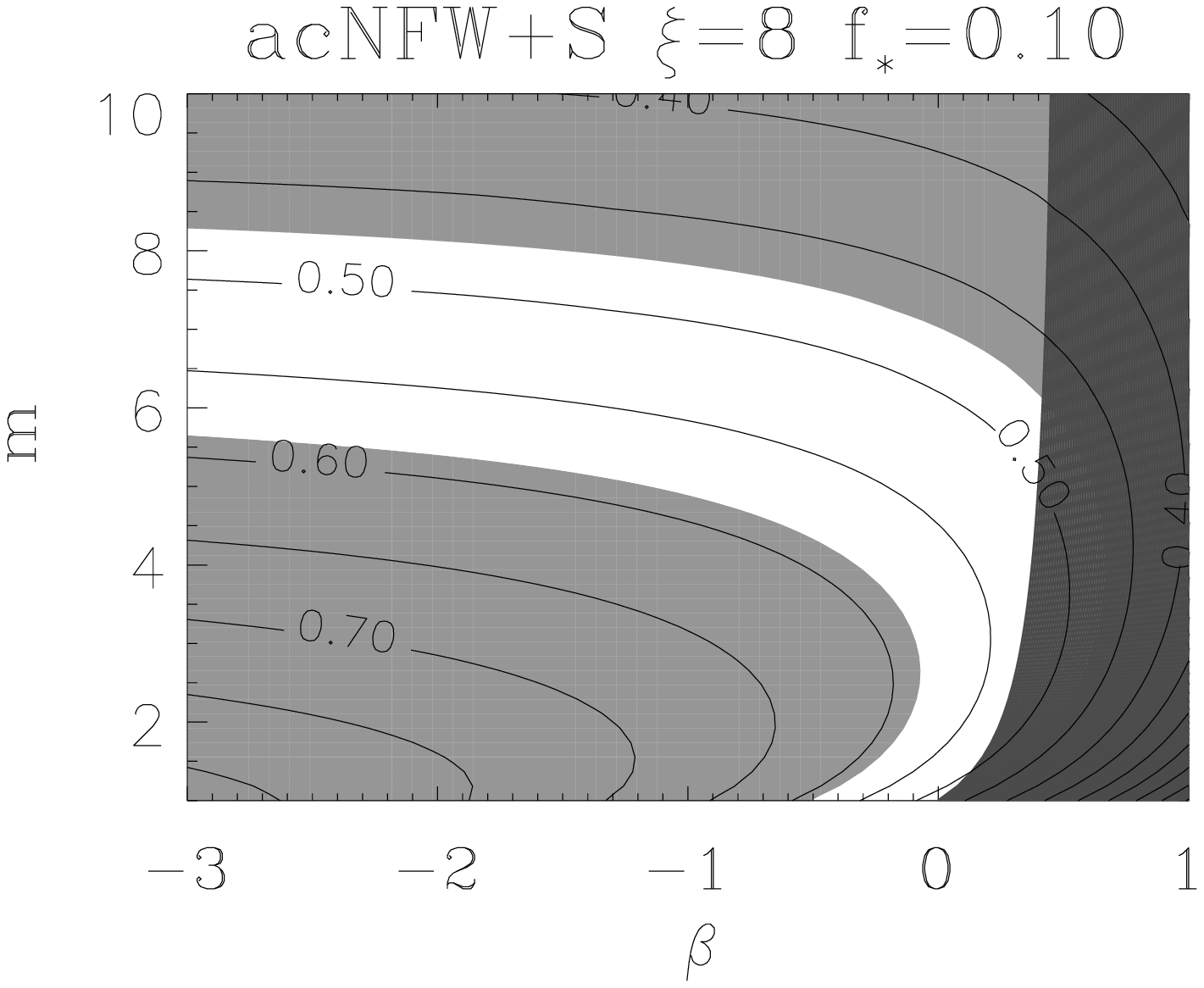}
\includegraphics[width=0.4\textwidth]{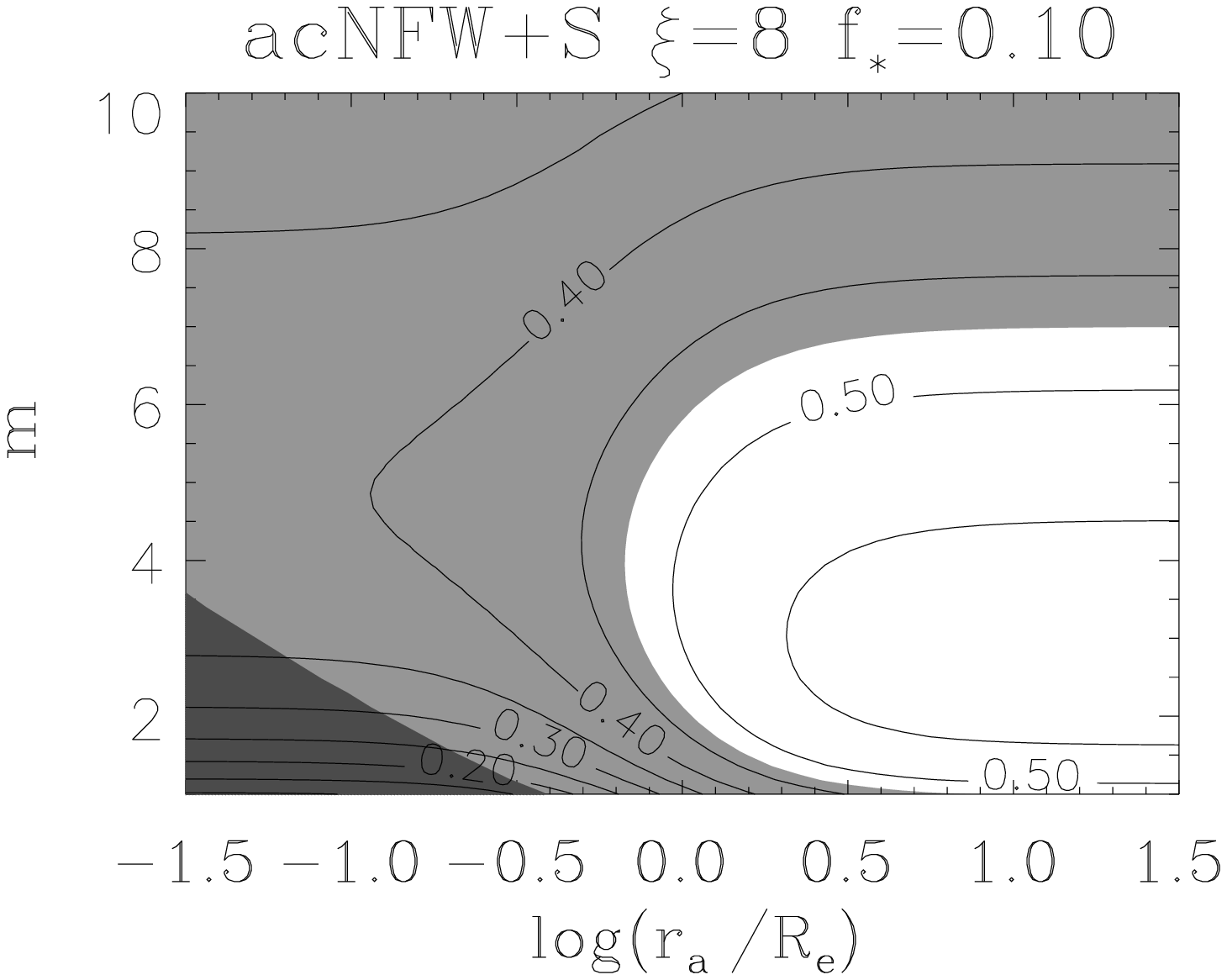}\\
\caption{Same as Fig.~\ref{figcsis}, but for adiabatically contracted NFW plus stars models with
$\fstar=0.1$.}
\label{figcacnfw10}
\end{figure*}

We numerically compute $\cet$ for $\gamma$ models with
$0\leq\gamma\leq 2$ and \Sersic models with $1\leq m\leq 10$ for
different total mass distributions and orbital anisotropy. Here we
discuss how the obtained values of $\cet$ compare with the observed
range $\langle \log \cet\rangle=0.53\pm0.057$ \citep{Bol08b}. For each
combination of family of stellar systems and total density
distribution, we represent our results using contour plots of $\log
\cet$ in planes of anisotropy versus stellar profile parameters
(Figs.~\ref{figcsis}-\ref{figcacnfw10}): $\beta$-$\gamma$ for
$\beta=\const$ $\gamma$ models, $\beta$-$m$ for $\beta=\const$ \Sersic
models, $\log (\ra/\Re)$-$\gamma$ for OM $\gamma$ models, and
$\log(\ra/\Re)$-$m$ for OM \Sersic models. In such plots, ``allowed''
regions are white, regions corresponding to non-consistent models are
dark-shaded, while regions outside the observed range are
light-shaded. 

OM models that are unphysical because non-consistent (dark-shaded
areas in the right-hand columns of
Figs.~\ref{figcsis}-\ref{figcacnfw10}) are also outside the observed
range. On the other hand, there are non-consistent $\beta=\const$
models with $\cet$ in the observed range. This different behaviour of
OM and $\beta=\const$ models is not surprising, because the latter are
known to be less realistic than OM models (see Section~\ref{secani}),
and, in particular, some radially anisotropic $\beta=\const$ models
turn out to be unphysical because they are radially anisotropic down
to the very centre of the system.  This finding stresses the
importance of investigating consistency when modeling observational
data.

At a first level of interpretation, the plots in
Figs.~\ref{figcsis}-\ref{figcacnfw10} show that a relatively wide
class of models have values of the structure parameter $\cet$ in the
observed range. Thus, the fact that the observed values of $\cet$ lie
in a small range does not necessarily imply that early-type galaxies
are structurally and dynamically homologous.  The reason for this is
that $\cet$ is not dramatically sensitive to the mass distribution and
orbital anisotropy. However, a more detailed analysis of the diagrams
indicates that there is also a wide class of models that lie outside
the region allowed by the observations. Here we summarize the
behaviour of families of models with different total mass
distribution:
 
\begin{itemize}

\item SIS models (Fig.~\ref{figcsis}): SIS models are consistent with
  observational constraints for a wide class of stellar density
  profile and anisotropy. All isotropic $\gamma$ models and isotropic
  \Sersic models with $m\lsim 5.5$ have $\cet$ within the observed
  range. Higher-$m$ \Sersic models can be reconciled with the
  observations if their stars have radially-biased orbits. Strong
  radial and tangential anisotropy is excluded. However, very strong
  radial anisotropy should be excluded on the basis of consistency
  arguments, while, as briefly discussed in Section~\ref{secani}, very
  strong tangential anisotropy is not expected. One might speculate
  that the OM models excluded by the observed range trace roughly the
  region of radially unstable models (see Section~\ref{seccons}).

\item LTM models (Fig.~\ref{figcltm}): isotropic (and mildly radially
  anisotropic) $\gamma$ models with LTM potential are consistent with
  the observational constraints, while isotropic (and mildly radially
  anisotropic) \Sersic models with LTM potential are acceptable only
  for $m\lsim 5.5$. In contrast with the case of SIS models, there is
  no way of reconciling $m\gsim 6.5$ \Sersic models with the
  observations. As in the case of SIS models, the lower limit on
  $\ra/\Re$ for OM models might be determined by radial-orbit
  instability.  More tangential anisotropy than in SIS models is
  allowed, though with some fine-tuning with the stellar profile
  parameters $m$ and $\gamma$. Curiously, if one considers
  $\gamma=2$ models under the assumption of LTM and constant
  anisotropy, the observational constraints would favour
  tangential  with respect to radial anisotropy.

\item NFW+S models (Figs.~\ref{figcnfw02} and \ref{figcnfw10}): for
  some NFW+S models only remarkably small regions of the parameter
  space are allowed, in contrast with the case of SIS models. The
  worst case is that of more dark-matter dominated models
  ($\fstar=0.02$, $\xi=6$): remarkably, isotropic $\gamma$ models are
  excluded, and isotropic \Sersic models are allowed only for $m \gsim
  7.5$).  Better is the most baryon dominated case ($\fstar=0.1$,
  $\xi=8$), which---as expected---behaves similarly to the LTM case,
  so OM \Sersic models can be reconciled with the observations only
  for $m\lsim 6$. In some cases (e.g. OM \Sersic NFW+S models) the
  anisotropy and stellar-profile parameters must be fine-tuned in
  order to have $\cet$ in the observed range. For NFW+S models with
  acceptable $\cet$ we find $0.1\lsim\fdm\lsim0.6$ and
  $0.2\lsim\fdmp\lsim0.7$, where $\fdm$ and $\fdmp$ are, respectively,
  the intrinsic and projected dark-matter-to-total mass ratios within
  $\Re$.

\item acNFW+S models (Figs.~\ref{figcacnfw02} and \ref{figcacnfw10}):
overall, adiabatically contracted models behave similarly to
non-contracted models, but allowed regions in the parameter space are
slightly more extended in acNFW+S models than in the corresponding
NFW+S models. As in the case of NFW+S, baryon-dominated models (larger
values of $\fstar$ and $\xi$) are more successful than dark-matter
dominated models (smaller values of $\fstar$ and $\xi$). However,
acNFW+S have some of the same undesirable features present in NFW+S
models, such as a wide class of unacceptable isotropic models
(especially when $\fstar=0.02$).  For dark-matter dominated OM \Sersic
acNFW+S models to have $\cet$ in the observed range, the anisotropy
and stellar-profile parameters must be fine-tuned. The acNFW+S models
with $\cet$ within the observed range have $0.3\lsim\fdm\lsim0.65$ and
$0.35\lsim\fdmp\lsim0.75$.

\end{itemize}

Summarizing, our results indicate that the tightness of the MP
requires some degree of ``fine tuning'' in the internal properties of
early-type galaxies. Although no family of models is strictly ruled
out, some families of models allow for more freedom in the remaining
parameters describing, e.g., the luminous profile and stellar orbits.
The tightness of the MP is well consistent with the hypothesis that
the total mass profile is that of a SIS, allowing a broad range of
values for the other parameters. The LTM hypothesis is acceptable if
early-type galaxies have intrinsic stellar density profiles described
by $\gamma$ models or \Sersic projected stellar density profiles with
index $m\lsim 5.5$.  For either non-contracted or adiabatically
contracted NFW cases, baryon-dominated models require less fine-tuning
than dark-matter-dominated models. In general, isotropic or mildly
radially anisotropic velocity distribution is easier to reconcile with
the MP than tangential or extremely radial anisotropy.

\section{S\'ersic index and the tilt of the Mass Plane}
\label{secser}

In the previous Section we have interpreted the average value and
intrinsic scatter of the parameter $\cet$ describing the MP. In this
Section we consider how future observations of the MP over a broader
range of galaxy masses could be used to further constrain the internal
structure of early-type galaxies, based on the known structural
non-homology of their luminous component.

\begin{figure*}
\centering
\includegraphics[width=0.32\textwidth]{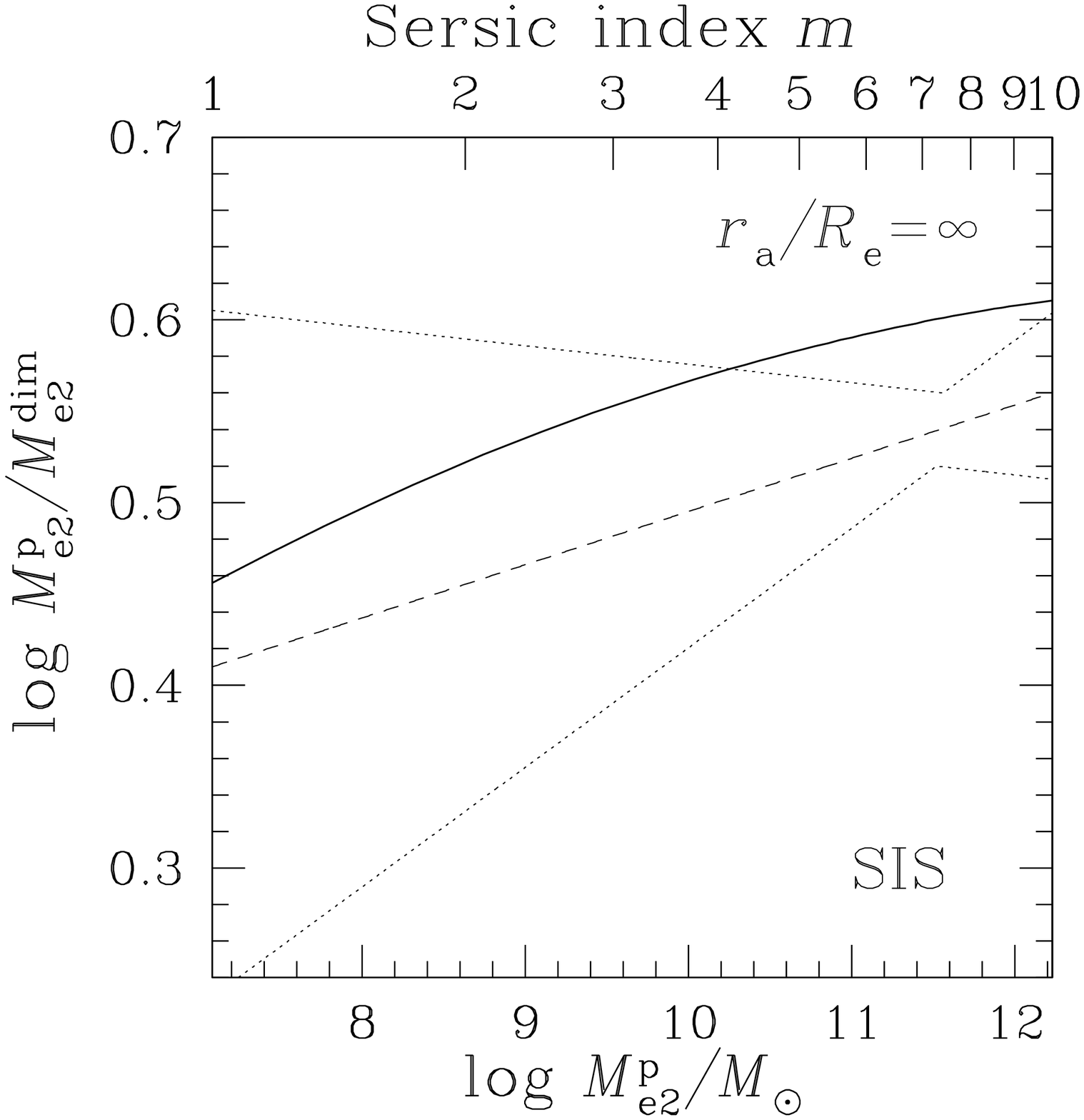}\includegraphics[width=0.32\textwidth]{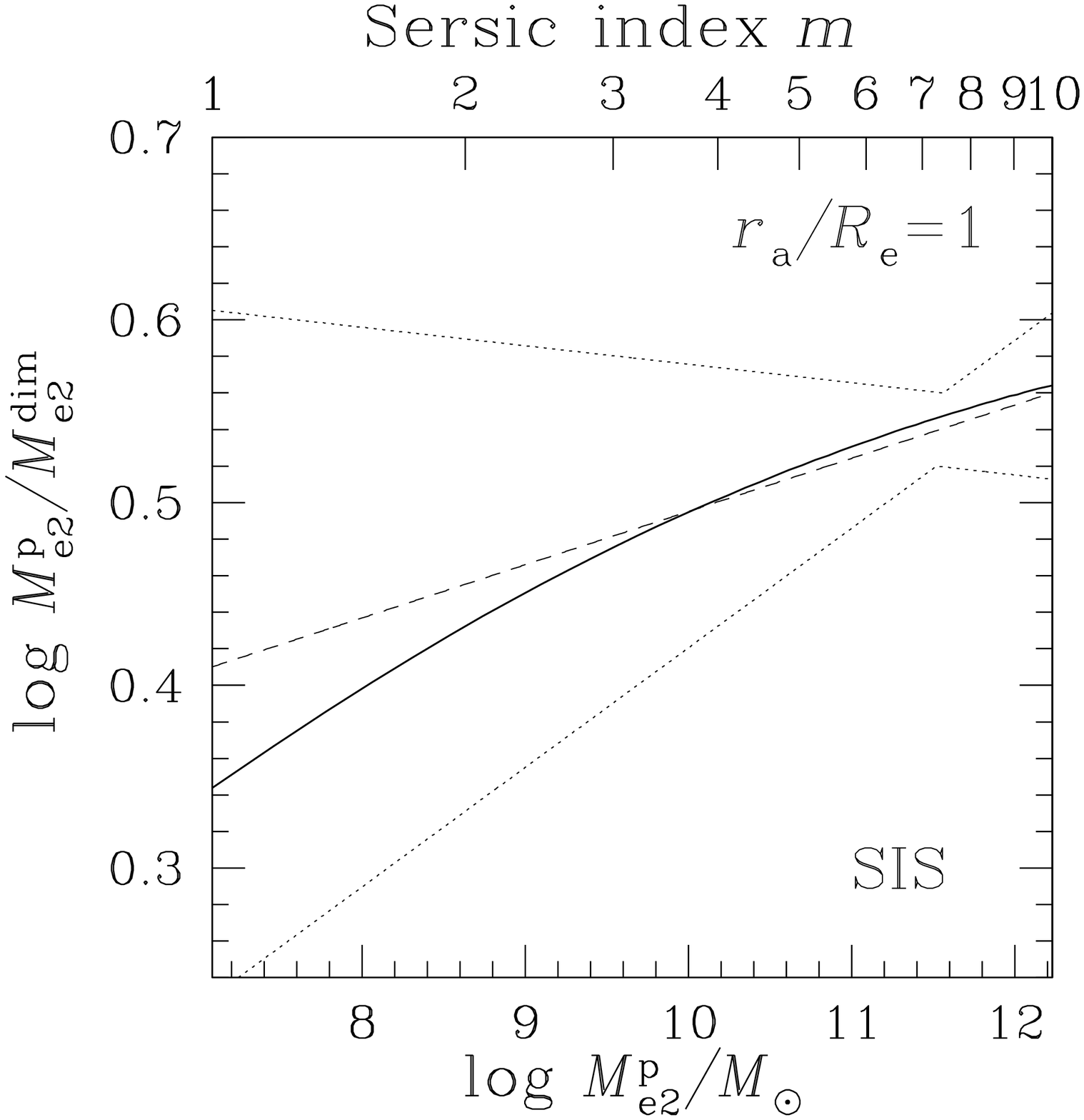}\includegraphics[width=0.32\textwidth]{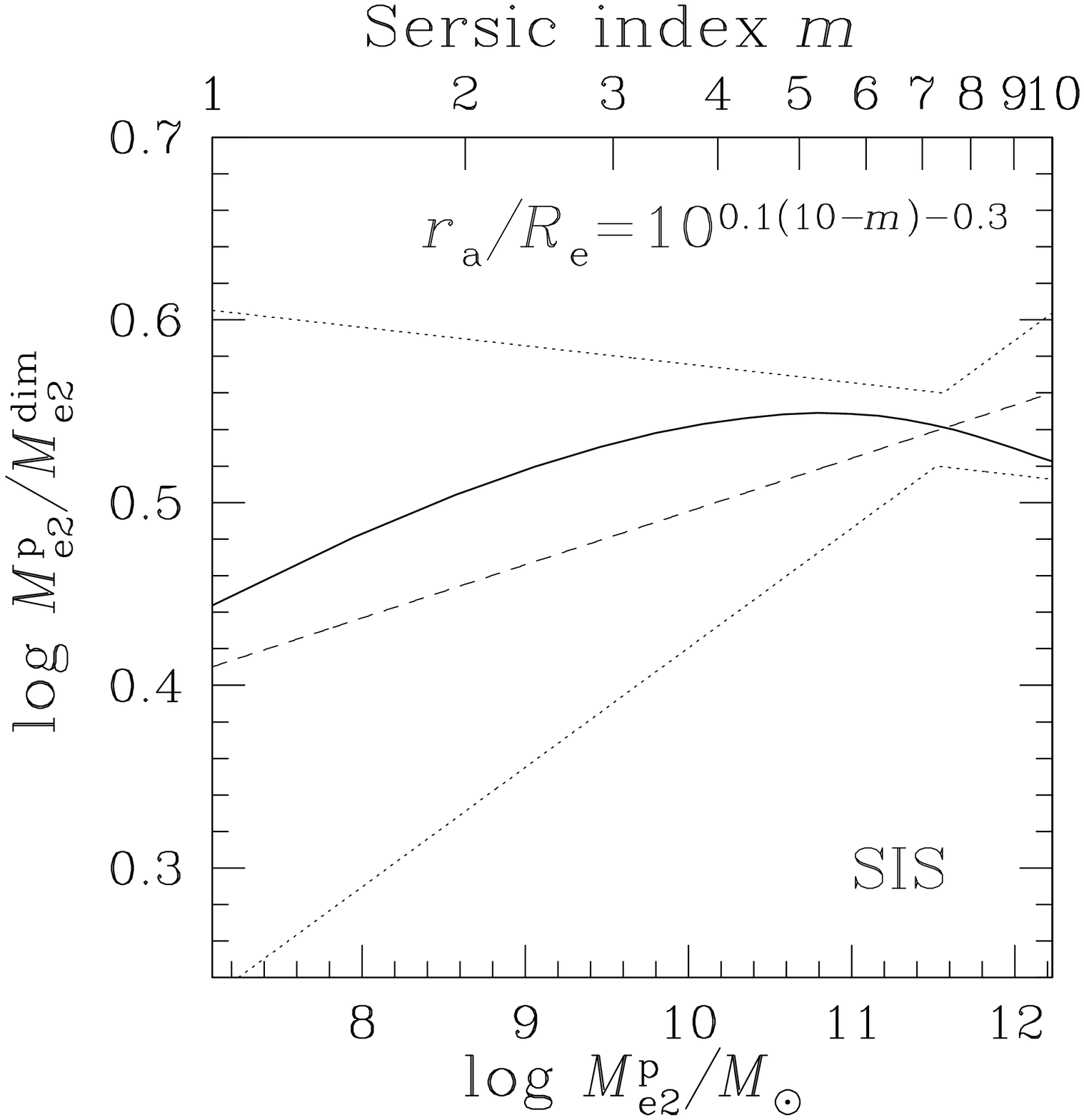}\\
\includegraphics[width=0.32\textwidth]{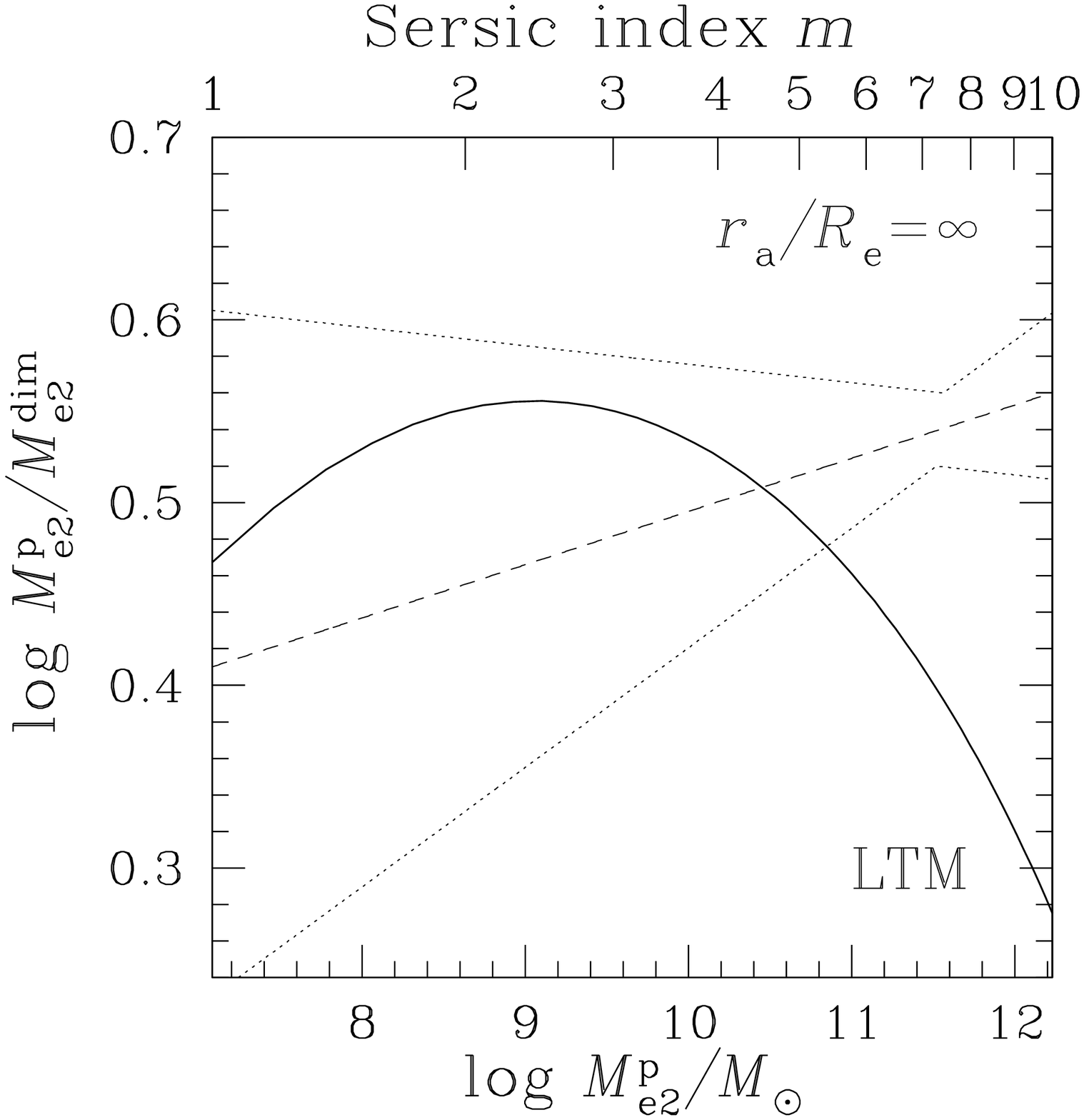}\includegraphics[width=0.32\textwidth]{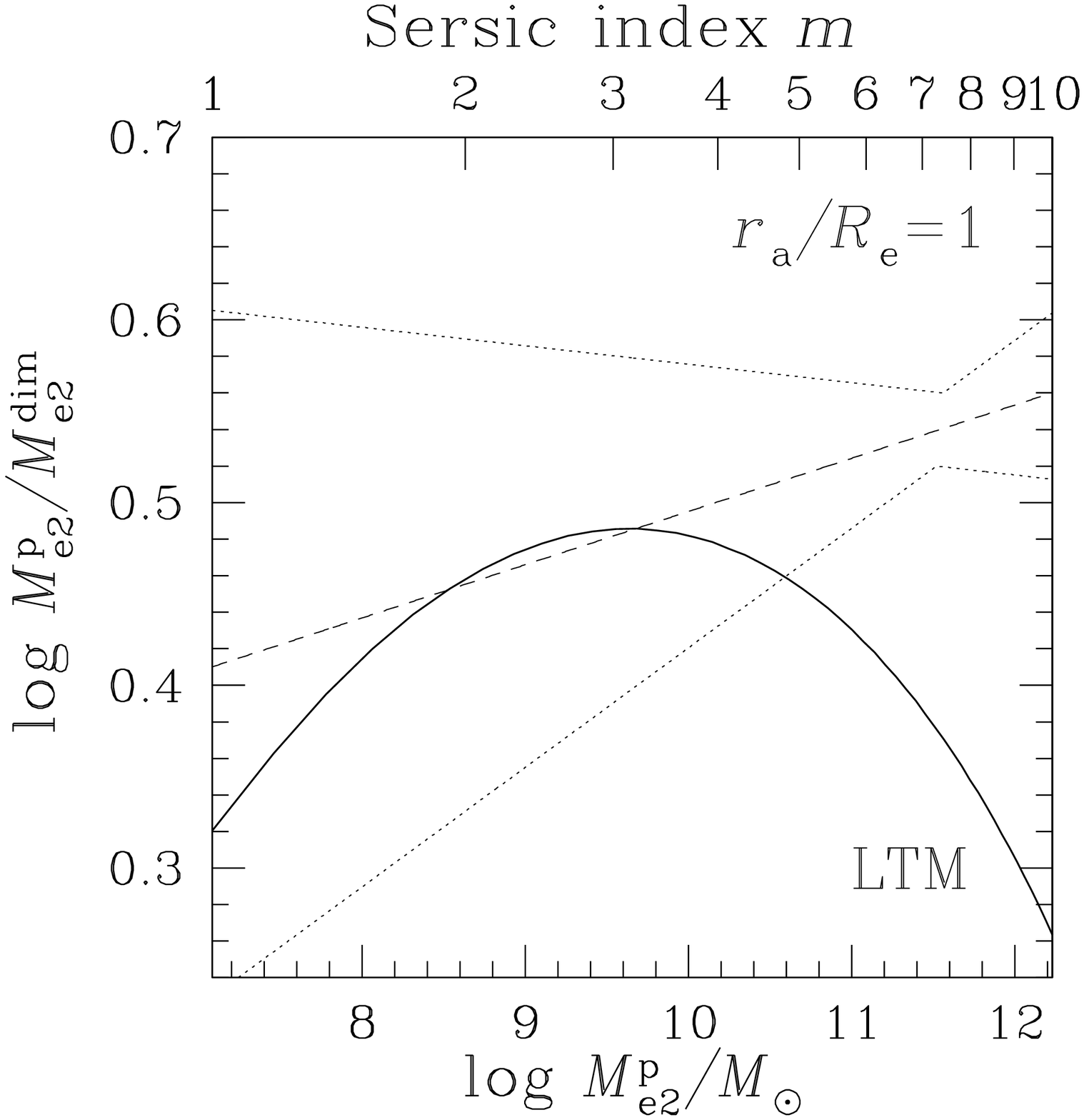}\includegraphics[width=0.32\textwidth]{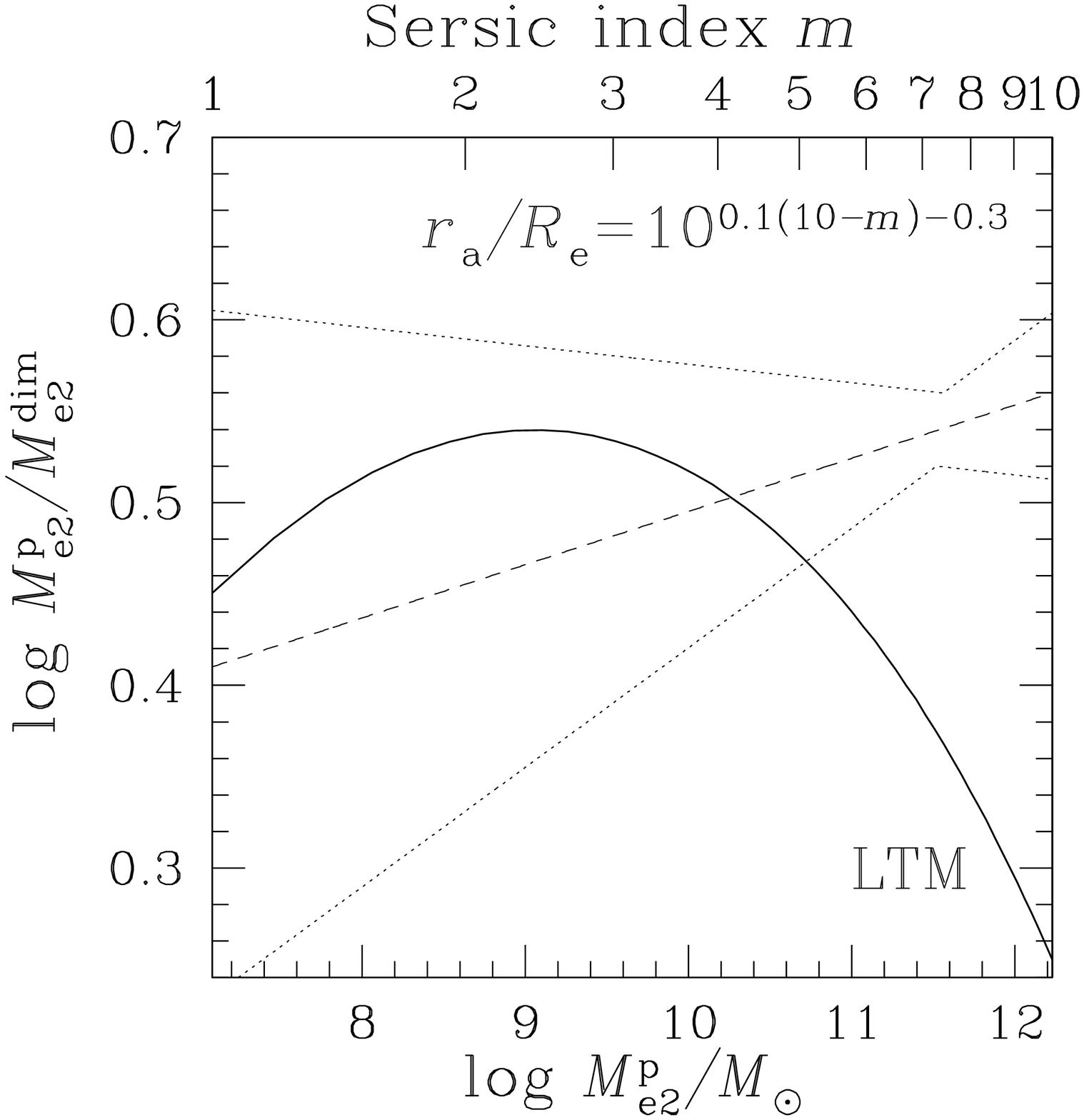}\\
\includegraphics[width=0.32\textwidth]{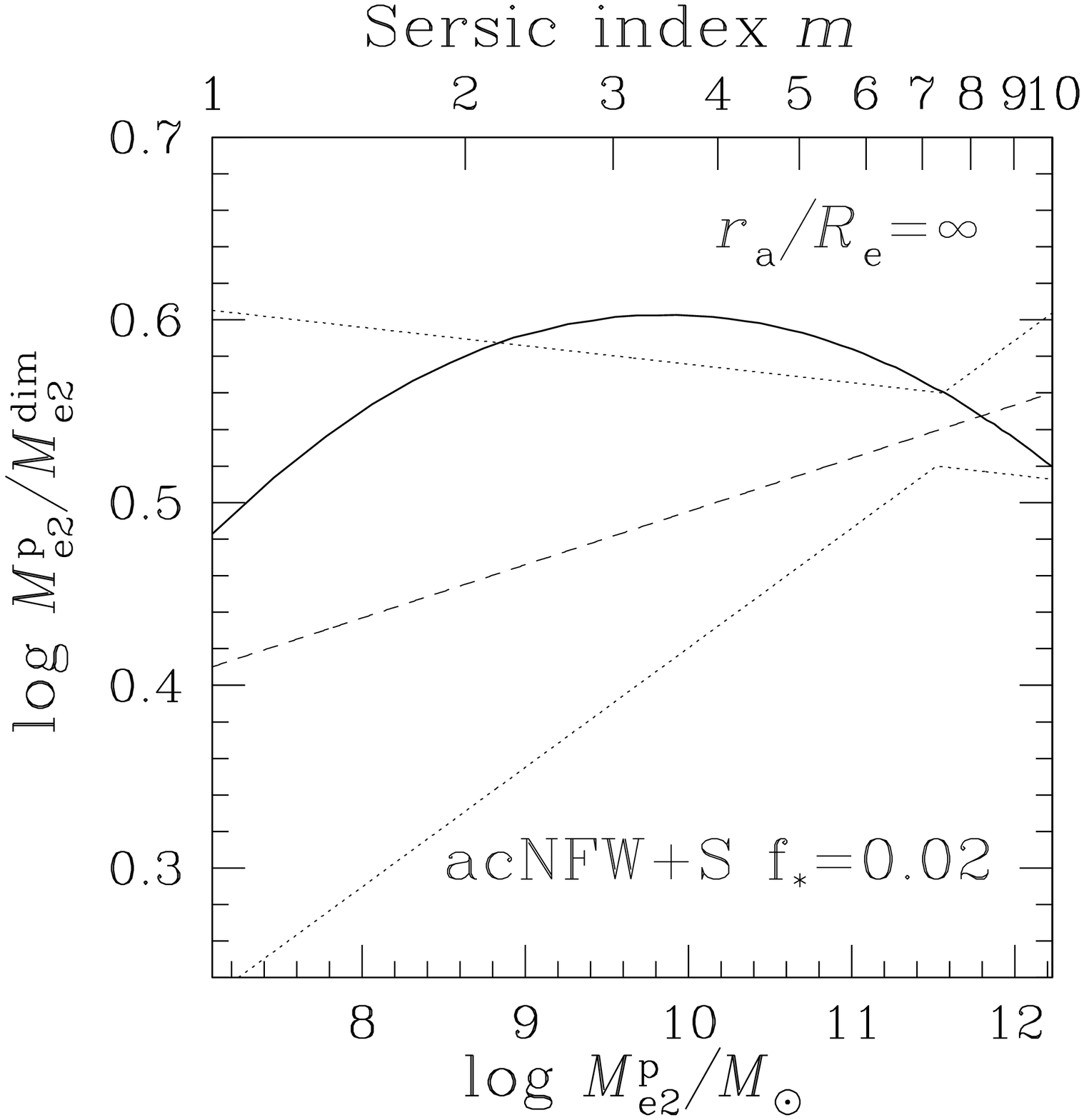}\includegraphics[width=0.32\textwidth]{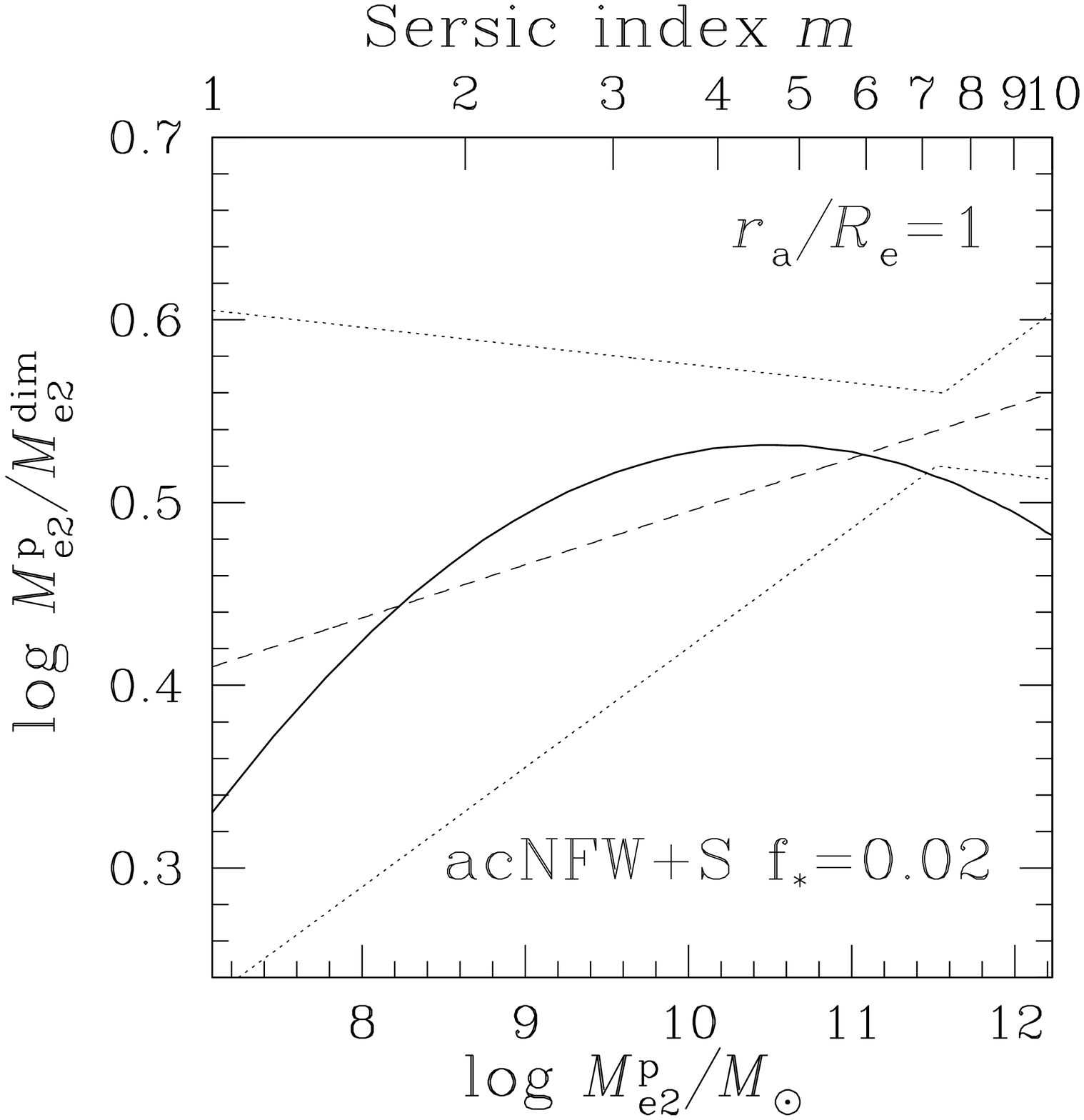}\includegraphics[width=0.32\textwidth]{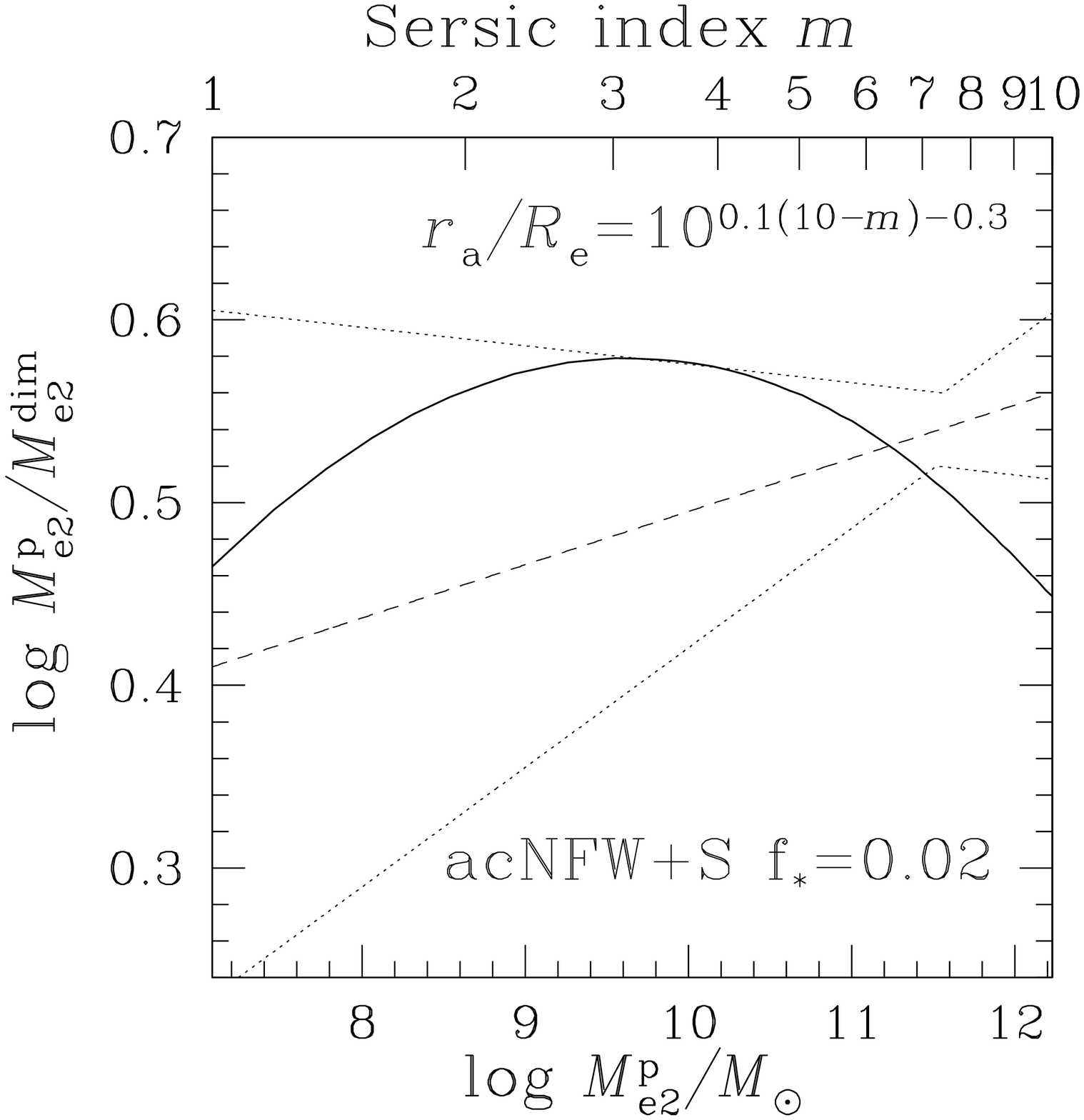}\\
\includegraphics[width=0.32\textwidth]{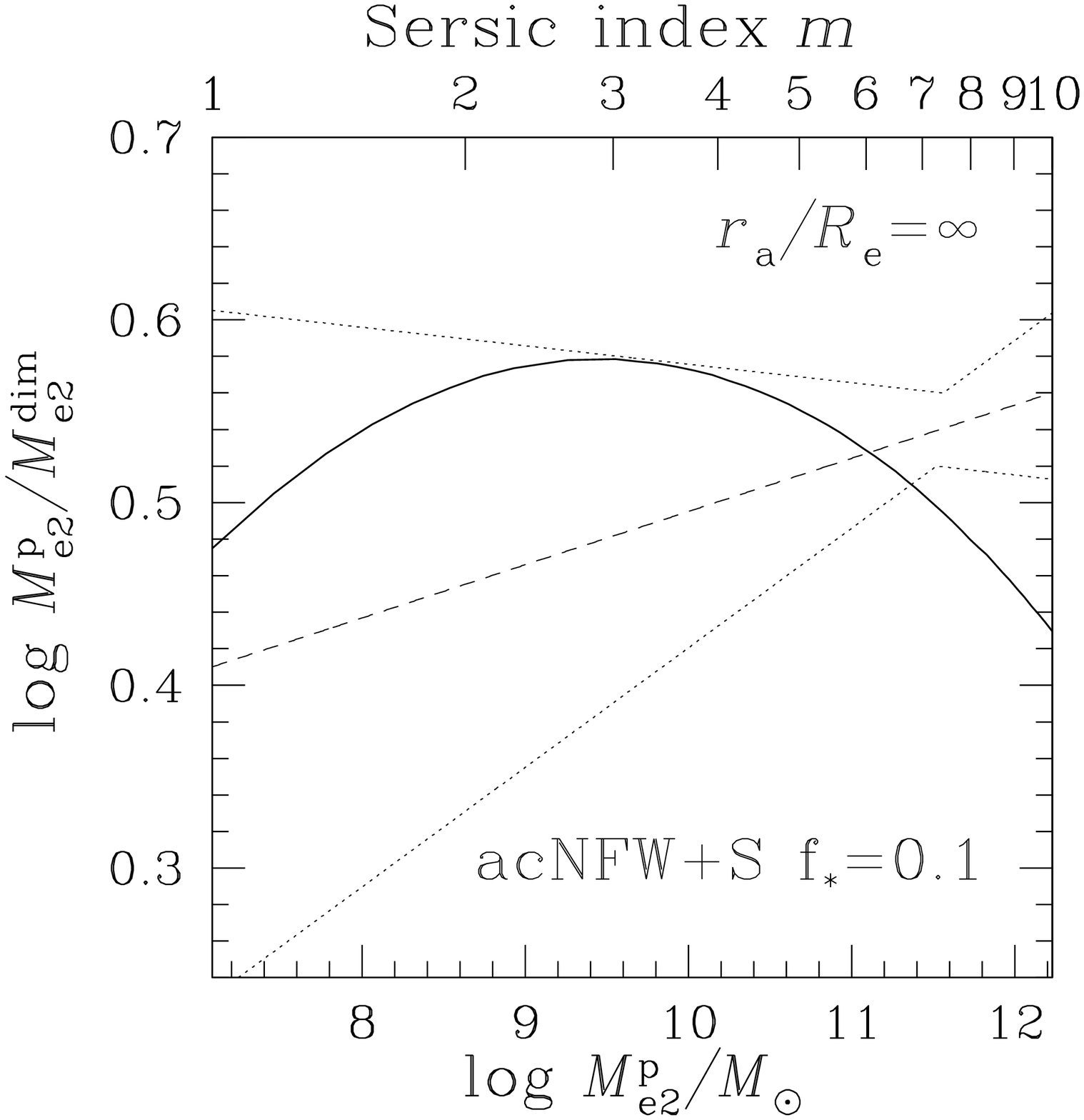}\includegraphics[width=0.32\textwidth]{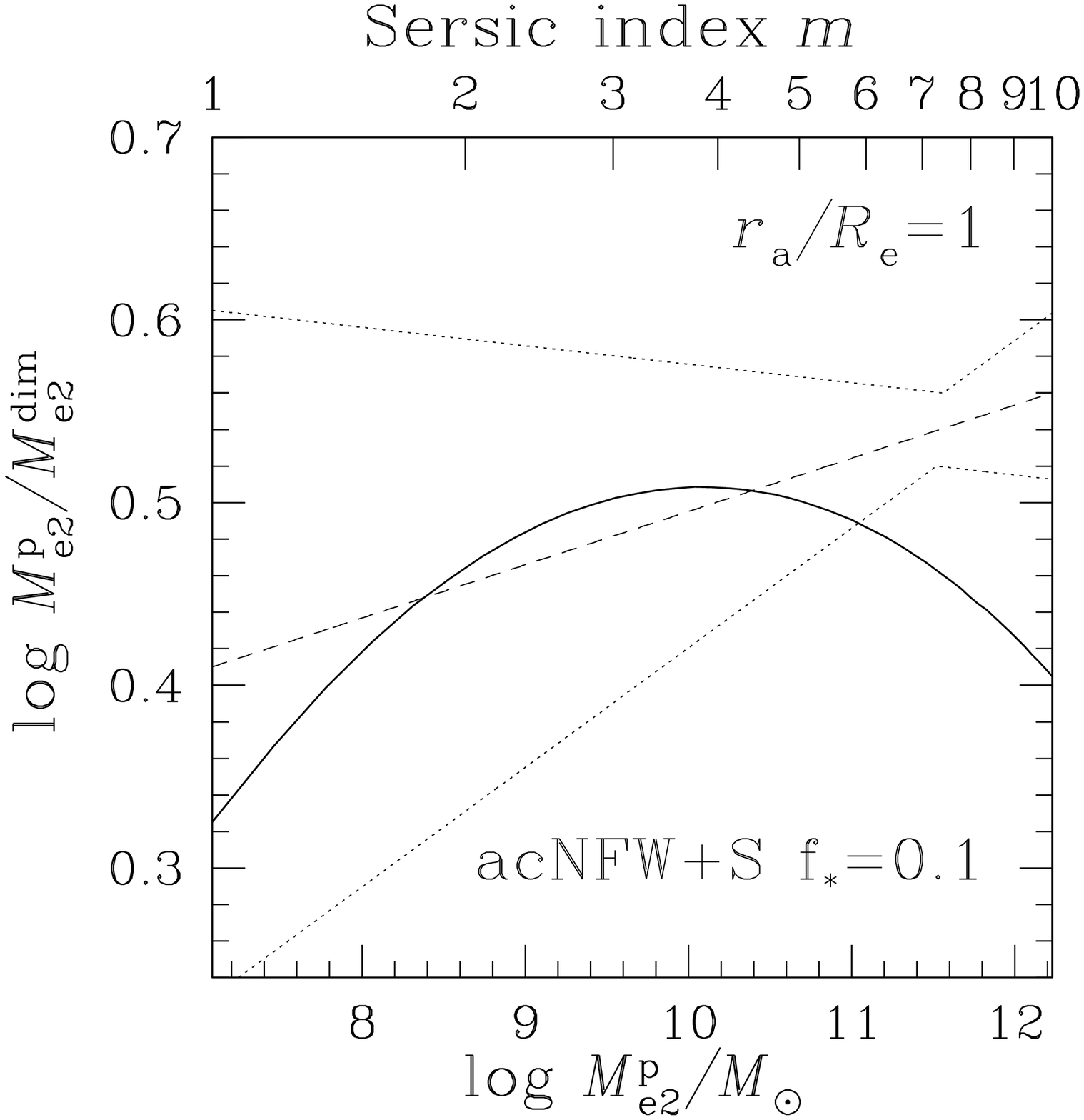}\includegraphics[width=0.32\textwidth]{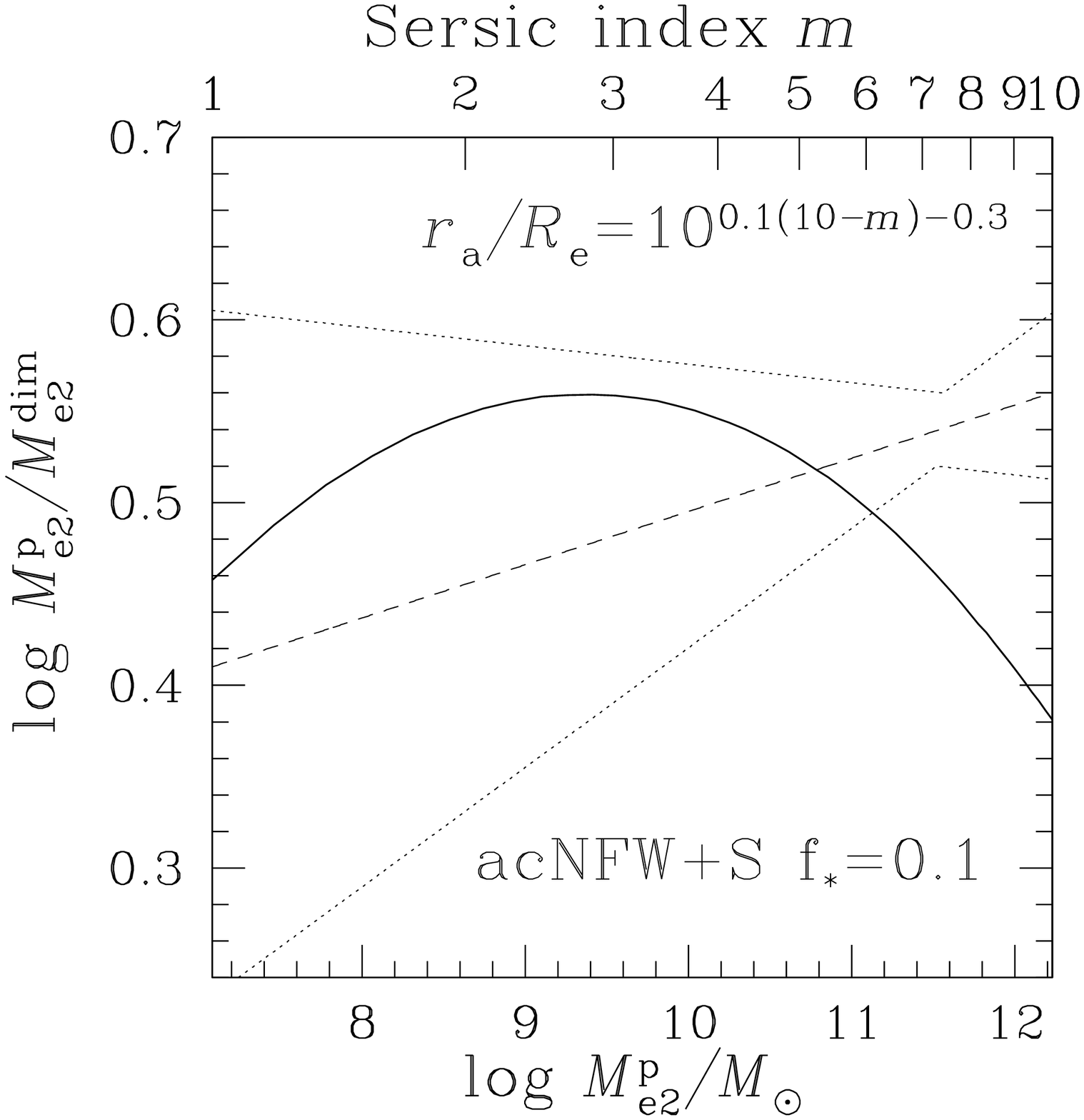}\\
\caption{Ratio between the total projected mass within $\Re/2$
  ($\Metp$) and the dimensional mass $\Metdim=\Re\sgetsq/(2G)$ as a
  function of $\Metp$ for $R^{1/m}$ \Sersic spherical OM galaxy models
  with different total mass distributions and values of $\ra/\Re$. In
  each panel the dashed line is the best-fit observed relation with
  the associated scatter (dotted lines) found by \citet[][see
  text for details]{Bol08b}.}
\label{figtilt}
\end{figure*}

The \Sersic index $m$ of early-type galaxies correlates with galaxy
size, in the sense that more extended galaxies have higher $m$
\citep{Cao93,Gra03,Fer06}: in particular, \citet{Cao93} found
\begin{equation}
\log m = 0.28 + 0.52 \log \left({\Re \over \kpc}\right).
\label{eqmre}
\end{equation}
As the effective radius increases with galaxy luminosity \citep[e.g.][]{Ber03a},
equation~(\ref{eqmre}) should imply a correlation between $m$ and
luminosity.  For example, using surface photometry from the
well-defined Virgo Cluster sample of \citet{Fer06}, we obtain the
following relation
\begin{equation}
\log m = (0.27\pm0.02) \log \left( {{\rm L}_{\rm B}\over {\rm L}_{{\rm B},\odot}}\right) - (2.07\pm0.02)
\label{eqmlum}
\end{equation}
where ${\rm L}_{\rm B}$ is the B-band luminosity obtained from the ACS
filters as described by \citet{Gal08}, and the uncertainties on
the best-fit coefficients are 1 $\sigma$ (the RMS scatter about the
relation is 0.16 dex). Such a systematic variation of $m$ with luminosity
could at least partly contribute to the observed tilt of the FP of
early-type galaxies \citep{Hjo95,CioL97,Gra97,Ber02,Tru04,Cap06}.  If
the stellar mass contributes significantly to the projected mass
within $\Re/2$---as is generally assumed and as supported by the
modeling results above---the dependence of $m$ on luminosity could
introduce a tilt in the MP\@.  This would in turn induce a tilt in the
relation between $\Metp\equiv \Mtotp(\Re/2)$ and
$\Metdim\equiv\Re\sgetsq/(2G)$, because the structure parameter $\cet$
depends on $m$ for given total mass distribution and orbital
anisotropy. For the massive galaxies studied in their sample,
\cite{Bol08b} found
\begin{equation}
\log \left({\Metp\over 10^{11}\Msun}\right)=\delta \left({\Metdim\over 10^{11}\Msun}\right)+ \log \cet, 
\label{eqtilt}
\end{equation}
with $\log \cet=0.54\pm0.02$ and $\delta =1.03 \pm 0.04$. Note that
here we are not considering the average value of $\log \cet$ as in
equation~(\ref{eqobsrange}), but rather are accounting for a possible
dependence of $\cet$ on mass.  A deviation of $\delta$ from unity is a
signature of tilt, so the observational data are consistent with
absence of tilt.  Actually \cite{Bol08b} found no correlation between
galaxy mass/luminosity and \Sersic index in their sample, consistent
with the fact that the SLACS sample is confined to a relatively small
range of galaxy luminosities towards the bright end of the luminosity
function of early-type galaxies: the mass range of Bolton et al.'s
sample is $10.3\lsim \log (\Metp/\Msun) \lsim 12$.  We thus compute
the tilt to determine whether it is measurable and could lead to
further discriminatory power if one considered a sample covering a
larger mass range.  For this purpose, we use our models to quantify
the tilt introduced in the MP by the expected dependence of $m$ on
mass. We consider our models with \Sersic stellar distributions and we
map $m$ into $\Metp$ by combining equation~(\ref{eqmlum}) with the
best-fit correlation \citep{Bol08b}
\begin{equation}
\log \left({\LV\over10^{11}\LVsun}\right)=0.73\log \left({\Metp\over10^{11}\Msun}\right)-0.24,
\label{eqlvmass}
\end{equation}
assuming $B-V=0.96$ \citep{Fuk95}.  In order to isolate the
effect of the structural non-homology of the stellar distribution, it
is useful to compare models under the same assumptions on the velocity
distribution. In this Section we focus on models with radial
anisotropy with OM parameterization.  We consider two families of OM
models with values of $\ra/\Re$ independent of $m$: isotropic models
($\ra/\Re=\infty$) and radially anisotropic models with $\ra=\Re$. As
for the FP, dynamical non-homology might contribute to produce a tilt
of the MP, so it is important to consider also the case in which more
massive (higher-$m$) system are more radially anisotropic than less
massive (lower-$m$) systems. This choice is motivated by the fact that
higher-$m$ system can sustain more radial anisotropy than lower-$m$
systems (see Section~\ref{seccons} and
Figs.~\ref{figcsis}-\ref{figcacnfw10}) and by previous studies on the
FP \citep[e.g.][]{CioLR96,CioL97,Nip02}. Thus, we also consider a
family of models in which $\log \ra/\Re=0.1(10-m)-0.3$: with this
parameterization $m\sim 1$ systems are almost isotropic, while $m\sim
10$ systems are strongly radially anisotropic.  In Fig.~\ref{figtilt}
we plot $\Metp/\Metdim$ as a function of $\Metp$ (and of $m$) for OM
$R^{1/m}$ models with different total mass distribution and for the
three different choices of the value of $\ra/\Re$ (isotropic in the
left-hand panels, $\ra=\Re$ in the central panels and $\ra/\Re$
depending on $m$ in the right-hand panels). In each panel the dashed
line and the dotted lines are, respectively, the best-fit observed
relation (equation~\ref{eqtilt} with $\log \cet=0.54$ and
$\delta=1.03$), and the associated scatter in $\delta$ and $\log
\cet$. In Fig.~\ref{figtilt} we report plots for the SIS, LTM and
acNFW+S models. We note that the acNFW+S models considered in this
case have fixed value of $\fstar$ ($\fstar=0.1$ or $\fstar=0.02$), but
the value of $\xi$ is a function of $m$: for given $m$ we obtain $\Re$
from the observed relation~(\ref{eqmre}) and then $\xi$ from
equation~(\ref{eqxi}), fixing $\Conc=8$.

From the diagrams in Fig.~\ref{figtilt} it is apparent that the SIS
models behave very differently from the LTM and acNFW+S models (NFW+S
models, which are not plotted, behave similarly to acNFW models).  Let
us focus first on the case with $\ra/\Re$ independent of $m$
(left-hand and central panels in Fig.~\ref{figtilt}).  For the SIS
model the ratio $\Metp/\Metdim$ gradually increases with $\Metp$,
while for all other models the ratio $\Metp/\Metdim$ increases with
$\Metp$ at low masses and decreases with $\Metp$ at high masses (in
sharp contrast with the tilt predicted for SIS models), and the
variations with $\Metp$ are stronger than in the case of the SIS
models.  Quantitatively, at high masses, the predicted slope $\delta$
of equation~(\ref{eqtilt}) is $\delta>1$ for SIS models and $\delta<1$
for the other models.  When $\ra/\Re$ depends on $m$ (right-hand
panels in Fig.~\ref{figtilt}), the dynamical non-homology introduces
additional tilt, in the sense that the predicted value of $\delta$ at
high masses becomes smaller, giving $\delta<1$ also for SIS
models. However, also in this case significantly less tilt is
predicted for the SIS models with respect to the other models.


In conclusion, the effect of structural non-homology on the tilt of
the MP is of order of a tenth of a dex in the mass ratio and thus
measurable if one had a sample comparable in size and quality to
SLACS, covering a further decade down in galaxy masses. Interestingly,
the tilt of the MP is measurably different depending on whether or not
the total mass profile is well represented by a SIS. The tilt of the
MP appears thus to be a powerful diagnostic of the internal structure
of early-type galaxies.

\section{Summary and conclusions}  
\label{seccon}

In the range $\sget=175-400$ km s$^{-1}$ the MP of early-type galaxies
has no significant tilt and small associated scatter. This means that
the dimensionless structure parameter $\cet$ defined in
equation~(\ref{eqcet}) is a nearly universal constant. In other words,
the range of values of $\cet$ ``allowed'' by the observational data is
remarkably small. Even for spherical galaxy models, $\cet$ is expected
to depend on the stellar density profile, orbital anisotropy of stars,
and total (dark plus luminous) mass distribution.  In this paper we
explored the constraints posed by the existence of the MP on several
relevant families of galaxy models.\footnote{ Throughout the present
paper we considered the MP in the standard context of Newtonian
gravity with dark matter. See \cite{San08} for an interpretation of
the MP in the context of Modified Newtonian Dynamics.}

Limiting to spherically symmetric models, we found that $\cet$ is not
very strongly dependent on galaxy structure and kinematics, so a
relatively wide class of models have values of $\cet$ within the
observed range.  Therefore, strictly speaking, the massive early-type
galaxies lying on the MP are not necessarily structurally and
dynamical homologous.  However, not all the studied models behave in
the same way when compared to the observational data.  Models in which
the total density profile is a SIS are consistent with the observed
range of $\cet$ for a wide class of stellar density profiles, and only
models with extremely radial or tangential anisotropies are
excluded. The light-traces-mass hypothesis is not excluded by the
observational constraints here considered, apart for the case of
high-$m$ \Sersic models, which cannot be reconciled with the MP within
the observed scatter.  (However, LTM models are known to fail other
observational constraints: see Section~\ref{sectot}). We also
considered cosmologically-motivated models with NFW dark-matter halos
(with or without adiabatic compression), finding that they are
consistent with the MP only for a relatively limited range of
values of their parameters, so a degree of fine-tuning between light
profile and anisotropy is required.  Among these NFW models, those
with adiabatically contracted halos and those that are baryon
dominated seem to require slightly less fine tuning than those with
non-contracted halos and those that are dark-matter dominated.

This work has focused on the average value of $\cet$ in the SLACS
sample, along with its intrinsic scatter.  With the exception of
Section~\ref{secser}, we have not explored the implications of the
fact that this intrinsic scatter is not correlated with either mass or
with the ratio of Einstein radius to $\Re$ \citep{Bol08b}.  These
observational results indicate a degree of structural homogeneity
across a range in mass.  In future works, we will explore these
mass-dependent
results in the context of mass-dynamical models such as those
considered here.  We also plan to refine these analyses based on the
results of forthcoming velocity-dispersion measurements of higher
signal-to-noise ratio and in smaller and more uniform spatial
apertures.

We also explored the possibility that the observed dependence of the
\Sersic index $m$ on the galaxy luminosity could tilt the MP when a
sufficiently large mass range is considered. In this respect, SIS
models behave differently from all other models: a slightly tilted MP
is predicted in the early type galaxies have SIS total density
distribution, while a ``bent'' MP is predicted in all the other
explored cases.  The effect is large enough to be measurable with
sample of lenses comparable to SLACS in size and quality and extending
a further decade in galaxy mass.

In conclusion, our results are consistent with the hypothesis
that massive early-type galaxies have isothermal ($\propto r^{-2}$)
total mass density distribution, though alternative hypotheses cannot
be excluded on the basis of the existence of the MP alone, although in
some cases they require a degree of fine tuning.  In any case, the
process of formation of early-type galaxies lead to systems with a
combination of total mass distribution, luminosity profile, and
orbital anisotropy such that they lie close to the MP.
It will be interesting to quantify whether the observed fine tuning is
quantitatively consistent with the range of simulated properties of
early-type galaxies in the standard hierarchical model of galaxy
formation.

\section*{Acknowledgments}

We acknowledge helpful discussions with Jin An, Giuseppe Bertin, Luca
Ciotti, and Leon Koopmans. T.T.  acknowledges support from the NSF
through CAREER award NSF-0642621, by the Sloan Foundation through a
Sloan Research Fellowship, and by the Packard Foundation through a
Packard Fellowship.  Support for the SLACS project (programs \#10174,
\#10587, \#10886, \#10494, \#10798) was provided by NASA through a
grant from the Space Telescope Science Institute, which is operated by
the Association of Universities for Research in Astronomy, Inc., under
NASA contract NAS 5-26555.


\begin{thebibliography}{}

\bibitem[An \& Evans(2006)]{AnE06}An J.H., Evans N.W., 2006, ApJ, 642,
  752
\bibitem[Barnes(1992)]{Bar92}Barnes J.E., 1992, ApJ, 393, 484
\bibitem[Bender, Burstein \& Faber(1992)]{Ben92} Bender R., Burstein
D., Faber S.M., 1992, ApJ, 399, 462
\bibitem[Bernardi et al.(2003a)]{Ber03a}Bernardi M., et al. 2003a, AJ, 125, 1849
\bibitem[Bernardi et al.(2003b)]{Ber03b}Bernardi M. et al., 2003b, AJ, 125, 1866
\bibitem[Bertin \& Stiavelli(1989)]{Ber89}Bertin G., Stiavelli M., 1989, ApJ, 338, 723
\bibitem[Bertin et al.(1994)]{Ber94}Bertin G. et al., 1994, A\&A, 292, 381
\bibitem[Bertin, Ciotti \& Del Principe(2002)]{Ber02}Bertin G., Ciotti L., Del Principe M., 2002, A\&A, 386, 1491
\bibitem[Binney \& Mamon(1982)]{Bin82}Binney J., Mamon G.A., 1982, MNRAS, 200, 361
\bibitem[Binney \& Tremaine(2008)]{Bin08}Binney J., Tremaine S., 2008, 
         Galactic Dynamics 2nd Ed., Princeton University Press, Princeton 
\bibitem[Blumenthal et al.(1986)]{Blu86}Blumenthal G.R., Faber S.M., Flores R.,
  Primack J.R., 1986, ApJ, 301, 27
\bibitem[Bolton et al.(2006)]{Bol06}Bolton A.S., Burles S., Koopmans
 L.V.E., Treu T., Moustakas L.A., 2006, ApJ, 638, 703
\bibitem[Bolton et al.(2007)]{Bol07}Bolton A.S., Burles S., Treu T.,
Koopmans L.V.E., Moustakas L.A., 2007, ApJ, 665, L105
\bibitem[Bolton et al.(2008a)]{Bol08a}Bolton A.S., et al. 2008a, ApJ,
  in press (arXiv:0805.1931) 
\bibitem[Bolton et al.(2008b)]{Bol08b}Bolton A.S., et al. 2008b, ApJ,
  in press (arXiv:0805.1932) 
\bibitem[Borriello, Salucci \& Danese(2003)]{Bor03} Borriello A., Salucci P., Danese L., 2003, MNRAS, 341, 1109 
\bibitem[Caon, Capaccioli \& D'Onofrio(1993)]{Cao93}Caon N., Capaccioli M., D'Onofrio M., 1993, 265, 1013 
\bibitem[Cappellari et al.(2006)]{Cap06}Cappellari M. et al., 2006,
  MNRAS, 366, 1126
\bibitem[Cappellari et al.(2007)]{Cap07}Cappellari M. et al., 2007,
  MNRAS, 379, 418
\bibitem[Ciotti(1991)]{Cio91}Ciotti L., 1991, A\&A, 249, 99
\bibitem[Ciotti(1994)]{Cio94}Ciotti L., 1994, Celestial Mechanics \&
  Dynamical Astronomy, 60,  401
\bibitem[Ciotti \& Bertin(1999)]{CioB99}Ciotti L., Bertin G., 1999, A\&A, 352, 447
\bibitem[Ciotti \& Lanzoni(1997)]{CioL97}Ciotti L., Lanzoni B., 1997, A\&A, 321, 724 
\bibitem[Ciotti \& Pellegrini(1992)]{CioP92}Ciotti L., Pellegrini S.,
  1992, MNRAS, 255, 561
\bibitem[Ciotti, Lanzoni \& Renzini(1996)]{CioLR96}Ciotti L., Lanzoni B., Renzini A., 1996, MNRAS, 282, 1 
\bibitem[Dehnen(1993)]{Deh93}Dehnen W., 1993, MNRAS, 265, 250
\bibitem[de Vaucouleurs(1948)]{Dev48}de Vaucouleurs G., 1948, Ann. d'Astroph., 11,247
\bibitem[Djorgovsky \& Davis(1987)]{Djo87}Djorgovsky S., Davis M., 1987, ApJ, 313, 59
\bibitem[Dressler et al.(1987)]{Dre87}Dressler A., Faber S.M.,
  Burstein D., Davies R.L., Lynden-Bell D., Terlevich R.J., Wegner, G., 1987, ApJ, 313, 37
\bibitem[{{Dye} {et~al.}(2007){Dye}, {Smail}, {Swinbank}, {Ebeling}, \& {Edge}}]{Dye07}
{Dye} S., {Smail} I., {Swinbank} A.M., {Ebeling} H., {Edge} A.~C.,
  2007, MNRAS, 379, 308
\bibitem[{{Dye} \& {Warren}(2005)}]{Dye05}
{Dye} S., {Warren} S.~J. 2005, ApJ, 623, 31
\bibitem[El-Zant et al.(2004)]{Elz04}El-Zant A., Hoffman Y., Primack J., Combes F., Shlosman I., 2004, ApJ, 607, L75
\bibitem[Faber et al.(1987)]{Fab87}Faber S.M., Dressler A., Davies
  R.L., Burstein D., Lynden-Bell D., 1987, in ``Nearly normal galaxies: From the Planck time to the present'', New York, Springer-Verlag, 1987, p. 175-183.
\bibitem[Ferrarese et al.(2006)]{Fer06}Ferrarese L., et al., 2006, ApJS, 164, 334
\bibitem[Fridman \& Polyachenko(1984)]{Fri84}Fridman A. M.,
  Polyachenko V.L., 1984, Physics of Gravitating Systems (Springer, New York)
\bibitem[Fukugita et al.(1995)]{Fuk95}Fukugita M., Shimasaku K., Ichikawa T., 1995, PASP, 107, 945
\bibitem[{{Gallo} {et~al.}(2008)}]{Gal08}Gallo E., Treu T., Jacob J., Woo J.-H., Marshall P.J., Antonucci R., 2008, ApJ, 680, 154
\bibitem[{{Gavazzi} {et~al.}(2007)}]{Gav07}{Gavazzi} R. {et~al.}, 2007, ApJ, 667, 176
\bibitem[{{Gavazzi} {et~al.}(2008)}]{Gav08}{Gavazzi} R. {et~al.}, 2008, ApJ, 677, 1046
\bibitem[Gerhard et al.(2001)]{Ger01}Gerhard O., Kronawitter A., Saglia R.P., Bender R., 2001, AJ, 121, 1936
\bibitem[Gnedin et al.(2004)]{Gne04}Gnedin O.Y., Kravtsov A.V., Klypin A.A., Nagai D., 2004, ApJ, 616, 16
\bibitem[Graham \& Colless(1997)]{Gra97}Graham A.W., Colless M., 1997, MNRAS, 287, 221
\bibitem[Graham \& Guzm\'an(2003)]{Gra03}Graham A.W., Guzm\'an R., 2003, AJ, 125, 2936
\bibitem[Hernquist(1990)]{Her90}Hernquist L., 1990, ApJ, 356, 359
\bibitem[Hernquist(1993)]{Her93}Hernquist L., 1993, ApJ, 409, 548
\bibitem[Hjorth \& Madsen(1995)]{Hjo95}Hjorth J., Madsen J., 1995, ApJ, 445, 55
\bibitem[Jaffe(1983)]{Jaf83} Jaffe W., 1983, MNRAS, 202, 995
\bibitem[Jiang \& Kochanek(2007)]{Jia07}Jiang G., Kochanek C.S., 2007, ApJ, 671, 1568
\bibitem[Keeton(2001)]{Kee01}Keeton C.R, 2001, 561, 46
\bibitem[Kochanek(1994)]{Koc94}Kochanek C.S., 1994, ApJ, 436, 56
\bibitem[{{Koopmans} \& {Treu}(2002)}]{Koo02}
{Koopmans} L.V.E., {Treu} T., 2002, ApJ, 568, L5
\bibitem[{{Koopmans} \& {Treu}(2003)}]{Koo03}
{Koopmans} L.V.E., {Treu} T., 2003, ApJ, 583, 606
\bibitem[Koopmans et al.(2006)]{Koo06}Koopmans L.V., Treu T., Bolton
  A.S., Burles S., Moustakas L.A., 2006, ApJ, 649, 599
\bibitem[Kronawitter et al.(2000)]{Kro00}Kronawitter A., Saglia R.P., Gerhard
  O., Bender R.,  2000, A\&AS, 144, 53
\bibitem[Lanzoni \& Ciotti(2003)]{Lan03}Lanzoni B., Ciotti L., 2003, A\&A,
  404, 819
\bibitem[Merritt(1985)]{Mer85}Merritt D., 1985, AJ, 90, 102
\bibitem[Merritt \& Aguilar(1985)]{MerA85}Merritt D., Aguilar L.A.,
  1985, MNRAS, 217, 787
\bibitem[Meza \& Zamorano(1997)]{Mez97}Meza A., Zamorano N., 1997, AJ, 490, 136
\bibitem[Navarro, Frenk \& White(1996)]{Nav96}Navarro J.F., Frenk
  C.S., White S.D.M. 1996, ApJ, 462, 563 (NFW)
\bibitem[Neto et al.(2007)]{Net07}Neto A.F. et al., 2007, MNRAS, 381,
  1450
\bibitem[Nipoti, Londrillo \& Ciotti(2002)]{Nip02}Nipoti C., Londrillo P., Ciotti L., 2002, MNRAS, 332, 901
\bibitem[Nipoti et al.(2004)]{Nip04}Nipoti C., Treu T., Ciotti L.,
  Stiavelli M., 2004, MNRAS, 355, 1119
\bibitem[Nipoti, Londrillo \& Ciotti(2006)]{Nip06}Nipoti C., Londrillo
  P., Ciotti L., 2006, MNRAS, 370, 681
\bibitem[Osipkov(1979)]{Osi79}Osipkov L.P., 1979, Soviet Astron. Lett., 5, 42
\bibitem[Pahre,  Djorgovski \& de Carvalho(1998)]{Pah98}Pahre M.A., Djorgovski S.G., de Carvalho R.R., 1998, AJ, 116, 1591
\bibitem[Prugniel \& Simien(1994)]{Pru94}Prugniel Ph., Simien F.,
  1994, A\&A, 282, L1
\bibitem[Renzini \& Ciotti(1993)]{RenC93} Renzini A., Ciotti L., 1993 ApJ, 416, L49
\bibitem[Richstone \& Tremaine(1984)]{Ric84}Richstone D.O., Tremaine S., 1984, ApJ, 286, 27
\bibitem[Riciputi et al.(2005)]{Ric05}Riciputi A., Lanzoni B., Bonoli S., Ciotti L., 2005, A\&A, 443, 133
\bibitem[{{Rusin} \& {Kochanek}(2005)}]{Rus05}
{Rusin} D., {Kochanek} C.S., 2005, ApJ, 623, 666
\bibitem[{{Rusin} {et~al.}(2003){Rusin}, {Kochanek}, \& {Keeton}}]{Rus03}
{Rusin} D., {Kochanek} C.S., {Keeton} C.R., 2003, ApJ, 595, 29
\bibitem[Saglia, Bender \& Dressler(1993)]{Sag93}Saglia R.P., Bender R.,
  Dressler A., 1993, A\&A, 279, 75
\bibitem[Saha(1991)]{Sah91}Saha P., 1991, MNRAS, 148, 494
\bibitem[Sanders \& Land(2008)]{San08}Sanders R.H., Land D.D., 2008,
  submitted to MNRAS (arXiv:0803.0468)
\bibitem[\Sersic(1968)]{Ser68}\Sersic J.L., 1968, Atlas de galaxias australes. Observatorio Astronomico, Cordoba
\bibitem[Shen et al.~(2003)]{She03}Shen S., Mo H.J., White S.D.M.,
  Blanton M.R., Kauffmann G., Voges W., Brinkmann J., Csabai I., 2003, MNRAS, 343, 978
\bibitem[Stiavelli \& Sparke(1991)]{Sti91}Stiavelli M., Sparke L.S., 1991, ApJ, 382, 466
\bibitem[Tremaine et al.(1994)]{Tre94}Tremaine S., Richstone D.O., Yong-Ik B., Dressler A., Faber S.M., Grillmair C., Kormendy J., Laurer T.R., 1994, AJ, 107, 634
\bibitem[{{Treu} \& {Koopmans}(2002)}]{Tre02}
{Treu} T., {Koopmans} L.V.E., 2002, ApJ, 575, 87
\bibitem[{{Treu} \& {Koopmans}(2003)}]{Tre03}
{Treu} T., {Koopmans} L.V.E., 2003, MNRAS, 343, L29
\bibitem[{{Treu} \& {Koopmans}(2004)}]{Tre04}
{Treu} T., {Koopmans} L.V.E., 2004, ApJ, 611, 739
\bibitem[Treu et al.(2006)]{Tre06}Treu T., Koopmans L.V., Bolton A.S., Burles
  S., Moustakas L.A.,  2006, ApJ, 640, 662
\bibitem[Treu et al.(2008)]{Tre08}Treu T., Gavazzi R., Gorecki A.,
  Marshall P.J., Koopmans L.V.E., Bolton A.S., Moustakas L.A., Burles
  S., 2008, ApJ, submitted (arXiv:0806.1056v1)
\bibitem[Trujillo, Burkert \& Bell(2004)]{Tru04}Trujillo I., Burkert A., Bell E.F., 2004, ApJ, 600, L39
\bibitem[van Albada(1982)]{Van82}van Albada T.S., 1982, MNRAS, 201, 939 
\bibitem[{{Wayth} {et~al.}(2005){Wayth}, {Warren}, {Lewis}, \&
    {Hewett}}]{Way05}{Wayth} R.B., {Warren} S.J., {Lewis} G.F.,
  {Hewett} P.C., 2005, MNRAS, 360, 1333
\end{thebibliography}
\end{document}